\documentclass[10pt]{article}

\usepackage{graphicx}
\usepackage{wrapfig}

\usepackage{subcaption}

\usepackage{tabulary}
\usepackage{multirow}
\usepackage[margin=1in]{geometry}
\usepackage{titlesec}
\usepackage[labelfont=bf,textfont=it,font=footnotesize]{caption}
\usepackage{parskip}
\usepackage{mathpazo}
\usepackage[comma,sort&compress,numbers]{natbib}
\usepackage{booktabs}
\usepackage{listings}
\usepackage{float}

\usepackage{amssymb}
\usepackage{amsthm}
\usepackage{amsmath}
\usepackage{nicefrac}        
\usepackage{siunitx}

\graphicspath{{Figures/}}
\DeclareGraphicsExtensions{.pdf,.png,.jpg}

\titlespacing\section{0pt}{6pt plus 1pt minus 0pt}{3pt plus 1pt minus 0pt}
\titlespacing\subsection{0pt}{4pt plus 1pt minus 0pt}{3pt plus 1pt minus 0pt}
\titlespacing\subsubsection{0pt}{4pt plus 1pt minus 1pt}{3pt plus 1pt minus 1pt}
\titleformat{\section}{\large\bfseries\sffamily}{\thesection}{1em}{}
\titleformat{\subsection}{\normalsize\bfseries\sffamily}{\thesubsection}{1em}{}
\titleformat{\subsubsection}{\normalsize\bfseries\sffamily}{\thesubsubsection}{1em}{}

\usepackage[ruled]{algorithm2e}

\usepackage{flushend}

\usepackage{tabulary}
\usepackage{multirow}

\usepackage{array}
\usepackage{enumitem}
\usepackage{todonotes}
\usepackage{url}
\usepackage[T1]{fontenc}

\usepackage{hyperref}
\hypersetup{
    colorlinks,
    linkcolor={magenta},
    citecolor={blue},
    urlcolor={blue!80!black},
    breaklinks=true,
	plainpages=true
}

\newcommand{\cref}[3]{\hyperref[#2]{#1~\ref*{#2}{#3}}}

\newcommand{\colref}[2]{\hyperref[#2]{#1~\ref*{#2}}}

\newcommand{\figref}[1]{\colref{Figure}{#1}}
\newcommand{\subfigref}[2]{\cref{Figure}{#1}{#2}}
\newcommand{\secref}[1]{\colref{Section}{#1}}
\newcommand{\tabref}[1]{\colref{Table}{#1}}
\newcommand{\coloredref}[2]{\hyperref[#2]{#1~\ref*{#2}}}
\newcommand{\coloredsubref}[3]{\hyperref[#2]{#1~\ref*{#2}{#3}}}
\newcommand{\comment}[1]{}

\newcommand{\degree}{\ensuremath{^\circ}}

\begin{document}

\begin{center}
{\usefont{OT1}{phv}{b}{n}\selectfont\Large{Direct Immersogeometric Fluid Flow and Heat Transfer Analysis

of Objects Represented by Point Clouds}}

{\usefont{OT1}{phv}{}{}\selectfont\normalsize
{Aditya~Balu$^1$, Manoj~R.~Rajanna$^1$, Joel~Khristy$^1$, Fei~Xu$^2$, Adarsh Krishnamurthy$^1$*, Ming-Chen~Hsu$^1$*}}

{\usefont{OT1}{phv}{}{}\selectfont\normalsize
{$^1$ Department of Mechanical Engineering, Iowa State University, Ames, Iowa 50011, USA\\
$^2$ Ansys Inc., Austin, Texas 78746, USA\\
}}
\end{center}

\section*{Abstract}
Immersogeometric analysis (IMGA) is a geometrically flexible method that enables one to perform multiphysics analysis \emph{directly} using complex computer-aided design (CAD) models. In this paper, we develop a novel IMGA approach for the simulation of incompressible and compressible flows around complex geometries represented by \emph{point clouds}. The point cloud object's geometry is represented using a set of unstructured points in the Euclidean space with (possible) orientation information in the form of surface normals. Due to the absence of topological information in the point cloud model, there are no guarantees for the geometric representation to be watertight or 2-manifold or to have consistent normals. To perform IMGA directly using point cloud geometries, we first develop a method for estimating the inside-outside information and the surface normals directly from the point cloud. We also propose a method to compute the Jacobian determinant for the surface integration (over the point cloud) necessary for the weak enforcement of Dirichlet boundary conditions. We validate these geometric estimation methods by comparing the geometric quantities computed from the point cloud with those obtained from analytical geometry and tessellated CAD models. In this work, we also develop thermal IMGA to simulate heat transfer in the presence of flow over complex geometries. The proposed framework is tested for a wide range of Reynolds and Mach numbers on benchmark problems of geometries represented by point clouds, showing the robustness and accuracy of the method. Finally, we demonstrate the applicability of our approach by performing IMGA on large industrial-scale construction machinery represented using a point cloud of more than 12 million points.

\subsection*{Keywords}
Immersogeometric analysis $|$
Point clouds $|$
Geometric algorithms $|$
Winding number $|$
Weakly enforced Dirichlet boundary conditions $|$
Nitsche's method

\vspace{0.1in}

\section{Introduction}
In immersogeometric analysis (IMGA), a solid object is immersed into a non-boundary-fitted discretization of the background fluid domain that is used to solve flow physics using finite-element-based computational fluid dynamics (CFD)~\cite{Kamensky15ch,Xu15ig}. IMGA alleviates the labor-intensive and time-consuming geometry cleanup process needed to create a boundary-conforming fluid mesh to perform traditional CFD. For example, in boundary-fitted CFD mesh generation, small and thin geometric features often need to be manually \emph{defeatured}, and the resulting gaps in the boundary representation (B-rep) of the solid model need to be filled to create a watertight surface representation of the solid object. This process can take up to several months for extensive industrial-scale simulations of flow over complex geometries such as tractors and trucks~\cite{Marcum00en,Wang02a,Beall04er,Lee09ku}. 

IMGA has been shown to be a practical method~\cite{Xu15ig,Hsu16fqa,Wang17gp,Xu21im} for the simulation of incompressible flow (both laminar and turbulent) around geometrically complex objects in the context of a tetrahedral finite cell approach~\cite{Varduhn16a}. Since the fluid domain is meshed independently, defeaturing or removing small geometric features from the immersed object is no longer necessary. The method was extended by \citet{Xu19gw,Xu20br} to handle moving particles, by \citet{Zhu20da} to consider free-surface flows, and by \citet{Saurabh21Indus,Saurabh21Scala} to perform industrial scale large eddy simulations using adaptive octree meshes. \citet{Xu19ct} recently proposed a compressible-flow version of the IMGA formulation for the simulation of aircraft aerodynamics, and \citet{Hoang19s} developed a skeleton-stabilized IMGA technique for the simulation of fluid flow through a porous medium. The immersogeometric approach has also been shown as an efficient method to solve computational fluid--structure interaction (FSI) problems for heart valve applications~\cite{Hsu14dh,Hsu15fb,Kamensky17im,Kamensky17pr,Xu18fr,Wu19fm,Johnson20ke,Xu21ci,Johnson22Effec}. It is also flexible enough to be automated and placed in an optimization loop that searches for an optimal design~\cite{Wu16kr}. A subset of the immersogeometric FSI functionality was recently implemented as the open-source library CouDALFISh \cite{Kamensky21op,Neighbor22Lever}.

We have previously developed the immersogeometric method to directly use the B-rep of the computer-aided design (CAD) model to perform fluid flow simulations. We were able to perform IMGA of flow over B-rep models represented using triangles~\cite{Hsu15fb}, trimmed non-uniform rational B-splines (NURBS)~\cite{Hsu16fqa}, and finally analytic surfaces~\cite{Wang17gp}. These approaches were able to handle complex geometry without needing any geometric defeaturing. However, they still worked based on the boundary representation of the well-defined CAD model (2-manifold, watertight, and with consistent normals). This paper extends the work to perform IMGA directly with point cloud representation of the solid geometry, which relaxes these restrictions (manifoldness and watertightness of the boundary representation).

In a point cloud representation, the boundary of the object is represented using a set of unstructured points in the Euclidean space with (possible) orientation information in the form of surface normals. Many geometric data acquisition methods, including optical laser-based scanners, LiDAR scanners, and even passive methods such as multi-view stereo, produce a point cloud representation of the geometry~\cite{berger2017survey}. Further, there has been a high interest in performing flow simulations over \emph{as-manufactured} or \emph{in-use} objects rather than ideal \emph{as-designed} objects for performing analysis (and to provide feedback) using digital twins~\cite{balu2022physics,rausch2021deploying,ghosh2021developing,jafari2017deformation,Kudela2020d}. In practice, we need the geometric information corresponding to the \emph{in-use} physical object to analyze the flow over them. While it is possible to perform IMGA by first reconstructing the boundary surfaces and then using them for the analysis, reconstructing the surfaces to generate watertight and manifold solid models is by itself very challenging. In addition, such an approach would relax the restrictions on the B-rep CAD models obtained directly after design. The geometry no longer needs to be cleaned up to ensure it is watertight; points can be directly sampled from boundary surfaces to perform IMGA.

There are two main geometry processing steps for performing IMGA directly on point cloud representations. First, we need to perform an inside-outside test to conduct a point membership classification (PMC) on the background fluid mesh based only on the point cloud representation of the solid model. Second, we need to perform surface integration over the point cloud to impose the weak Dirichlet boundary conditions~\cite{Bazilevs07c}. Surface integration requires estimating the surface normals and area for each point in the point cloud to be used as the Jacobian determinant for each surface element during integration. In addition, the fluid mesh needs to be adequately refined near the surface locations, i.e., the region around the points of the point cloud. In this paper, we exploit the prior art in geometric processing methods to achieve these steps.

Inside-outside evaluation for the background fluid mesh is necessary to identify points inside the solid geometry. Traditionally, a point membership classification is performed over the solid CAD model. However, these approaches are restricted to two-manifold solids with a watertight surface representation. Since point clouds are not manifold and do not have any inherent order, we propose using the generalized winding-number-based inside-outside testing approach to perform the point membership classification. This method relaxes the manifold and watertight requirements of the surface representation.

As mentioned earlier, the immersed objects used in previous immersogeometric methods consisted of tessellations, trimmed NURBS, or analytic surfaces. In the case of triangles, the Gaussian quadrature points were directly generated on the planar triangular surfaces~\cite{Xu15ig}. In the case of trimmed NURBS, the NURBS parameterization was used to generate the quadrature points~\cite{Hsu16fqa}. Similarly, the surface parametric equations were used to generate the Gaussian quadrature points for trimmed analytic surfaces~\cite{Wang17gp}. In this work, it is convenient for the integration points to be co-located with the points of the point cloud. However, this approach requires calculating the Jacobian of the integration for each point in the point cloud. In this paper, we have developed a Voronoi projection-based area to compute the Jacobian determinant (or the effective area) associated with each point of the point cloud. In addition, we also need the surface normal corresponding to each point in the point cloud. We use a local hyperplane fitting algorithm to compute the surface normals.

In many industrial applications, fluid flow analyses are employed to verify and validate the efficacy of thermal control systems. The thermal analysis predicts the surface and ambient temperatures of critical components in an industrial product assembly. Previous work demonstrated the capability to simulate heat transfer using a variational multiscale method with weakly enforced Dirichlet conditions on conforming boundaries~\cite{Xu19jd}. In this paper, we develop a thermal IMGA formulation to apply fixed temperature boundary conditions weakly on immersed point cloud surfaces, showing convection and conduction of thermal quantities in the flow. We first validate the proposed formulation using benchmark problems and later demonstrate the utility of thermal IMGA on a large vehicle assembly.

To summarize, the specific contributions in this paper include: 1) Methods to compute the surface Jacobian determinant and surface normals from the point cloud representation of a solid model. 2) Methods to compute the inside-outside information for classifying the quadrature points of the background fluid mesh. 3) A thermal IMGA formulation for modeling the heat transfer in flow over an immersed object. 4) Validation studies on the proposed methods to understand their efficacy in performing IMGA with a point cloud. 5) Demonstration of coupled fluid and thermal flow analysis on a large industrial-scale object represented using a point cloud.

This paper is organized as follows. We provide the mathematical formulation of IMGA for incompressible and compressible flows in \secref{Sec:IMGA}. In \secref{Sec:PointClouds}, we provide details on processing the point cloud for computing the normals, Jacobian determinant, and inside-outside information. We also perform validation studies for the point cloud processing methods. Next, we provide the details of the flow and thermal validation studies in \secref{Sec:Validation} and the application of the method to industrial-scale parts represented as point clouds in \secref{Sec:Loader}. Finally, in \secref{Sec:Conclusions}, we conclude our work and provide a few directions for future work.

\section{Immersogeometric analysis}
\label{Sec:IMGA}

The immersogeometric flow analysis methodology consists of three main components: 1) The thermal fluid system is modeled using stabilized finite element methods for incompressible~\cite{Bazilevs07b, Takizawa15bd, Xu19jd} and compressible~\cite{Xu2017c, Codoni20ci, Rajanna22a} flows. 2) The Dirichlet boundary conditions imposed on the immersed objects are enforced weakly in the sense of Nitsche's method~\cite{Nitsche71, Xu15ig, Xu19ct}. 3) To accurately capture the geometry of the flow domain, the concept of the Finite Cell Method (FCM) is employed in which the quadrature rules are adaptively refined~\cite{Duester08.1, Schillinger14.3, Xu15ig}. These numerical ingredients are presented in this section.

\subsection{Mathematical formulation}
\label{SubSec:Formulation}

\begin{figure}[!b]
    \centering
    \includegraphics[width=0.3\linewidth]{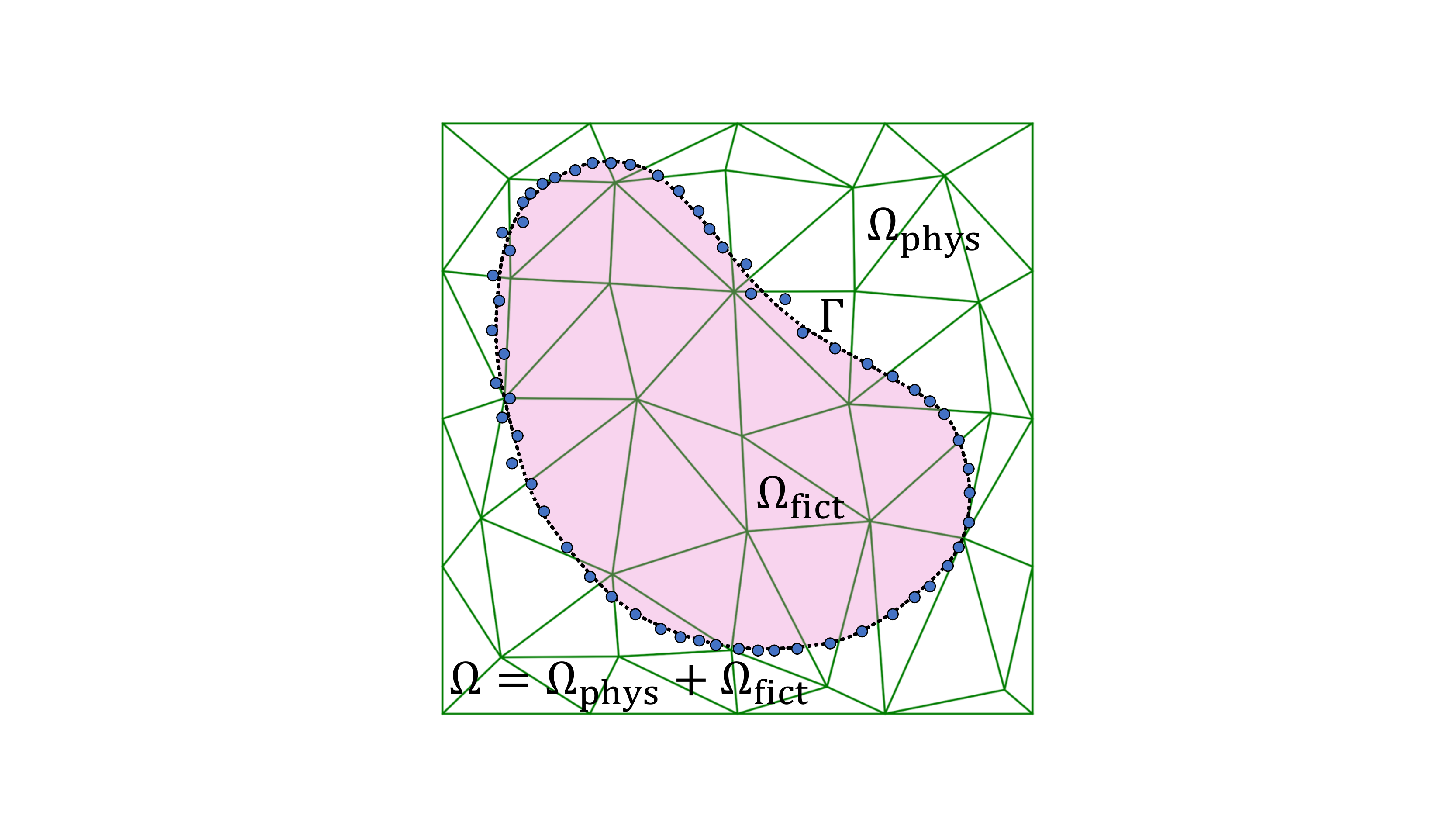}
    \caption{An example of flow over an object. The object with boundary $\Gamma$ is immersed into the domain $\Omega$. The immersed boundary separates the domain $\Omega$ into a physical part $\Omega_{\text{phys}}$ and a fictitious part $\Omega_{\text{fict}}$.}
    \label{fig:FCM_domains}
\end{figure}

\subsubsection{Variational multiscale formulation of the incompressible thermal fluid flow}
\label{SubSec:Imcompressible}

Let $\Omega$ (subsets of $\mathbb{R}^d, d\in\{2,3\}$) denote the spatial domain, and $\Gamma$ be its boundary. In the context of immersogeometric flow analysis, the computational domain $\Omega$ consists of two exclusive parts, the physical domain $\Omega_{\text{phys}}$, i.e., the fluid domain, and the fictitious domain $\Omega_{\text{fict}}$, i.e., the domain enclosed by solid objects. $\Omega_{\text{phys}}$ and $\Omega_{\text{fict}}$ are separated by the immersed boundary $\Gamma$, as shown in \figref{fig:FCM_domains}. Consider a collection of disjoint elements $\{\Omega^e\}$, $\cup_e\Omega^e\subset\mathbb{R}^d$, with closures covering the computational domain, $\Omega\subset\cup_e\overline{\Omega^e}$, and let $\Gamma$ be discretized into a collection of boundary elements $\{\Gamma^b\}$. In what follows, $\Omega^e_\text{phys} = \Omega^e \bigcap \Omega_\text{phys}$ and $\Omega^e_\text{fict} = \Omega^e \bigcap \Omega_\text{fict}$. Note that the finite-element discretization of the domain $\Omega$ is created without conforming to the geometry $\Gamma$, which greatly simplifies the mesh generation process. In what follows, a superscript $h$ indicates that the variable is evaluated in the discrete space. The variational multiscale (VMS) discretization of the incompressible thermal fluid flow problem can be stated as: find the pressure $p^h$, fluid velocity $\mathbf{u}^h$, and temperature $T^h$ in the discrete solution space $\mathcal{S}^h$, such that for all their corresponding test functions $w_p^h$, $\mathbf{w}^h$, and $w_T^h$ in the test function space $\mathcal{V}^h$,
\begin{align}
B^\text{IC}\left(\{w_p^h,\mathbf{w}^h, w_T^h\},\{p^h,\mathbf{u}^h, T^h\}\right)-F^\text{IC}\left(\{w_p^h,\mathbf{w}^h, w_T^h\}\right) = 0\text{,} 
\end{align}
where
\begin{align}
\nonumber & B^\text{IC}\left(\{w_p^h,\mathbf{w}^h, w_T^h\},\{p^h, \mathbf{u}^h,T^h\}\right) = \\
\nonumber &  \int_{\Omega_\text{phys}}\mathbf{w}^h\cdot\rho\left(\frac{\partial\mathbf{u}^h}{\partial t}+\mathbf{u}^h\cdot{\nabla} \mathbf{u}^h\right)~d\Omega  + \int_{\Omega_\text{phys}}\pmb{\varepsilon}(\mathbf{w}^h):\pmb{\sigma}(\mathbf{u}^h,p^h)~d\Omega + \int_{\Omega_\text{phys}}w_p^h{\nabla} \cdot\mathbf{u}^h~d\Omega\\
\nonumber&-\sum_e\int_{\Omega^e_\text{phys}}\left(\rho\mathbf{u}^h\cdot{\nabla} \mathbf{w}^h+{\nabla}w_p^h\right)\cdot\mathbf{u}'~d\Omega-\sum_e\int_{\Omega^e_\text{phys}}p'{\nabla} \cdot\mathbf{w}^h~d\Omega\\
\nonumber&+\sum_e\int_{\Omega^e_\text{phys}}\rho\mathbf{w}^h\cdot(\mathbf{u}'\cdot{\nabla} \mathbf{u}^h)~d\Omega-\sum_e\int_{\Omega^e_\text{phys}} \rho{\nabla}\mathbf{ w}^h:\left(\mathbf{u}'\otimes\mathbf{u}'\right)~d\Omega\\
\nonumber&+\sum_e\int_{\Omega^e_\text{phys}}\rho\left(\mathbf{u}'\cdot{\nabla} \mathbf{w}^h\right)\overline{\tau}\cdot\left(\mathbf{u}'\cdot{\nabla} \mathbf{u}^h\right)~d\Omega\\
\nonumber&+\int_{\Omega_\text{phys}} w_T^h \, \rho c \left(\frac{\partial T^h}{\partial t} + \mathbf{u}^h\cdot \nabla T^h \right) d\Omega+\int_{\Omega_\text{phys}}\nabla w_T^h \cdot \kappa  \nabla T^h d\Omega \\
\nonumber&+\sum_e\int_{\Omega^e_\text{phys}}\nabla w_T^h \cdot \kappa_\text{DC}^\text{IC}  \nabla T^h d\Omega \\
\nonumber&-\sum_e\int_{\Omega^e_\text{phys}} \rho c \left(  \mathbf{u}^h\cdot\nabla w_T^h \right) T' d \Omega +\sum_e\int_{\Omega^e_\text{phys}}  w_T^h \,\rho c   \left( \mathbf{u}' \cdot \nabla T^h\right) d \Omega \\
&- \sum_e\int_{\Omega^e_\text{phys}} \rho c\nabla w_T^h\cdot \left( \mathbf{u}' T' \right) d\Omega,
\end{align}
and
\begin{align}
F^\text{IC}\left(\{w_p^h, \mathbf{w}^h,w_T^h\}\right) =&\int_{\Omega_\text{phys}}\mathbf{w}^h\cdot\rho\mathbf{f}_\text{buoy}~d\Omega 
+ \int_{\Gamma_{\mathbf{u}}^{\text{N}}}\mathbf{w}^h\cdot\mathbf{h}~d\Gamma 
+ \int_{\Gamma_{T}^{\text{N}}} w_T^h h_T~d\Gamma\text{.}
\end{align}
In the above equations, $\rho$, $c$, and $\kappa$ are the density, specific heat capacity, and thermal conductivity of the fluid, respectively. $\pmb{\sigma} (\mathbf{u}^h, p^h) = -p^h\mathbf{I}+2\mu\pmb{\varepsilon}(\mathbf{u}^h)$ and $\pmb{\varepsilon}(\mathbf{u}^h) = \frac{1}{2}({\nabla}\mathbf{u}^h+(\nabla\mathbf{u}^h)^\text{T})$ are the Cauchy stress and strain-rate tensors of incompressible fluid, respectively, with $\mu$ being the dynamic viscosity. $\mathbf{I}$ is the $d\times d$ identity matrix if not otherwise specified. $\mathbf{f}_\text{buoy}$ is the buoyancy force per unit mass modeled using Boussinesq approximation as  $-  \beta \left( T -T_\text{ref} \right)\mathbf{g}$, with $\beta$ being the fluid thermal expansion coefficient, $T_\text{ref}$ being the reference temperature, and $\mathbf{g}$ being the gravitational acceleration. $\mathbf{h}$ and $h_T$ contain the prescribed traction and heat flux conditions, respectively, applied on $\Gamma_{\mathbf{u}}^{\text{N}}$ and $\Gamma_T^{\text{N}}$, which are the portions of $\Gamma$ where the corresponding Neumann boundary conditions are applied. The fine-scale velocity, pressure, and temperature are defined by $\mathbf{u}'=-\tau_\text{M}\mathbf{r}_\text{M}/\rho$, $p' = -\rho \tau_\text{C}r_\text{C}$, and $T' = - \tau_\text{E}r_\text{E}/(\rho c)$, respectively, where
\begin{align}
  &\mathbf{r}_\text{M}= \rho\left(\frac{\partial\mathbf{u}^h}{\partial t}+\mathbf{u}^h\cdot\nabla \mathbf{u}^h\right)- \nabla \cdot \pmb{\sigma}_1\left( \mathbf{u}^h, p^h\right)-\rho\mathbf{f}_\text{buoy}\text{ ,}\\
 &r_\text{C} = \nabla\cdot \mathbf{u}^h\text{,}\\
&r_\text{E} = \rho c\left(\frac{\partial T^h}{\partial t} +\mathbf{u}^h \cdot \nabla T^h\right)  -  \nabla\cdot\left( \kappa \nabla T^h \right)\text{ .}
\end{align}
In the above equations, the stabilization parameters are given by
\begin{align}
&\tau_\text{M} = \left(\frac{C_t}{\Delta t ^2} + \mathbf{u}^h\cdot \mathbf{G} \mathbf{u}^h +C_I \nu^2 \mathbf{G}:\mathbf{G}\right)^{-\frac{1}{2}}\text{,}\\
\label{tauC}
&\tau_\text{C} = \left(\tau_\text{M} \text{tr} \mathbf{G} \right)^{-1},\\
\label{tauM}
&\tau_\text{E} = \left(\frac{C_t}{\Delta t ^2} + \mathbf{u}^h\cdot \mathbf{G} \mathbf{u}^h +C_I \alpha^2 \mathbf{G}:\mathbf{G}\right)^{-\frac{1}{2}}\text{,}\\
\label{tauBar}
&\overline{\tau} = \left(\mathbf{u}'\cdot \mathbf{G}\mathbf{u}' \right)^{-\frac{1}{2}}\text{,}
\end{align}
where $\Delta t$ is the time step size, the constants $C_t = 4$ and $C_I =3$ are chosen from an appropriate element-wise inverse estimation~\cite{Johnson87a,BrSc02}, $\nu = \mu/\rho$ is the kinematic viscosity, $\alpha = \kappa/(\rho c)$ is the thermal diffusivity, 
and $\mathbf{G}$ contains the information about the element size derived from the element geometric mapping from the parametric parent element to physical coordinates $\mathbf{x}(\mathbf{\xi})$. The components of $\mathbf{G}$ are defined as  $G_{ij} = \frac{\partial \xi_k}{\partial x_i} \frac{\partial \xi_k}{\partial x_j}$. The term $\kappa_\text{DC}^\text{IC}$ formulates a discontinuity capturing (DC) operator, which provides additional numerical stability to locations where temperature gradients are large. The DC stabilization parameter~\cite{Xu21c} is given by
\begin{align}
\label{eq::dc_incomp}
\kappa_\text{DC}^\text{IC} = C_\text{DC}^\text{IC} \left(\nabla T^h \cdot \mathbf{G} \nabla T^h\right)^{-\frac{1}{2}} \left|r_\text{E}\right|\text{,}   
\end{align}
where $C_\text{DC}^\text{IC}$ is a positive constant scaling the strength of the DC operator and is set to 0.5 in this paper.

The standard way of imposing Dirichlet boundary conditions is to enforce them strongly by ensuring that these conditions are satisfied by all trial solution functions, which is not feasible in immersed methods. Instead, the strong enforcement is replaced by weakly enforced Dirichlet boundary conditions originally introduced by Bazilevs et al.~\cite{Bazilevs07c,BaMiCaHu07,Bazilevs09d} for incompressible isothermal flows and later extended in Refs.~\cite{Bazilevs15ALE,Takizawa16Comput,Xu19jd} for thermal fluid flows. Let $\Gamma^\text{D} = \Gamma^\text{D}_\mathbf{u}\cup\Gamma^\text{D}_T$ be the portions of $\Gamma$ on which the velocity and temperature Dirichlet boundary conditions are applied. The semi-discrete problem of the thermal fluid system can now be stated as follows:
\begin{align}\label{eq:incomp_wbc}
\nonumber & B^\text{IC}\left(\{w_p^h,\mathbf{w}^h, w_T^h\},\{p^h,\mathbf{u}^h, T^h\}\right)-F^\text{IC}\left(\{w_p^h,\mathbf{w}^h, w_T^h\}\right) \\
\nonumber&-\sum_b\int_{\Gamma^b \bigcap\Gamma^\text{D}_\mathbf{u}}\mathbf{w}^h\cdot\left(-p^h\,\mathbf{n}+2\mu\,\pmb{\varepsilon}(\mathbf{u}^h)\,\mathbf{n}\right) d\Gamma
 -\sum_b \int_{\Gamma^b \bigcap\Gamma^\text{D}_\mathbf{u}}\left(w_p^h\,\mathbf{n} + \tilde{\gamma}\,2\mu\,\pmb{\varepsilon}(\mathbf{w}^h)\,\mathbf{n}\right)\cdot\left(\mathbf{u}^h - \mathbf{u}_\text{D}\right) d\Gamma\\
\nonumber&-\sum_b \int_{\Gamma^b \bigcap\Gamma^{\text{D},-}_\mathbf{u}}\mathbf{w}^h\cdot\rho\left(\mathbf{u}^h\cdot\mathbf{n}\right)\left(\mathbf{u}^h-\mathbf{u}_\text{D}\right) d\Gamma +\sum_b\int_{\Gamma^b\bigcap\Gamma^\text{D}_\mathbf{u}}\mathbf{w}^h\cdot\tau_\mu^\text{IC}\left(\mathbf{u}^h - \mathbf{u}_\text{D}\right) d\Gamma\\
\nonumber&-\sum_e\int_{\Gamma^b\bigcap\Gamma^\text{D}_T} w_T^h\, \kappa \nabla T^h \cdot \mathbf{n} \,d\Gamma - \sum_b\int_{\Gamma^b\bigcap\Gamma^\text{D}_T}\tilde{\gamma}\,\kappa \nabla w_T^h \cdot \mathbf{n} \left(T^h - T_\text{D} \right) d\Gamma\\
&-\sum_b\int_{\Gamma^b\bigcap\Gamma^{\text{D},-}_T}  w_T^h\,\rho c \left(\mathbf{u}^h\cdot\mathbf{n}\right)\left(T^h-T_\text{D}\right) d\Gamma+\sum_b\int_{\Gamma^b\bigcap\Gamma^\text{D}_T} w_T^h\,\tau_\kappa^\text{IC}  \left(T^h-T_\text{D}\right) d\Gamma = 0\text{.}
\end{align}
In the above, $\mathbf{u}_\text{D}$ is the prescribed velocity on $\Gamma^\text{D}_\mathbf{u}$, $T_\text{D}$ is the prescribed temperature on $\Gamma^\text{D}_T$, $\mathbf{n}$ is the unit outward normal vector, and $\Gamma^{\text{D},-}$ is the \emph{inflow} part of $\Gamma^\text{D}$, on which $\mathbf{u}^h\cdot\mathbf{n} < 0$. The value of $\tilde{\gamma}$ can be selected as 1 or $-1$ which determines whether Eq.~\eqref{eq:incomp_wbc} is a symmetric or non-symmetric type of Nitsche's method, respectively~\cite{Bazilevs07c,Schillinger16bs,Xu19ct}. Finally, $\tau_\mu^\text{IC}$ and $\tau_\kappa^\text{IC}$ are stabilization parameters that need to be estimated element-wise as a compromise between the conditioning of the stiffness matrix, variational consistency, and the stability of the formulation. The choice of $\tilde{\gamma}$ influences the performance of the weak boundary condition operator and the selection of $\tau_\mu^\text{IC}$ and $\tau_\kappa^\text{IC}$, which will be discussed in detail in Section~\ref{SubSec:selection_tau}.

An important advantage of using weakly enforced Dirichlet boundary conditions is the release of the point-wise no-slip condition at the boundary of the fluid domain. This, in turn, allows the flow to slip slightly on the solid surface and imitates the presence of the thin boundary layer that typically needs to be resolved with spatial refinement. It was shown in~\citet{Bazilevs09d} and~\citet{HsuAkk12b} that weak boundary conditions allow for an accurate overall flow solution even if the mesh size in the wall-normal direction is relatively large. In the immersogeometric method, the fluid mesh is arbitrarily cut by the object boundary, leaving a boundary layer discretization of inferior quality compared to the boundary-fitted counterpart. However, it was shown in~\citet{Xu15ig} that accurate flow solutions were obtained using the immersogeometric method with a mesh resolution and refinement pattern comparable to the boundary-fitted mesh used to obtain the reference values. We believe this is partially due to the use of weak-boundary-condition formulation.

\subsubsection{Compressible flow formulation}
\label{SubSec:compressible}
The compressible-flow governing equations are discretized using a streamline upwind Petrov--Galerkin (SUPG) formulation~\cite{Brooks82a, Hughes84a, Shakib91a, LeBeau92a, Aliabadi92a, HauHug94, Tezduyar94b, Wren95a, Wren97a, Mittal98a, Ray00a, hauke2001simple, Hughes08a, Takizawa17ec, Kanai18a} augmented by a DC operator~\cite{Tezduyar86a, Hughes86c, Hughes86e, almeida1996adaptive, HauHug98, Tezduyar04o, Tezduyar05f, Tezduyar05d, Rispoli07a, Rispoli08a, Rispoli15a, Takizawa17e}. In what follows, Roman indices take on values $\{1,...,d\}$, and summation convention on repeated indices is applied. In addition, we use $(\cdot)_{,t}$ to denote a partial time derivative, and we use $(\cdot)_{,i}$ to denote the spatial gradient. Let $\mathbf{Y}=[p\; \mathbf{u}\; T]^\text{T}$ denote the solution vector of pressure, velocity and temperature, and $\mathbf{W}=[w_p\; \mathbf{w}\; w_T]^\text{T}$ denote the test function vector of their respective test functions. The problem can be stated as follows: find $\mathbf{Y}^h\in\mathcal{S}^h$ such that for all  $\mathbf{W}^h\in\mathcal{V}^h$,
\begin{align}\label{eq:weak_comp}
B^\text{C}\left(\mathbf{W}^h,\mathbf{Y}^h\right)-F^\text{C}\left(\mathbf{W}^h\right
)=0\text{ ,}
\end{align}
where
\begin{align}\label{eq:weak_B_comp}
 \nonumber B^\text{C}\left(\mathbf{W}^h,\mathbf{Y}^h\right) = &\int_{\Omega_\text{phys}}\mathbf{W}^h\cdot\left( \mathbf{A}_0\mathbf{Y}^h_{,t}+\mathbf{A}^{\text{adv} \backslash \text{p}}_i\mathbf{Y}^h_{,i} +\mathbf{A}_i^\text{sp}\mathbf{Y}^h_{,i}\right)~d\Omega\\
 - & \nonumber \int_{\Omega_\text{phys}}\mathbf{W}^h_{,i}\cdot \left( \mathbf{A}^p_i \mathbf{Y}^h -\mathbf{K}_{ij}\mathbf{Y}^h_{,j} \right)~d\Omega \\
 +& \nonumber\sum_{e}\int_{\Omega^e_\text{phys}} \left(\left(\mathbf{A}_i+\mathbf{A}_i^\text{sp}\right)^{\text{T}}\mathbf{W}^h_{,i}\right)\cdot \left(\mathbf{A}_0^{-1}\hat{\pmb{\tau}}_{\text{SUPG}}\right) \mathbf{Res}\left(\mathbf{Y}^h\right)~d\Omega \\
 +&\sum_{e}\int_{\Omega^e_\text{phys}} \mathbf{W}^h_{,i}\cdot \left(\hat{\kappa}_\text{DC}^\text{C} \mathbf{A}_0\right)\mathbf{Y}^h_{,i}~d\Omega\text{,}
\end{align}
and
\begin{align}
F^\text{C}\left(\mathbf{W}^h\right) =  &\int_{\Omega}\mathbf{W}^h\cdot \mathbf{S}~d\Omega+\int_{\Gamma_\mathbf{H}}\mathbf{W}^h\cdot\mathbf{H}~d\Gamma\text{.}
\end{align}
In the above, $\mathbf{A}$'s and $\mathbf{K}_{ij}$ are the Euler Jacobian matrices and the diffusivity matrix, respectively, whose specific definitions can be found in~\ref{App:AppendixA}. Note that the superscript $h$ for $\mathbf{A}$'s and $\mathbf{K}_{ij}$ are dropped for the clarity of notation, even though they are evaluated in the discrete space. $\mathbf{H}$ contains the prescribed fluid traction and heat flux boundary conditions, and $\Gamma_\mathbf{H}$ is the subset of $\Gamma$ where $\mathbf{H}$ is specified. $\mathbf{Res}$ is the residual of the compressible-flow equations defined as
\begin{align}
\label{primitive-residual}
\mathbf{Res}\left(\mathbf{Y}^h\right) = \mathbf{A}_0\mathbf{Y}^h_{,t}+\left(\mathbf{A}_i+\mathbf{A}_i^{\text{sp}}\right)\mathbf{Y}^h_{,i} - \left(\mathbf{K}_{ij}\mathbf{Y}^h_{,j}\right)_{,i} - \mathbf{S}\text{ .}
\end{align}
The stabilization matrix $\hat{\pmb{\tau}}_{\text{SUPG}}$ is defined as
\begin{align}
\label{tauhat}
\hat{\pmb{\tau}}_{\text{SUPG}} = \left(\frac{C_t}{\Delta t ^2} \mathbf{I} + G_{ij}\left(\hat{\mathbf{A}}_i+\hat{\mathbf{A}}_i^{\text{sp}}\right)\left(\hat{\mathbf{A}}_j+\hat{\mathbf{A}}_j^{\text{sp}}\right) + C_IG_{ij} G_{kl} \hat{\mathbf{K}}_{ik}\hat{\mathbf{K}}_{lj}\right)^{-\frac{1}{2}}\text{.}
\end{align}
Note that in Eq~\eqref{tauhat}, the identity matrix $\mathbf{I}$ is $(d+2)\times(d+2)$, and the notation $\hat{(\,\cdot\,)}$ on top of the matrices indicates that they are evaluated based on the governing equations using conservation variables. The term associated with $\hat{\kappa}_\text{DC}^\text{C}$ in Eq.~\eqref{eq:weak_B_comp} is a DC operator, where the DC stabilization parameter is given by
\begin{align}
\hat{\kappa}_\text{DC}^\text{C} = C_\text{DC}^\text{C}\left(G_{ij} {\mathbf{U}^h_{,i}}^\text{T}\mathbf{\tilde{A}}_0^{-1}\mathbf{U}^h_{,j}\right)^{-\frac{1}{2}}\left({\mathbf{Res}(\mathbf{Y}^h)}^\text{T}\mathbf{\tilde{A}}_0^{-1}\mathbf{Res}(\mathbf{Y}^h)\right)^{\frac{1}{2}}\text{.}
\label{eq:kappadc}
\end{align}
In the above, $C_\text{DC}^\text{C} $ is a $\mathcal{O}(1)$ positive constant ($C_\text{DC}^\text{C} =0.5$ in this work), and $\tilde{\mathbf{A}}_0^{-1}$ is the inverse of the zeroth Euler Jacobian of the transformation between the conservation and entropy variables (see \citet[Appendix~A]{Rajanna22Fluid}). Equation~\eqref{eq:kappadc} is an extension of the $\delta_{91}$ definition designed by Tezduyar and colleagues \cite{LeBeau91a,LeBeau92a}, where only the convective part of the full residual operator $\mathbf{Res}(\mathbf{Y}^h)$ was employed. While the SUPG terms provide the necessary stability across a wide range of Reynolds numbers, the DC operator provides the necessary additional dissipation in the shock regions. Finally, the compressible-flow version of the weak-boundary-condition formulation is added to the weak form:
\begin{align}\label{eq:comp_wbc}
\nonumber & B^\text{C}\left(\mathbf{W}^h,\mathbf{Y}^h\right)-F^\text{C}\left(\mathbf{W}^h\right) \\
&\nonumber-\sum_{b}\int_{\Gamma^b\cap\Gamma^\text{D}_\mathbf{u}}\mathbf{w}^h\cdot\left(-p^h\mathbf{n}+\left(\lambda \nabla\cdot\mathbf{u}^h\right)\mathbf{n} + 2\mu\pmb{\varepsilon}(\mathbf{u}^h)\mathbf{n}\right)\text{d}\Gamma\\ &\nonumber-\sum_{b}\int_{\Gamma^b\cap\Gamma^\text{D}_\mathbf{u}} \left(\rho^h w_p^h \mathbf{n}+\tilde{\gamma}\left(\left(\lambda \nabla\cdot\mathbf{w}^h\right)\mathbf{n} + 2\mu\pmb{\varepsilon}(\mathbf{w}^h)\mathbf{n}\right)\right)\cdot\left(\mathbf{u}^h-\mathbf{u}_\text{D}\right)~\text{d}\Gamma\\
&\nonumber -\sum_{b}\int_{\Gamma^b\cap\Gamma^{\text{D},{-}}_\mathbf{u}}\mathbf{w}^h\cdot\rho^h\left(\mathbf{u}^h\cdot\mathbf{n}\right)\left(\mathbf{u}^h-\mathbf{u}_\text{D}\right) ~\text{d}\Gamma \\
&\nonumber+ \sum_{b}\int_{\Gamma^b\cap\Gamma^\text{D}_\mathbf{u}}\mathbf{w}^h\cdot \tau_\mu^\text{C} (\mathbf{u}^h-\mathbf{u}_\text{D})~\text{d}\Gamma
+ \sum_{b}\int_{\Gamma^b\cap\Gamma^\text{D}_\mathbf{u}}\left(\mathbf{w}^h\cdot \mathbf{n}\right) \tau_\lambda^\text{C} \left(\left( \mathbf{u}^h- \mathbf{u}_\text{D}\right)\cdot\mathbf{n}\right)~\text{d}\Gamma \\
&\nonumber- \sum_{b}\int_{\Gamma^b\cap\Gamma^\text{D}_T} w_T^h\, \kappa \nabla T^h\cdot\mathbf{n}~\text{d}\Gamma 
-\sum_{b}\int_{\Gamma^b\cap\Gamma^\text{D}_T}\tilde{\gamma}\, \kappa \nabla w_T^h \cdot\mathbf{n}\left(T^h-T_\text{D}\right)~\text{d}\Gamma\\ &-\sum_{b}\int_{\Gamma^b\cap\Gamma^{\text{D},{-}}_T}w_T^h\, \rho^h c_\text{v} \left(\mathbf{u}^h\cdot\mathbf{n}\right)\left(T^h-T_\text{D}\right)~\text{d}\Gamma
+ \sum_{b}\int_{\Gamma^b\cap\Gamma^\text{D}_T}w_T^h\, \tau_\kappa^\text{C} (T^h-T_\text{D})~\text{d}\Gamma = 0\text{ ,}
\end{align}
where $\lambda$ is the second coefficient of viscosity ($\lambda = - 2 \mu /3$ based on Stokes' hypothesis), and $\tau_\mu^\text{C}$, $\tau_\lambda^\text{C}$, and $\tau_\kappa^\text{C}$ are the stabilization parameters of Nitsche's method.  In this work, the compressible gas is assumed to be calorically perfect and the specific heats at constant volume and constant pressure can be defined as $c_\text{v} = R/(\gamma-1)$ and $c_\text{p} = \gamma R/(\gamma-1)$, respectively, where $R$ is the ideal gas constant and $\gamma$ is the heat capacity ratio. The pressure, density, and temperature are related through the ideal gas equation of state, $p = \rho R T$.  In addition, thermal conductivity can be calculated by $\kappa = c_\text{p} \mu/Pr$, where $Pr$ is the Prandtl number.

\subsubsection{Stabilization parameter selection for weak boundary conditions}\label{SubSec:selection_tau}
Eqs.~\eqref{eq:incomp_wbc} and~\eqref{eq:comp_wbc} can be either the symmetric or non-symmetric form of Nitsche's method, depending on the value of $\tilde{\gamma}$ being selected as 1 or $-1$, respectively. The symmetric Nitsche method provides excellent accuracy and robustness when the stabilization parameters are properly estimated, which sometimes requires the solution of a local eigenvalue problem~\cite{Embar10, RueSchOzc14a, jiang2015r, dePrenter18note, Divi22Resid}. However, the arbitrarily intersected elements in immersed methods make obtaining accurate solutions to the eigenvalue problem challenging. An intuitive way to circumvent this difficulty is to use a uniform value that can be determined by a trial-and-error approach to simultaneously satisfy the requirements for numerical stability and conditioning of the system matrix~\cite{Kamensky15ch, Xu15ig}. Some delicately designed algorithms have been proposed for the local selection of stabilization parameters in intersected elements~\cite{harari2015u, jiang2015r} and for preconditioning of the linear system \cite{dePrenter17Condi, dePrenter19Preco, dePrenter20Multi}. However, even with the added algorithmic complexity, careful estimations of $\tau$'s do not necessarily result in improved accuracy in $L^2$ errors when compared with the uniform-valued stabilization parameters~\cite{Schillinger16bs}. Therefore, in our IMGA CFD for incompressible flows, we set $\tau_\mu^\text{IC} = 10^3$ and $\tau_\kappa^\text{IC} = c \tau_\mu^\text{IC}$ to take advantage of this most straightforward but effective strategy.

Our previous numerical experiments suggest that, compared with incompressible flows, IMGA CFD for compressible flows is more sensitive to the oscillations around immersed boundaries caused by the uniform, large-value stabilization parameters. With this observation, the non-symmetric Nitsche type weak boundary condition operator~\cite{OdeBab98, RiWhGi01, ArnEtal02, Hartmann07Adjoi, Schillinger16bs, Xu19ct} becomes an attractive alternative. The non-symmetric Nitsche formulation can be parameter-free \cite{BauOd99, Baumann99vc, Burman12Penal, Boiveau16penal, Dettmer16cp}, or with a stabilization added to reduce the oscillations near interfaces and to improve the $L^2$ accuracy~\cite{atkins1999a, kirby2005s, Heimann13unfit, Guo16eg}. Since the stabilization parameter in this case is not required to be larger than a specific lower-bound value to ensure stability, the estimation of the parameter becomes much simpler. In addition, the value of stabilization does not need to be very large, so it is less likely to overshadow the consistency term. Inspired by~\citet{Wu16kr}, we scale the stabilization parameters as $\tau_{\mu}^\text{C} = \tau_{\lambda}^\text{C} = {4\rho h^e_n}/{\Delta t}$  and $\tau_{\kappa}^\text{C} = c_\text{v} \tau_{\mu}^\text{C}$ for compressible-flow IMGA with non-symmetric Nitsche-type weak boundary conditions. Note that here $h^e_n=(\mathbf{n}\cdot\mathbf{G}\mathbf{n})^{-1/2}$ is calculated from the \emph{full element}, which greatly simplifies the evaluation in intersected elements.

\subsection{Adaptive quadrature near point cloud surface}
\label{SubSec:AQ}

Accurate numerical integration over the physical domain $\Omega_\text{phys}$ with the presence of intersected elements is crucial for obtaining accurate simulation results in immersogeometric flow analysis. This is achieved in this paper via a subdivision-based adaptive quadrature rule in the context of the tetrahedral finite cell method~\cite{Xu15ig}. Without changing the original background elements that support the basis functions, the method adaptively refines the quadrature distributions to create an aggregation of quadrature points in the vicinity of the immersed boundary. We refer to the elements used for generating adaptive quadrature rules as cells to avoid confusion with elements used for constructing basis functions. An intersected cell is recursively split into sub-cells until the sub-cell is completely inside $\Omega_\text{phys}$ or $\Omega_\text{fict}$, or when the subdivision reaches a prescribed refinement level. Standard quadrature rules are applied to determine the set of quadrature points and weights within the sub-cells.

\begin{figure}[!h]
    \centering
    \includegraphics[width=0.35\linewidth]{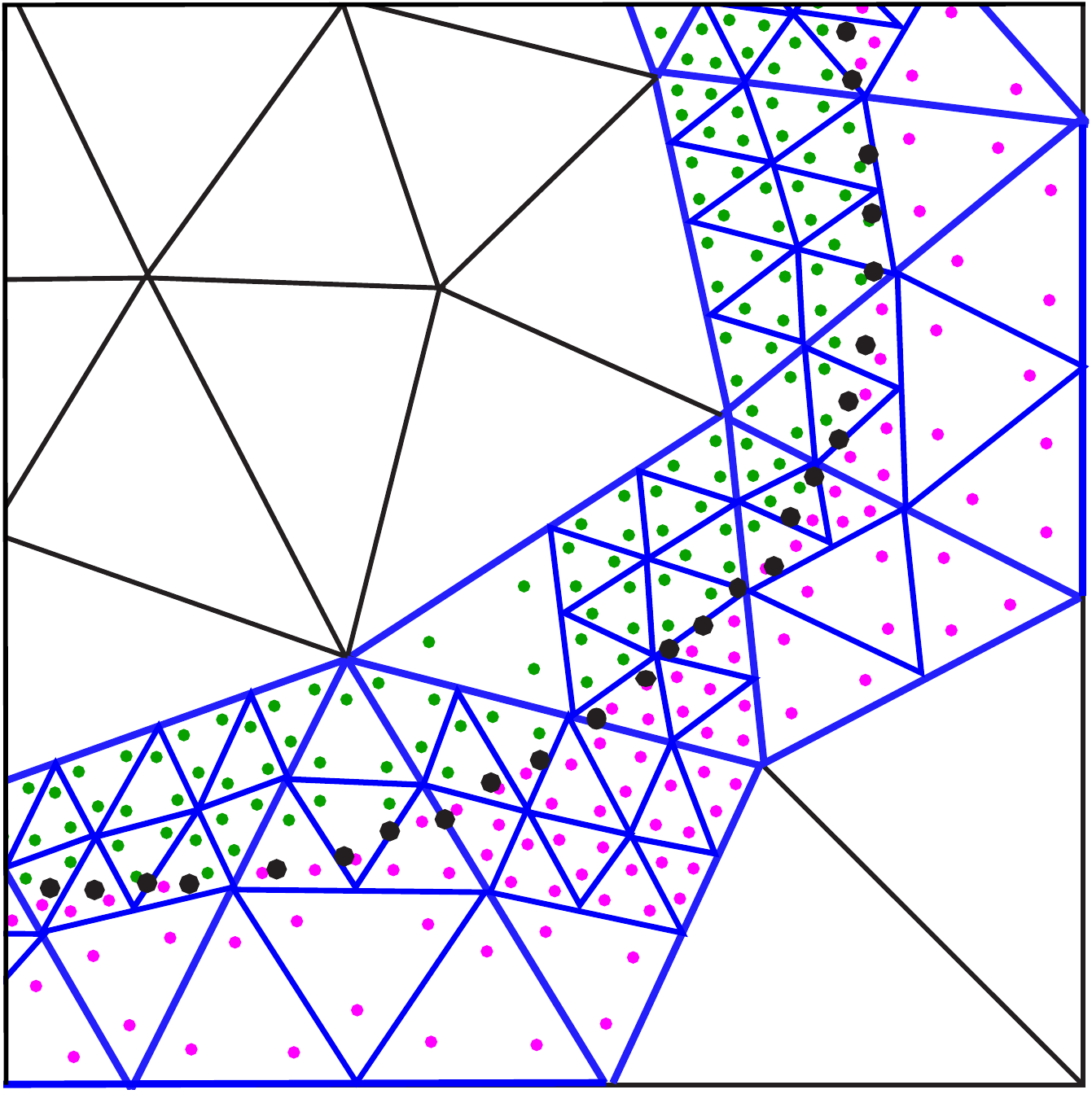}
    \caption{Adaptive quadrature of mesh cells cut by the point cloud object boundary. Black points define the object's point cloud, green points are (inactive) quadrature points inside of the object, and the pink points are (active) quadrature points outside of the object. Two levels of subdivisions are shown here.}
    \label{fig:aq_levels}
\end{figure}

For clarity of visualization, \figref{fig:aq_levels} shows an illustration of the method in 2D based on adaptive sub-cells for triangles. The adaptive quadrature scheme needs two phases of point membership classification: the first classification is performed on the cell vertices to determine whether this cell is intersected, and the second classification is performed on the quadrature points to determine whether they should be active (inside $\Omega_\text{phys}$) or inactive (inside $\Omega_\text{fict}$). In the context of point cloud surface representation, a point membership classification method based on winding number will be presented in~\secref{SubSec:PMC}. For more details on the adaptive quadrature algorithms, see \citet[Section 3.2]{Xu15ig}. 

\section{Point cloud processing}
\label{Sec:PointClouds}
Before we extend the IMGA for point cloud representation, we begin with a few notations for representing the point cloud and its associated differential quantities. A solid geometry is represented as a smooth manifold $\mathcal{S}$ isometrically embedded in the Euclidean space $\mathcal{R}^d$, and the boundary surface of the solid geometry is represented as $\Gamma$ (interacting with the domain $\Omega$). A point cloud $\mathcal{P} \subset \Gamma $ could be defined as a set of points sampled from the boundary surface, i.e. $\mathcal{P} = \{\mathbf{p} \in \Gamma\}$. We assume that $\Gamma$ is two-manifold (to ensure that $\mathcal{S}$ is solid) and piecewise smooth (for estimating the normals and curvatures). Further, $\mathcal{P}$ is sampled randomly from $\Gamma$, with the number of points denoted by $|\mathcal{P}|$. For performing IMGA using point clouds, we present the following methods for (i) estimating the normal or the orientation of each point in the point cloud, (ii) enforcing Dirichlet boundary conditions, and (iii) performing point membership classification for any given domain point with respect to the boundary surface represented by point clouds for adaptive quadrature computations.

\begin{figure}[!b]
    \centering
    \includegraphics[width=0.49\linewidth,trim={3in 1.2in 3in 1.2in},clip]{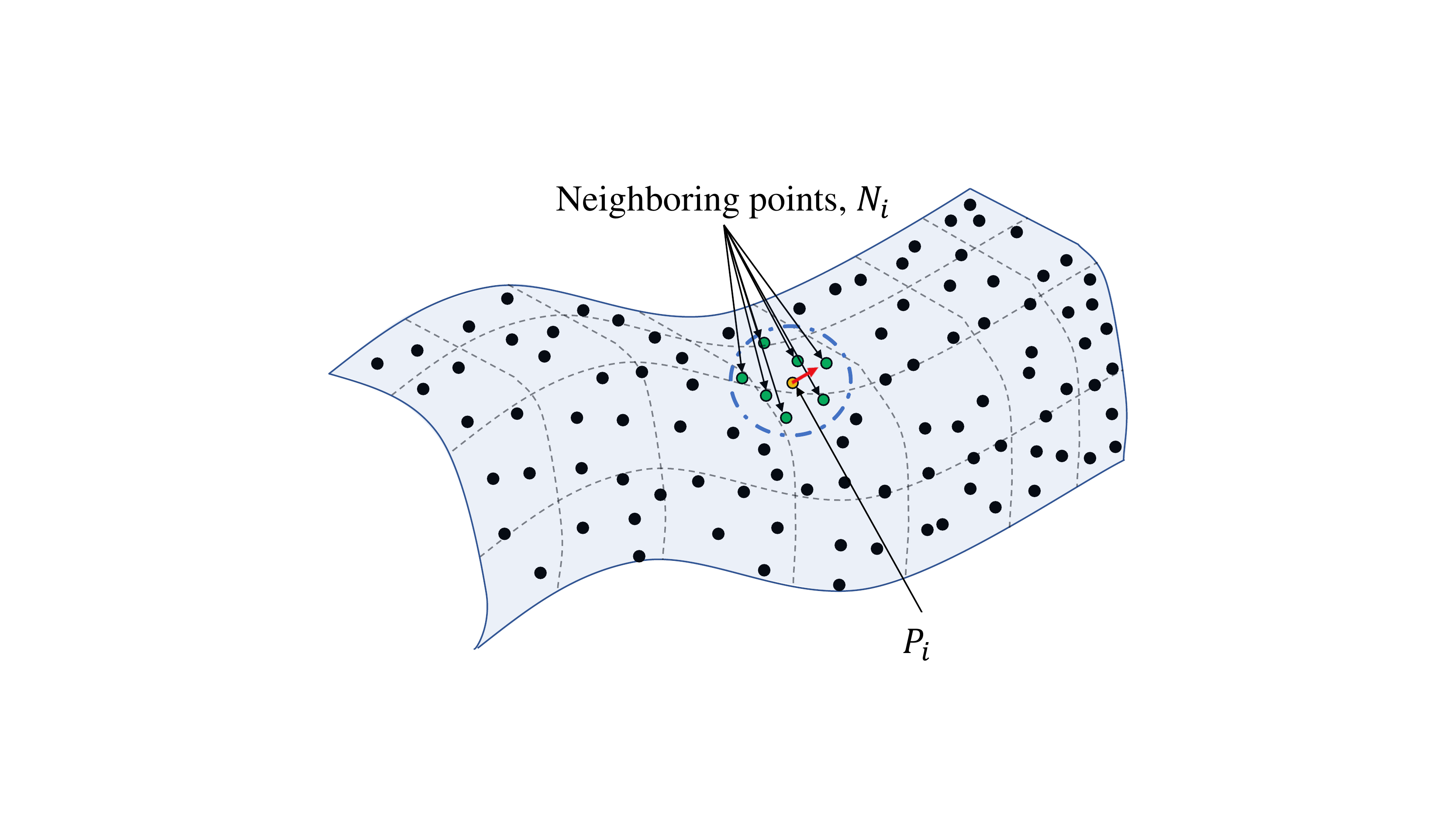}
    \includegraphics[width=0.49\linewidth,trim={3in 1.2in 3in 1.2in},clip]{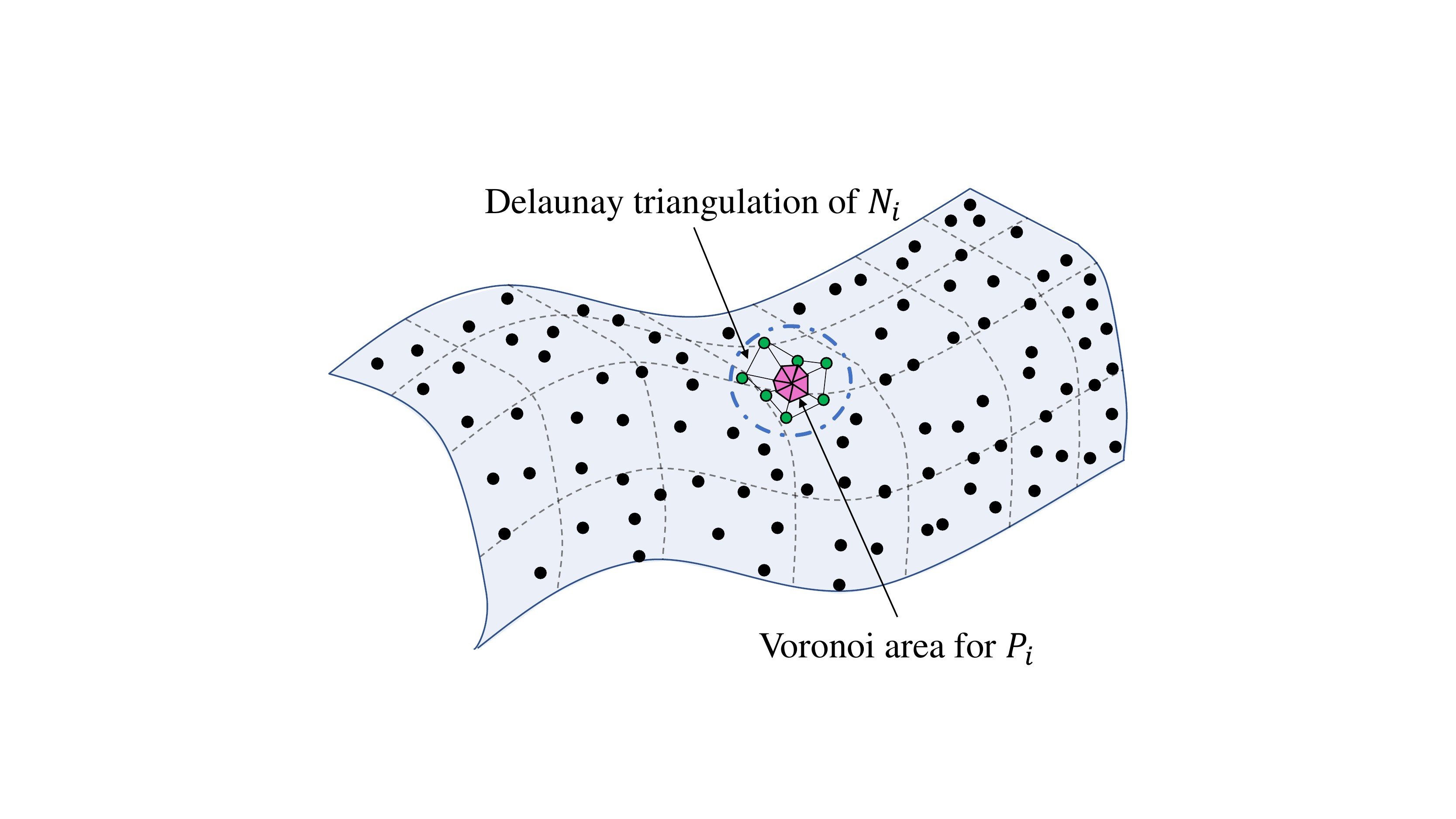}
    \caption{Estimation of the normal and Jacobian determinant for each point on the surface. The image on the left shows the estimation of normals. First, we identify the $k$-nearest neighbors for each point. Then we fit a local surface using the neighbors, and the computed normal of the local surface is considered as the normal for that corresponding point. On the right, we show the estimation of the area (Jacobian) corresponding to each point. Using the $k$-nearest neighbors computed earlier, we first use Delaunay triangulation to obtain the local triangulation surrounding the query point. Then using the barycentric coordinates, we compute the Voronoi area corresponding to each point.}
    \label{fig:normal_estimation}
\end{figure}

\subsection{Normal computation}
\label{SubSec:Normal}

The raw point cloud $\mathcal{P}$ does not contain any information of the orientation (or the normal vector for each point in the point cloud). We estimate the normals from the raw (unoriented) point cloud by constructing a local surface proxy for each point and then estimating the normals for the surface proxy as illustrated in \figref{fig:normal_estimation}. For constructing this surface, we first identify the neighborhood of each point. For a given query point $\mathbf{p}_i \in \mathcal{P}$, where $i\in \left[1,|\mathcal{P}|\right]$, we define the neighborhood $N_i$ to be the $k$-nearest neighbors of the point in the set $\mathcal{P}$. For fitting the local surface for each point $\mathbf{p}_i$, we first transform all the neighboring points, $\mathbf{q} \in N_i$, to a local coordinate system with the origin at $\mathbf{p}_i$. Considering the points lie in $\mathcal{R}^3$, we now define a height function for $f(x,y)$:
\begin{equation}
    f(x,y) = J_{\mathbf{B},n}(x,y) + \mathcal{O}\left(||(x,y)||^{\,n+1}\right),
\end{equation}
with 
\begin{equation}
    J_{\mathbf{B},n}(x,y) = \sum_{r=0}^n H_{\mathbf{B},r}(x,y), \qquad H_{\mathbf{B},r}(x,y) = \sum_{j=0}^r B_{r-j,j}x^{r-j}y^{j}.
\end{equation}
Here, $B_{(\cdot,\cdot)}$ are different constants for a $n$-degree surface fit for the height function, \mbox{$z = f(x,y)$}. The neighboring points are fit with this height function by minimizing objective function of $\sum_{l=0}^{k} \left(f(x_l,y_l) - z_l\right)^2$. To simplify this problem, we write it in matrix representation, where the objective function is to minimize $||\mathbf{M}\mathbf{B} - \mathbf{Z}||^2$. Here, $\mathbf{M}$ is the Vandermonde matrix $\left[1, x_l, y_l, x_l^2, \dots, x_ly_l^{n-1}, y_l^n\right]_{l=1,\dots,k}$ and $\mathbf{B}$ is the matrix with constants of the surface fit $\left[B_{0,0}, B_{0,1}, B_{1,0},  \dots , B_{1,n-1}\right]$. We then find the solution to the linear system $\mathbf{M}\mathbf{B}=\mathbf{Z}$, which is equivalent to fitting the best-fit surface. The guarantee for the existence of a unique solution and the accuracy of the solution to $\mathbf{B}$ as $\mathcal{O}(h^{n-j+1})$ are proven in \citet{cazals2005estimating}. Once the solution for $\mathbf{B}$ is obtained, the normal can be computed as follows:
\begin{equation}
    \mathbf{n} = \frac{\left(-B_{1,0}, -B_{0,1}, 1\right)}{\sqrt{1 + B_{1,0}^2 + B_{0,1}^2}}\text{ .}
\end{equation}

Note that when fitting a surface, we use a different and simpler approach by constructing a covariance matrix $\mathbf{C}$, which can be defined as:
\begin{equation}
    \mathbf{C} = \left[ \begin{array}{c}
   \mathbf{q}_1 - \bar{\mathbf{p}} \\
   \mathbf{q}_2 - \bar{\mathbf{p}} \\
   \vdots \\
   \mathbf{q}_k - \bar{\mathbf{p}}\\ 
  \end{array}  \right]^\text{T}     \left[ \begin{array}{c}
   \mathbf{q}_1 - \bar{\mathbf{p}} \\
   \mathbf{q}_2 - \bar{\mathbf{p}} \\
   \vdots \\
   \mathbf{q}_k - \bar{\mathbf{p}}\\ 
  \end{array}  \right].
\end{equation}
Here, $\bar{\mathbf{p}}$ represents the centroid of the neighboring points cluster $N_i$. We can estimate the normal vector for the plane at point $\mathbf{p}_i$ by computing the eigenvalues and eigenvectors of the covariance matrix $\mathbf{C}$. The normal is defined as the eigenvector corresponding to the smallest eigenvalue~\cite{hoppe1992surface}. Since the eigenvector corresponding to the smallest eigenvalue is also a principal component for the point cloud, this method is also known as principal component analysis (PCA)-based normal estimation. 

Note that the method above does not guarantee that the calculated normals consistently point inward or outward with respect to the point cloud geometry. For a given point, we construct a minimum spanning tree consisting of the nearest neighbors and flip the calculated normal vector if its orientation is inconsistent with the neighbors~\cite{hoppe1992surface}.

While the PCA-based normal estimation performs well (using plane fitting), for better accuracy, we can fit a quadratic surface to obtain better curvature and smoothness~\cite{cazals2005estimating,cazals2008algorithm} (also commonly known as 2-jet fitting, or simply, jet fitting). In this paper, we compare both methods (PCA and jet fitting) to understand the advantages and disadvantages of both methods.

\subsection{Point cloud surface integration rules}
\label{SubSec:Area}
To weakly enforce the Dirichlet boundary conditions on $\Gamma^\text{D}$, we first discretize the boundary surface into elements. We then compute the weak-boundary-condition (or Nitsche) terms in Eq.~\eqref{eq:incomp_wbc} using the points of the point cloud and their corresponding Jacobian for each discrete element of the surface. For a point cloud represented using $\mathcal{P}$, we do not need any additional discretization, and hence we consider each point $\mathbf{p}_i \in \mathcal{P}$ as a surface element\footnote{In this work, a surface element is represented by a point of the point cloud, which we sometimes refer to as a point cloud element.} and its associated (single) quadrature point for performing the integration of the Nitsche terms over $\Gamma^\text{D}$. However, computing the Jacobian determinant for each quadrature point for performing the integration is not straightforward.  While the surface area or the Jacobian determinant is not defined in a strict sense for a point-cloud-based representation, we define an effective surface area $a_i$ for each point in the point cloud (i.e., point area or Voronoi area). This area can be associated with the geodesic Voronoi area of the point $\mathbf{p}_i$ on the boundary surface $\Gamma$. For some point cloud acquisition processes, this is already available from the sensors~\cite{fuhrmann2014floating}, and for others, this can be approximated as shown in \citet{belkin2009constructing} and \citet{barill2018fast}. Similar to the approach used for computing the normals of the point cloud, we construct a local neighborhood for each point $\mathbf{p}_i \in \mathcal{P}$ to obtain the geodesic Voronoi area $a_i$ as shown in \figref{fig:normal_estimation}{b}. 

The complete process can be detailed in three steps: (i) For each point $\mathbf{p}_i$ in the set of point cloud $\mathcal{P}$, we compute the neighborhood $N_i$ to be the $k$-nearest neighbors. (ii) Using each set of neighborhood points $N_i$, we perform Delaunay triangulation to result in a local surface $T_i$. (iii) The dual graph of a Voronoi diagram $V_i$ is the Delaunay triangulation $T_i$. Therefore, computing the Voronoi area of $\mathbf{p}_i$ can be easily performed by taking the contribution area of each triangle in $T_i$ to $a_i$. Once we obtain the area $a_i$ corresponding to each quadrature point, we can integrate and obtain the weak-boundary-condition terms in Eq.~\eqref{eq:incomp_wbc}. 

\subsection{Point membership classification}
\label{SubSec:PMC}

For IMGA, the attribution of a given point $\mathbf{q}$ in the domain $\Omega$ to the physical domain $\Omega_\text{phys}$ or fictitious domain $\Omega_\text{fict}$ is fundamental. In other words, we would have to determine if the point $\mathbf{q}$ is inside the solid $\mathcal{S}$ or outside it. Traditional B-rep CAD representations perform inside-outside testing by using ray-intersection-based methods. However, these approaches assume that the surface $\Gamma$ representing the boundary of solid $\mathcal{S}$ is 2-manifold and watertight. When the CAD geometry is represented using a non-watertight surface representation or badly oriented normals, the ray-intersection-based methods fail to determine the membership correctly. Particularly, ray-intersection-based methods do not work for representations such as point clouds (which are non-manifold). 

\begin{figure}[!t]
    \centering
    \hspace*{0.5cm}\includegraphics[width=0.7\linewidth]{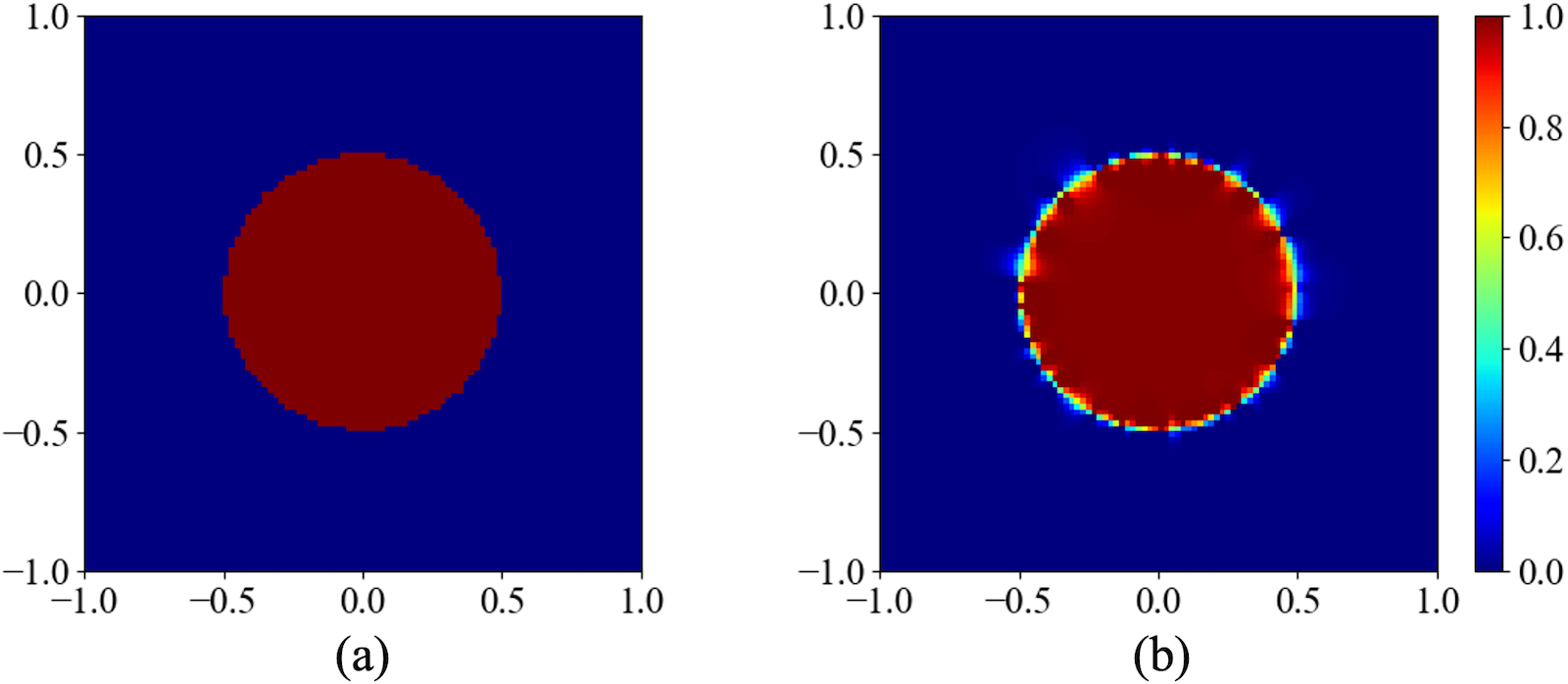}
    \caption{Illustration of winding number for (a) a manifold representation of a circle and (b) a non-manifold representation of a circle. The circle is centered at (0.0, 0.0) and with a radius of 0.5.  The non-manifold point cloud shown on the right is represented using 1000 points.}
    \label{fig:winding}
\end{figure}

There has been recent work on defining inside-outside tests for arbitrary non-manifold geometries such as triangle soups and point clouds using generalized winding numbers~\cite{barill2018fast,jacobson2013robust}. The winding number $w(\mathbf{q},\Gamma) \in \mathcal{R}$ can be computed as:
\begin{equation}
    w(\mathbf{q},\Gamma) = \frac{1}{4\pi}\int_{\Gamma} dA.
\end{equation}
Here, $A$ refers to the total solid angle subtended by $\Gamma$ on $\mathbf{q}$, $dA$ refers to the differential solid angle subtended by each element of $\Gamma$ on $\mathbf{q}$. We then project the surface onto a unit sphere to obtain the surface integral and further simplify it for a discrete point cloud represented by $\mathcal{P}$ as follows:
\begin{equation}
    w(\mathbf{q},\Gamma) = \frac{1}{4\pi}\int_{\Gamma} dA = \int_{\Gamma} \frac{(\mathbf{x} - \mathbf{q})\cdot \mathbf{n}}{4\pi||\mathbf{x} - \mathbf{q}||^3} dA = \sum_{i=1}^{|\mathcal{P}|} a_i \frac{(\mathbf{p}_i - \mathbf{q})\cdot\mathbf{n}_i}{||\mathbf{p}_i - \mathbf{q}||^3}.
\end{equation}
$\mathbf{p}_i$, $\mathbf{n}_i$, and $a_i$ refer to the point coordinates, the normal, and the geodesic Voronoi area, respectively, for the $i$th point in the point cloud $\mathcal{P}$. Although the winding number is defined as a real number, for a 2-manifold, watertight representation of the geometry, the possible winding number value at any given point $\mathbf{q} \in \mathcal{R}^3$ are positive integers. Further, if the surface does not have self-intersections, the winding number is in the binary set of $\{0,1\}$. In other words, for any point $\mathbf{q} \in \mathcal{R}^3$:
\begin{equation}
     w(\mathbf{q},\Gamma) =  \begin{cases}
                        1 \qquad \mathbf{q} \in \Omega_\text{phys} \\
                        0 \qquad \mathbf{q} \not \in \Omega_\text{phys}
                    \end{cases}.
\end{equation}
However, in the case of a non-manifold boundary representation such as a point cloud representation $\mathcal{P}$, the winding number $w(\mathbf{q}, \Gamma)$ is still zero for all $\mathbf{q} \not \in \Omega_\text{phys}$ with a continuous value between $0$ and $s+1$, where $s$ is the number of self-intersections of the surface $\Gamma$. We illustrate how the winding number field for a non-manifold and manifold boundary representation differs in \figref{fig:winding}. Therefore, for attribution of a given point $\mathbf{q}$ to be inside or outside of $\Omega_\text{phys}$, we threshold the winding number at $0.5$, and any point with a smaller winding number is considered outside the physical domain, $\Omega_\text{phys}$.

\section{Validation}
\label{Sec:Validation}
  
\subsection{Validation of geometric quantities}
\label{SubSec:GeomConvergence}
Before performing IMGA using point clouds, we first validate the estimated geometric quantities to understand their accuracy in representing the geometry. We consider three different geometries represented by point clouds for our study: (i) an icosahedron-based sampling of a sphere; (ii) a randomly sampled sphere; and (iii) a Fandisk model (a benchmark CAD geometry) containing sharp and smooth features~\cite{hoppe1994piecewise}.

\subsubsection{Validation on icosphere and randomly sampled sphere}
While our proposed method for estimating the geometric quantities required for IMGA works independent of the point sampling, for the first study, we consider the case of an icosphere. An icosphere is a set of points obtained by tessellating an icosahedron inscribed inside a sphere of unit radius~\cite{kahler2018creating}. Icosphere provides an ideal set of points sampled on a sphere such that there is a uniform distance between the points lying on the sphere. This geometric setup provides an ideal example for understanding the behavior of different parameters used to estimate geometric quantities. A more realistic geometry is a randomly sampled point cloud. We randomly generate points on a sphere by generating a set of random points and then projecting them to the sphere by dividing each point by its distance to the sphere center. Using this method, we can obtain an arbitrarily large number of points on the spherical surface.

\begin{figure}[!t]
    \centering
    \raisebox{-0.5\height}{\includegraphics[width=0.68\linewidth, trim={0pt, 0pt, 0pt, 0pt}, clip]{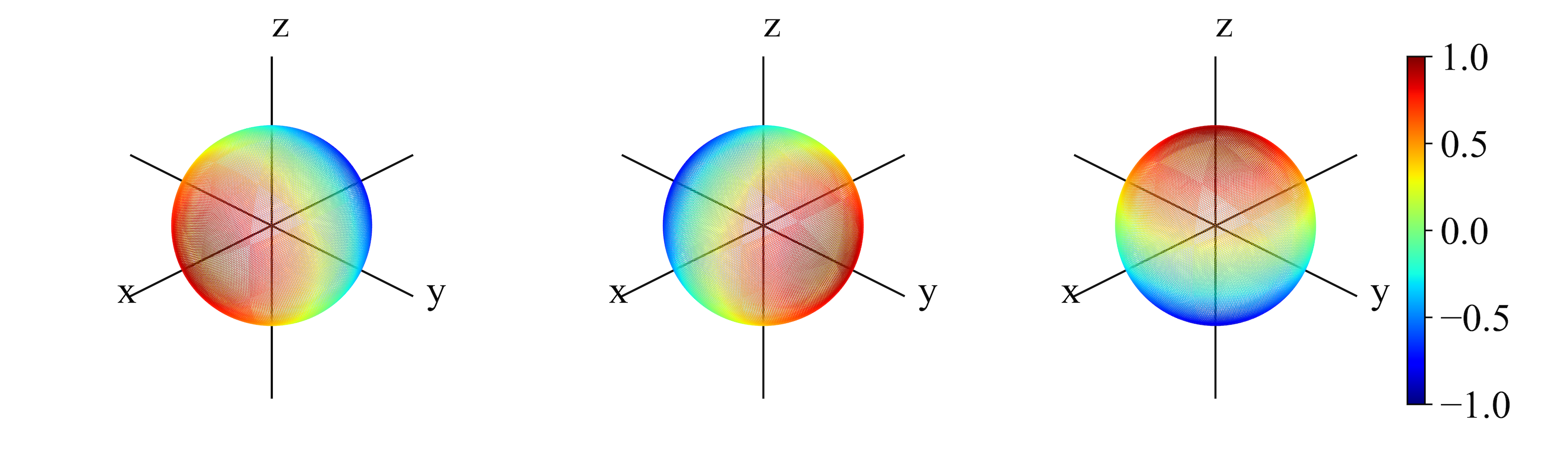}}\hfill\raisebox{-0.5\height}{\includegraphics[width=0.32\linewidth, trim={0pt, 0pt, 0pt, 0pt}, clip]{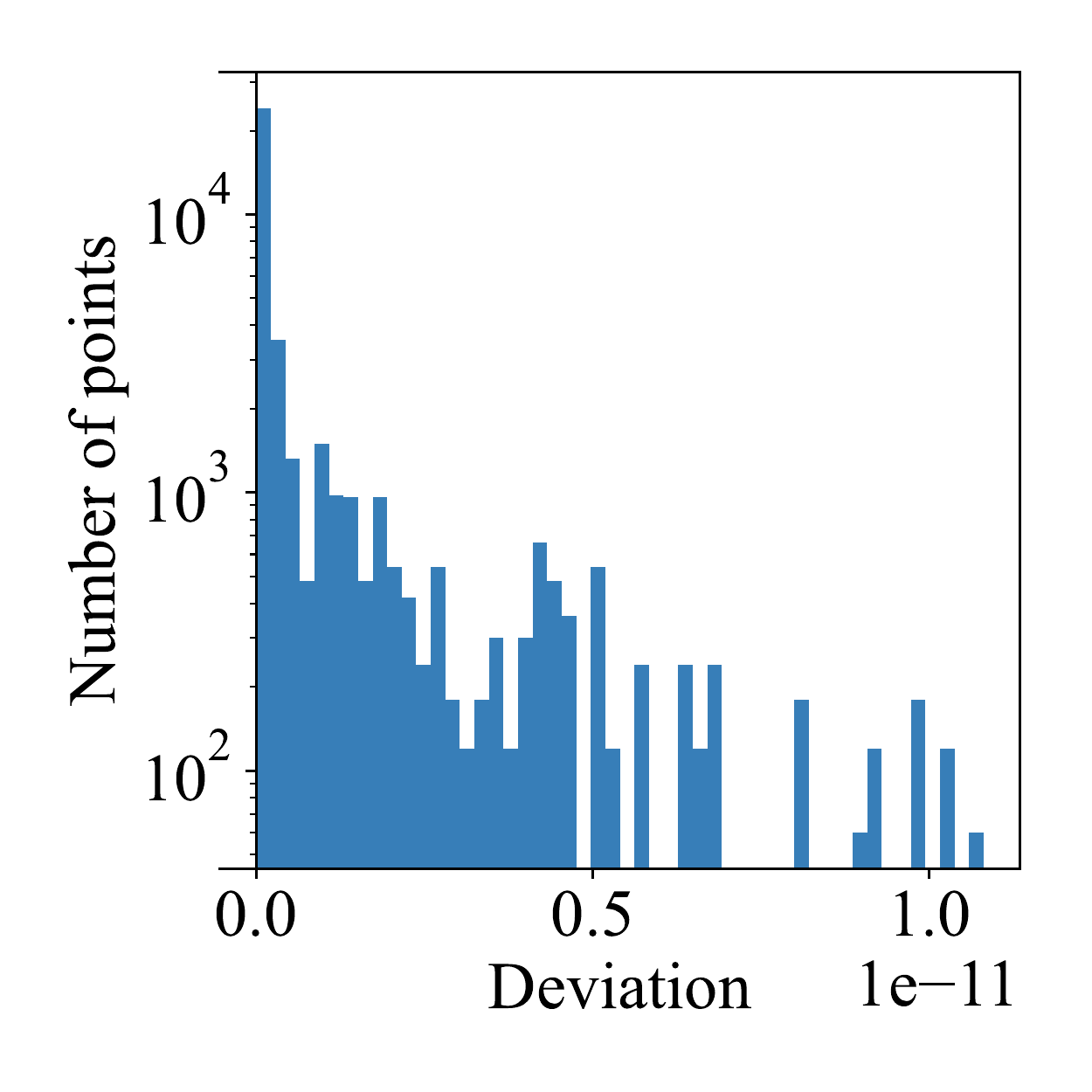}}
    \caption{The $x$, $y$, and $z$ components of the normals estimated from 163,842 points sampled from an icosphere and the histogram of the deviation in the estimated normals (right).}
    \label{fig:icosphere_normals}
\end{figure}

\begin{figure}[!t]
    \centering
    \raisebox{-0.5\height}{\includegraphics[width=0.68\linewidth, trim={0pt, 0pt, 0pt, 0pt}, clip]{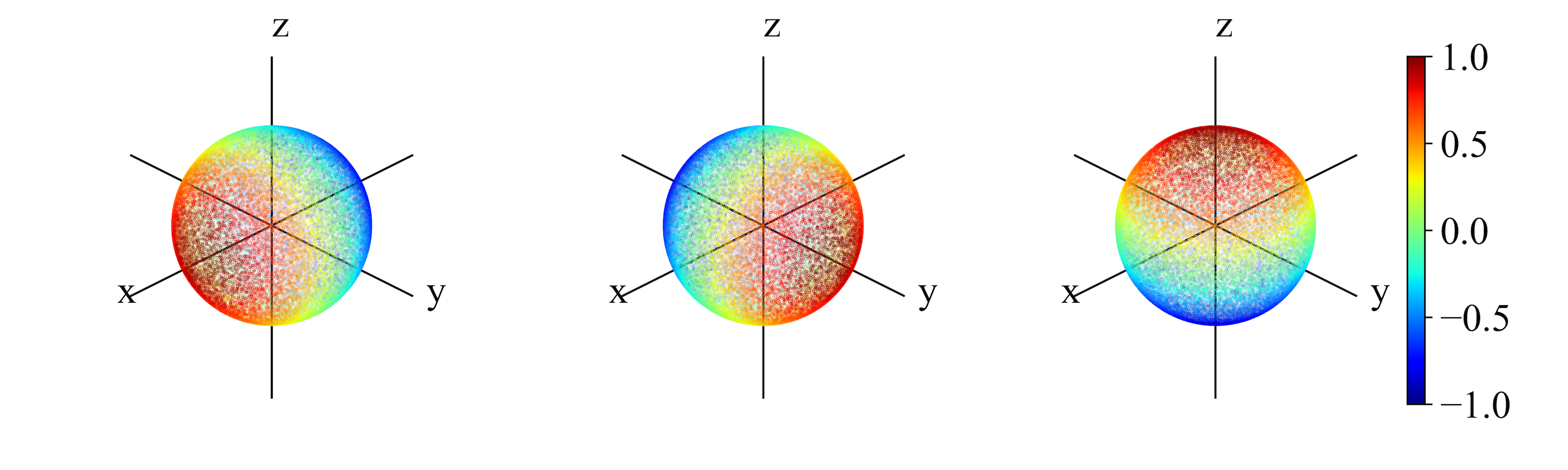}}\hfill\raisebox{-0.5\height}{\includegraphics[width=0.32\linewidth, trim={0pt, 0pt, 0pt, 0pt}, clip]{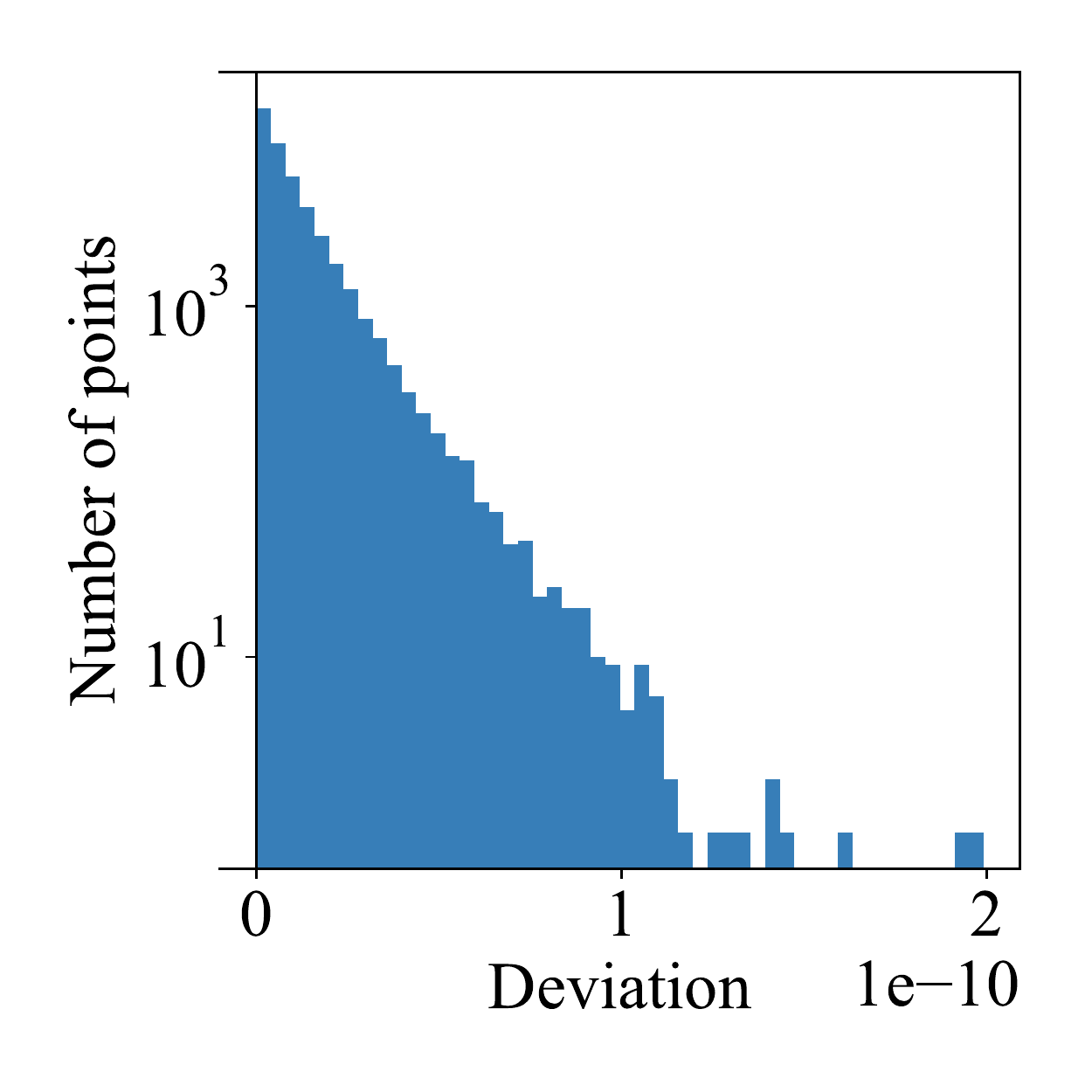}}
    \caption{The $x$, $y$, and $z$ components of the normals estimated from 200,000 randomly sampled points and the histogram of the deviation in the estimated normals (right).}
    \label{fig:randomsphere_normals}
\end{figure}

For some methods used for acquiring the point cloud, the normal information is not available. Hence, we estimate the normals using the 2-jet fitting (mentioned above) for the point cloud. It takes 4.06\,s to estimate normals for 163,842 points on the icosphere. \figref{fig:icosphere_normals} shows the visualization of the estimated normals for the icosphere. Here, we visualize the three components of the normal vector individually. We observe that each normal component is aligned with the corresponding primary axes. A similar visualization for a randomly sampled sphere with 200,000 points is shown in \figref{fig:randomsphere_normals}.

We also compute the deviation between the estimated and theoretical spherical normals. The theoretical normal of a given point on a unit sphere can be calculated by normalizing the vector from the sphere center to the point. The deviation is computed by subtracting the dot product between the estimated normal vector $\mathbf{n}_\text{esti}$ and the theoretical normal vector $\mathbf{n}_\text{theo}$ from unity \mbox{$(1 - \mathbf{n}_\text{theo} \cdot \mathbf{n}_\text{esti})$}. The maximum deviation in estimating the normals for 163,842 points is $1.15 \times 10^{-11}$. A similar behavior can be observed for 200,000 randomly sampled in \figref{fig:randomsphere_normals} with a slightly larger deviation of $2.00 \times 10^{-10}$.

\begin{figure}[!t]
    \centering
    \includegraphics[width=0.46\linewidth, trim={0pt, 0pt, 0pt, 0pt}, clip]{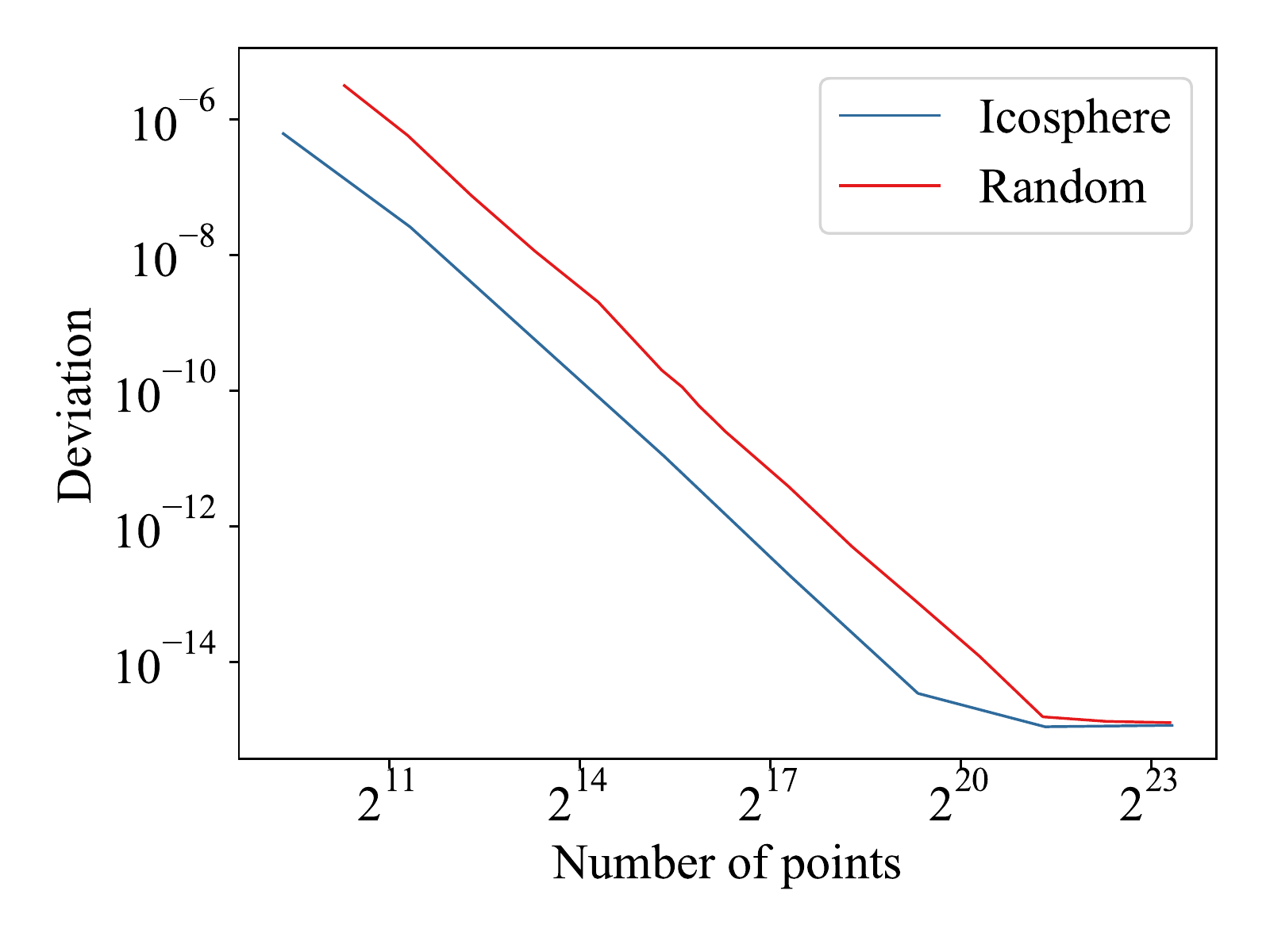}
    \includegraphics[width=0.45\linewidth, trim={0pt, 0pt, 0pt, 0pt}, clip]{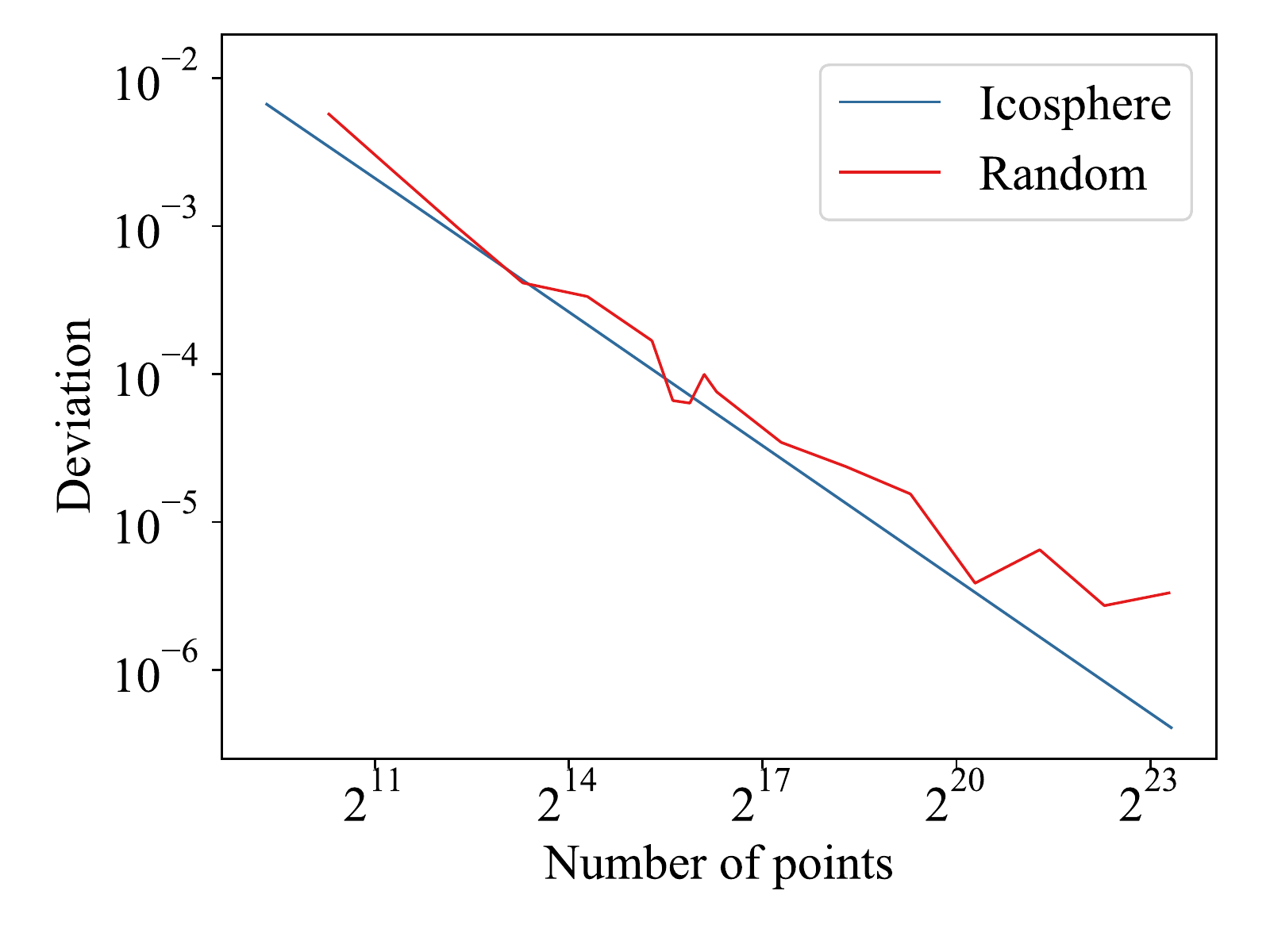}
    \caption{Trends of deviation in normal estimation (left) and area estimation (right) with the increase in the number of points.}
    \label{fig:sphere_points_comparison}
\end{figure}

Since the method for normal estimation relies on the local geometry, naturally, this deviation reduces with the number of points used to represent the geometry (hence capturing more intricate features of the geometry). In \figref{fig:sphere_points_comparison}, we compare the mean deviation in normal estimation and deviation in total area estimation. For normal estimation, we use $1 - \mathbf{n}_\text{theo} \cdot \mathbf{n}_\text{esti}$ for estimating the deviation at each point and then use the mean deviation for comparison. However, we compare the sum of each point area for area estimation. The analytical value of the surface area of a sphere is used as the baseline for comparing the deviation. We see a linear trend for deviation of the geometric quantities with the increase in the number of points on the log scale. Also, the difference in mean deviation of normals obtained from the icosphere and randomly sampled sphere reduces with an increase in the number of points. The total area deviation for randomly sampled points is almost similar to the icosphere. 

\begin{figure}[!t]
    \centering
    \includegraphics[width=0.46\linewidth, trim={0pt, 0pt, 0pt, 0pt}, clip]{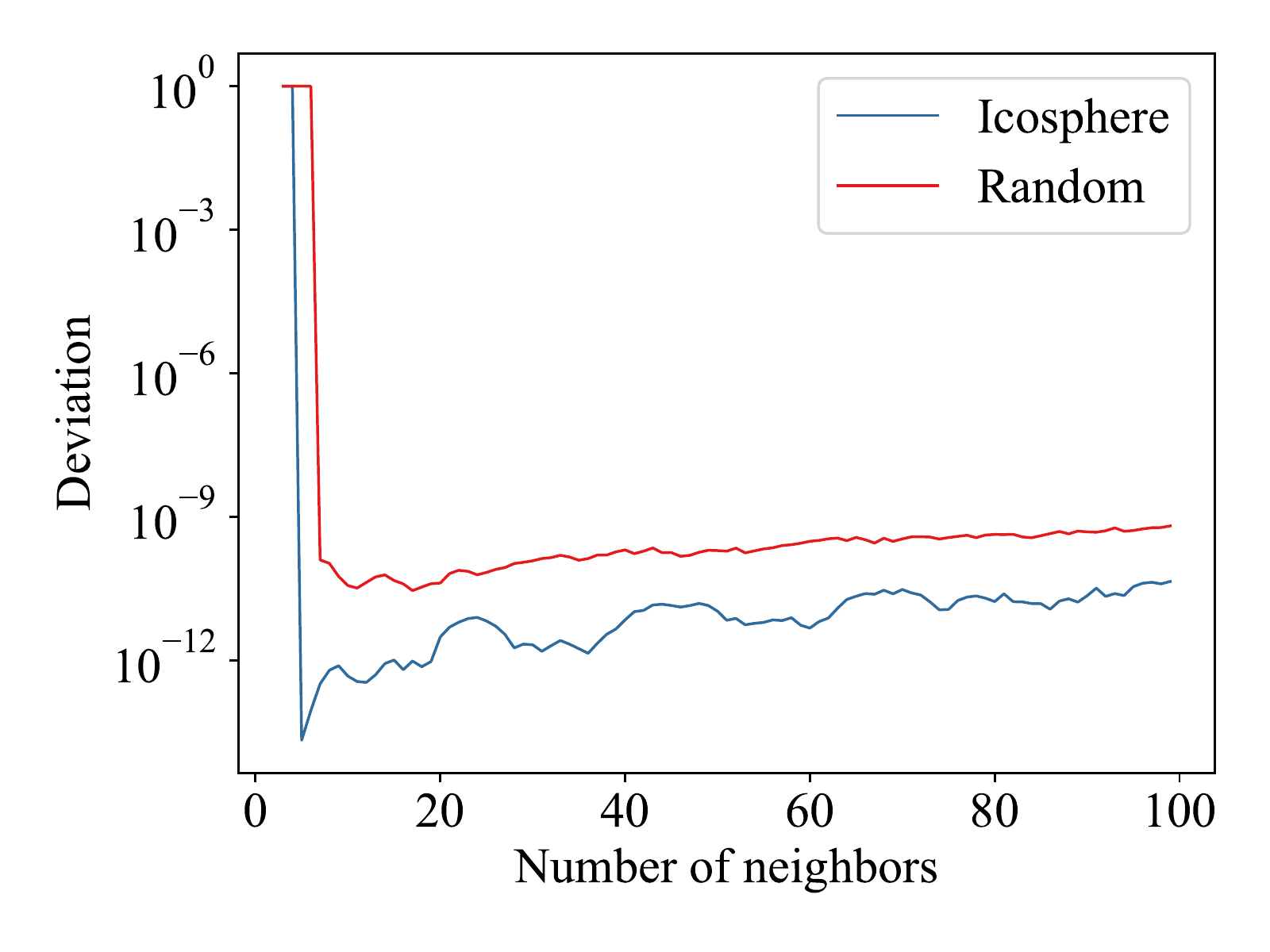}
    \includegraphics[width=0.46\linewidth, trim={0pt, 0pt, 0pt, 0pt}, clip]{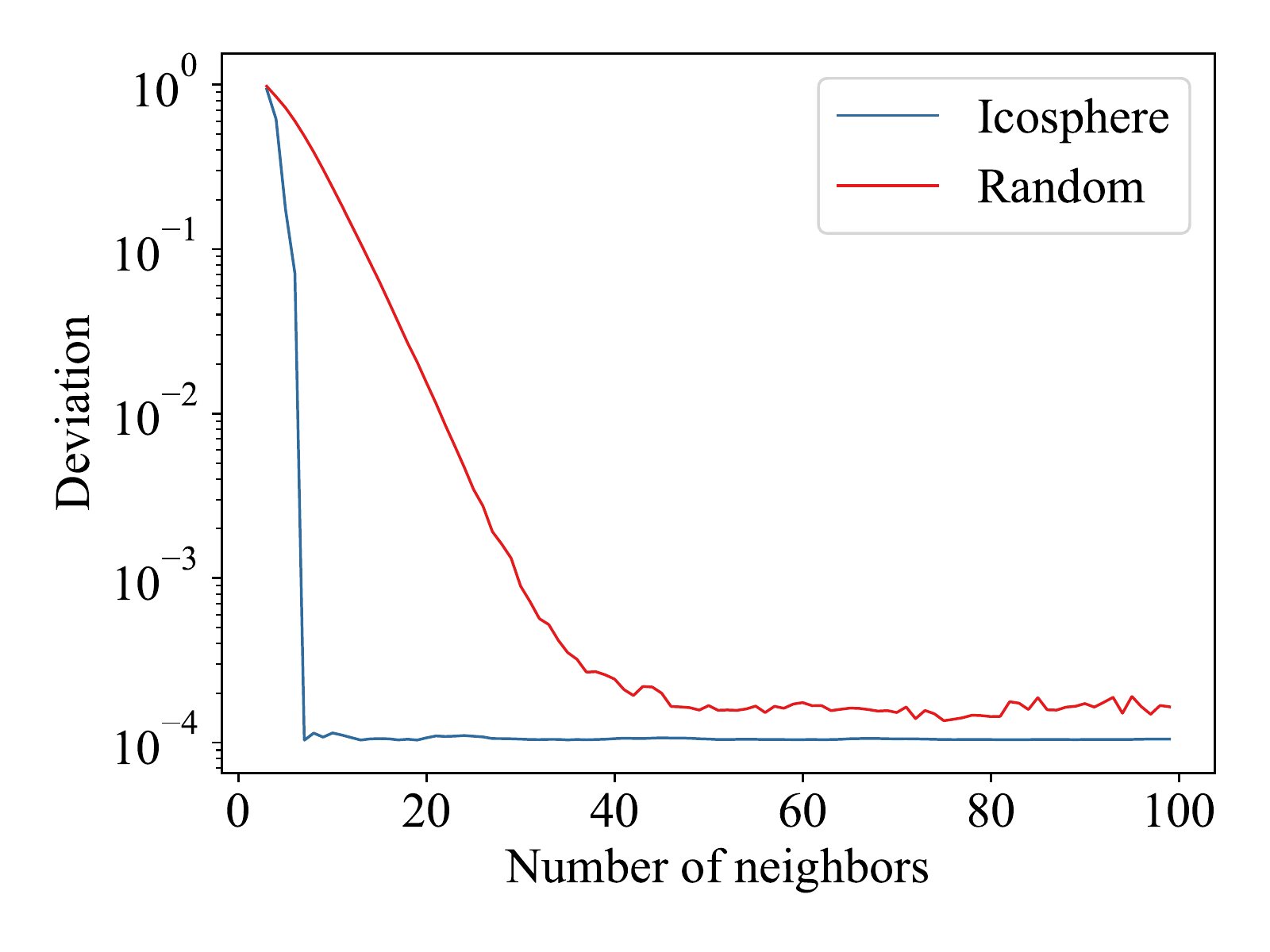}
    \caption{Trends of deviation in normal estimation (left) and area estimation (right) with the increase in the number of neighboring points used for estimation.}
    \label{fig:sphere_neighbors_comparison}
\end{figure}

Another parameter affecting the estimation of geometric quantities is the number of neighbors used to estimate them. \figref{fig:sphere_neighbors_comparison} shows the trend in the mean deviation of the estimated normals and the deviation in the estimated total surface area of the point cloud for different numbers of neighbors. The deviation is the minimum with an optimal number of neighbors (empirically close to 18--20 neighboring points, subject to change based on the number of points in the point cloud) and remains almost constant with a slight increase in the mean deviation of the estimated normals. This increase in deviation with an increase in the number of neighbors is because of adding more points that do not meaningfully contribute to the normal estimation. While a similar behavior is observed in normal estimation for a randomly sampled number of points, we also observe that random sampling increases the optimal number of neighbors to account for non-uniform distances between the points, specifically in point area estimation. 

\begin{figure}[!t]
    \centering
    \includegraphics[width=0.4\linewidth, trim={0pt, 24pt, 0pt, 0pt}, clip]{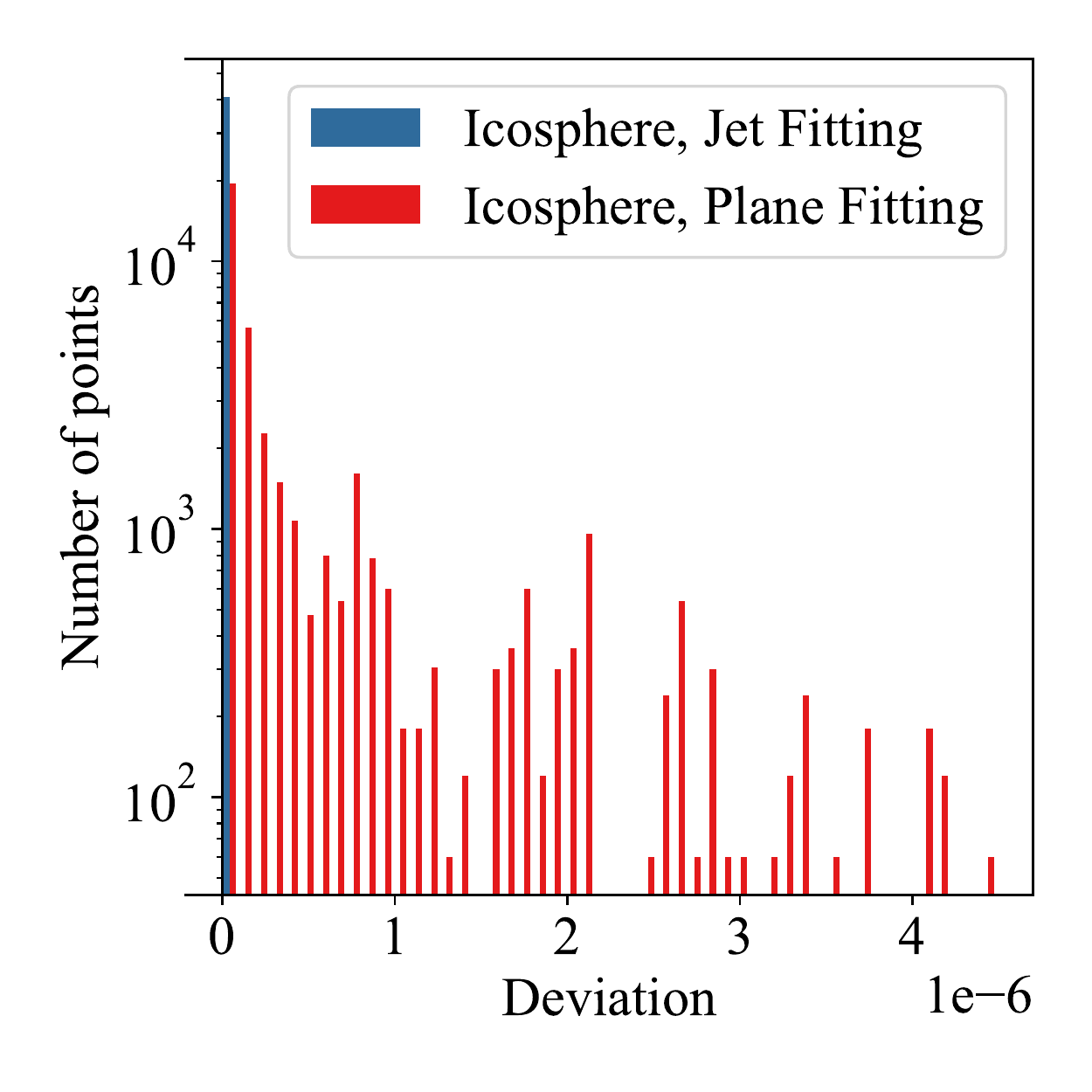}
    \includegraphics[width=0.4\linewidth, trim={0pt, 24pt, 0pt, 0pt}, clip]{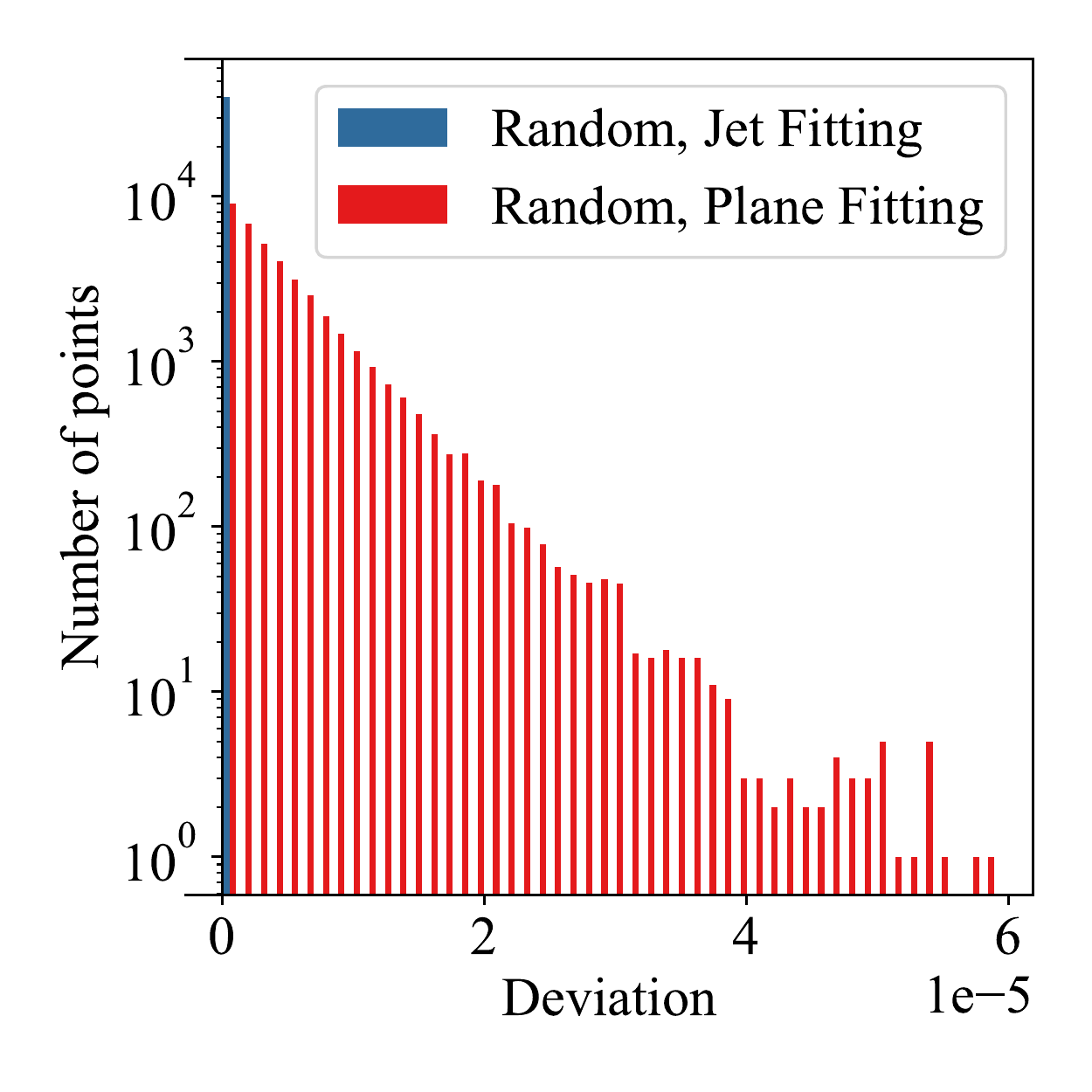}
    \caption{Trends in the histogram of deviation in normals estimation from PCA-based estimation and 2-jet-based fitting for icosphere (left) and randomly sampled sphere (right).}
    \label{fig:sphere_comparison_w_pca}
\end{figure}

Another parameter to study is the degree of the surface used for fitting from the nearest neighbors. Recall that the linear plane fitting is equivalent to PCA-based fitting and is faster than the jet-fitting approach. While in terms of speed, the PCA-based plane fitting takes 0.79\,s for normal estimation (compared to 4.06\,s for 2-jet fitting), there is a difference of an order of magnitude in the estimation of the normals and consequently even in the area estimation. \figref{fig:sphere_comparison_w_pca} shows the histogram of deviation in normals estimation for icosphere and randomly sampled points using both PCA and 2-jet fitting. While there is a significant difference in the computational time and the deviation, the maximum deviation in the estimation is still less than $10^{-4}$, and hence is still a viable method for estimating the quantities for IMGA. Both methods, with appropriately chosen neighbors and point cloud density, have very little deviation for the sphere.

\subsubsection{Validation of normals estimation on a point cloud sampled from the Fandisk model}
\begin{figure}[!t]
     \centering
     \begin{subfigure}[b]{0.26\textwidth}
         \centering
        \includegraphics[width=\linewidth, trim={5pt, 5pt, 5pt, 5pt}, clip]{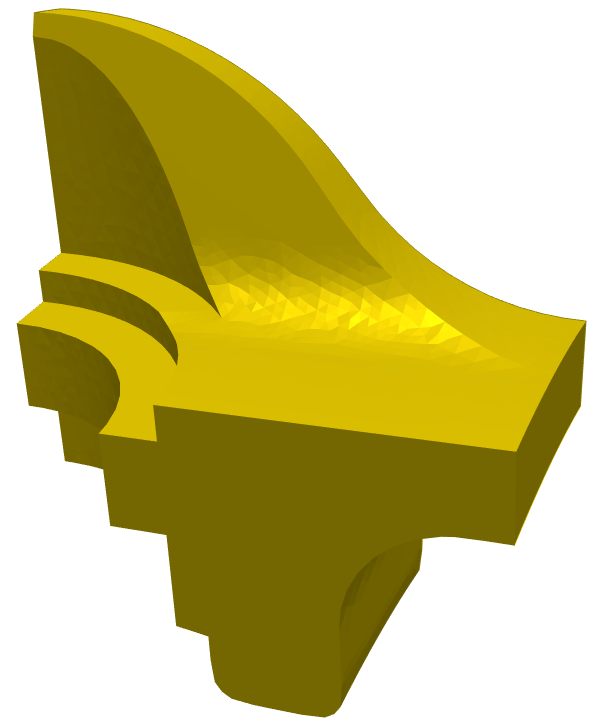}
         \caption{}
         \label{fig:y equals x}
     \end{subfigure}
     \hfill
     \begin{subfigure}[b]{0.35\textwidth}
         \centering
         \includegraphics[width=1\linewidth, trim={0pt, 0pt, 0pt, 0pt}, clip]{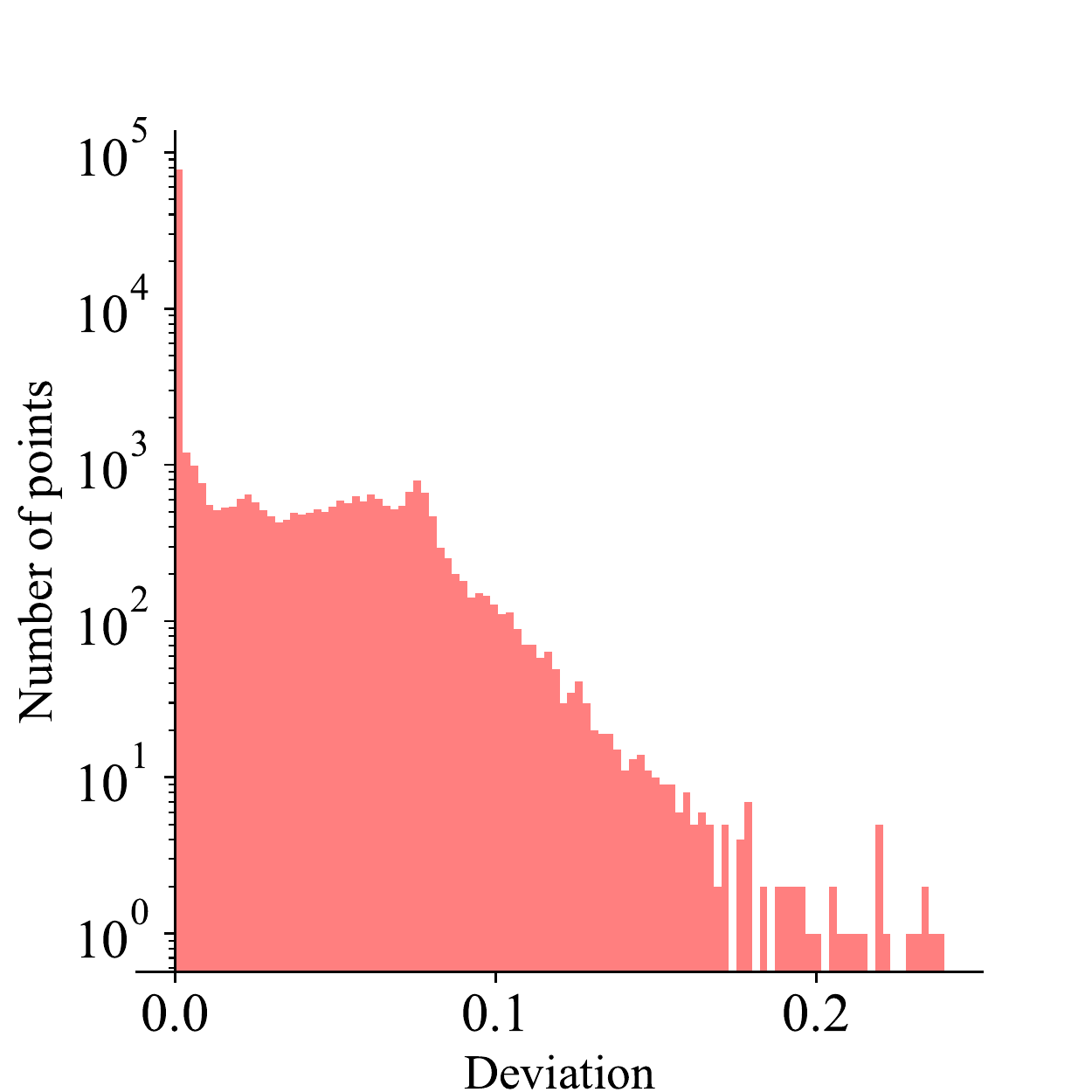}
         \caption{}
         \label{fig:three sin x}
     \end{subfigure}
     \hfill
     \begin{subfigure}[b]{0.33\textwidth}
         \centering
         \includegraphics[width=\linewidth, trim={0pt, 0pt, 0pt, 0pt}, clip]{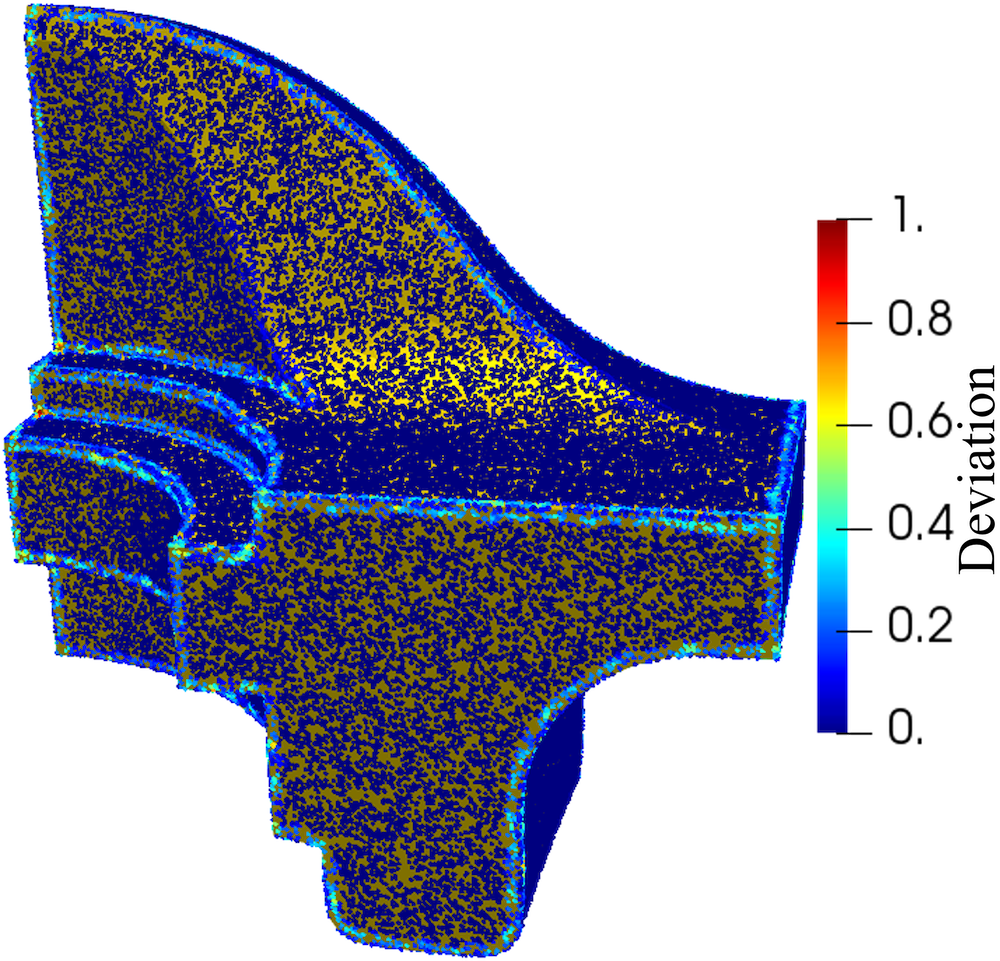}
         \caption{}
         \label{fig:five over x}
     \end{subfigure}
        \caption{Normal estimation on Fandisk model. (a) Surface representation of Fandisk model. (b) Variation of the maximum deviation in estimation of normals with the number of nearest neighbors used for fitting the surface. (c) Visualization of deviations of the normals at each point on the surface.}
    \label{fig:Fandisk_normal_estimation}
\end{figure}

As mentioned above, the normal estimation depends on the nearest neighbors for a given point. This means there could be significant deviation when the geometry has sharp edges or very fine features. To test this, we use a benchmark CAD model (Fandisk model, shown in \subfigref{fig:Fandisk_normal_estimation}{a}). The Fandisk model consists of smooth and sharp features, thus enabling us to test the robustness of the normal and area estimation methods. While the experiments elucidated above can also be performed for the Fandisk model, we show just a few key results that provide insights for brevity.

We use the normals computed from the triangle faces of the benchmark CAD model to set a baseline for the normal vectors. This baseline preserves sharp features since the normals are completely different for adjacent triangles. However, such sharp features cannot be replicated using point clouds due to a lack of connectivity information. In \subfigref{fig:Fandisk_normal_estimation}{b}, we show the histogram of deviations in normal estimation. While most points have very little deviations, about 20\% of the total points have some significant deviation. We visualize the deviations of the normals at each point on the surface in \subfigref{fig:Fandisk_normal_estimation}{c}. While most of the points at the center of each face have minimal deviation, all the points near the edges show significant normal deviation, which is expected. Note that the deviation of 0.30 (the maximum deviation for a point) refers to a deviation of $45 \degree$ between the two normal vectors, the average direction between perpendicular planes.

\subsubsection{Validation of winding number computation}
\begin{figure}[!t]
    \centering
    \begin{subfigure}[b]{0.48\linewidth}
        \includegraphics[width=\linewidth, trim={0pt, 0pt, 0pt, 0pt}, clip]{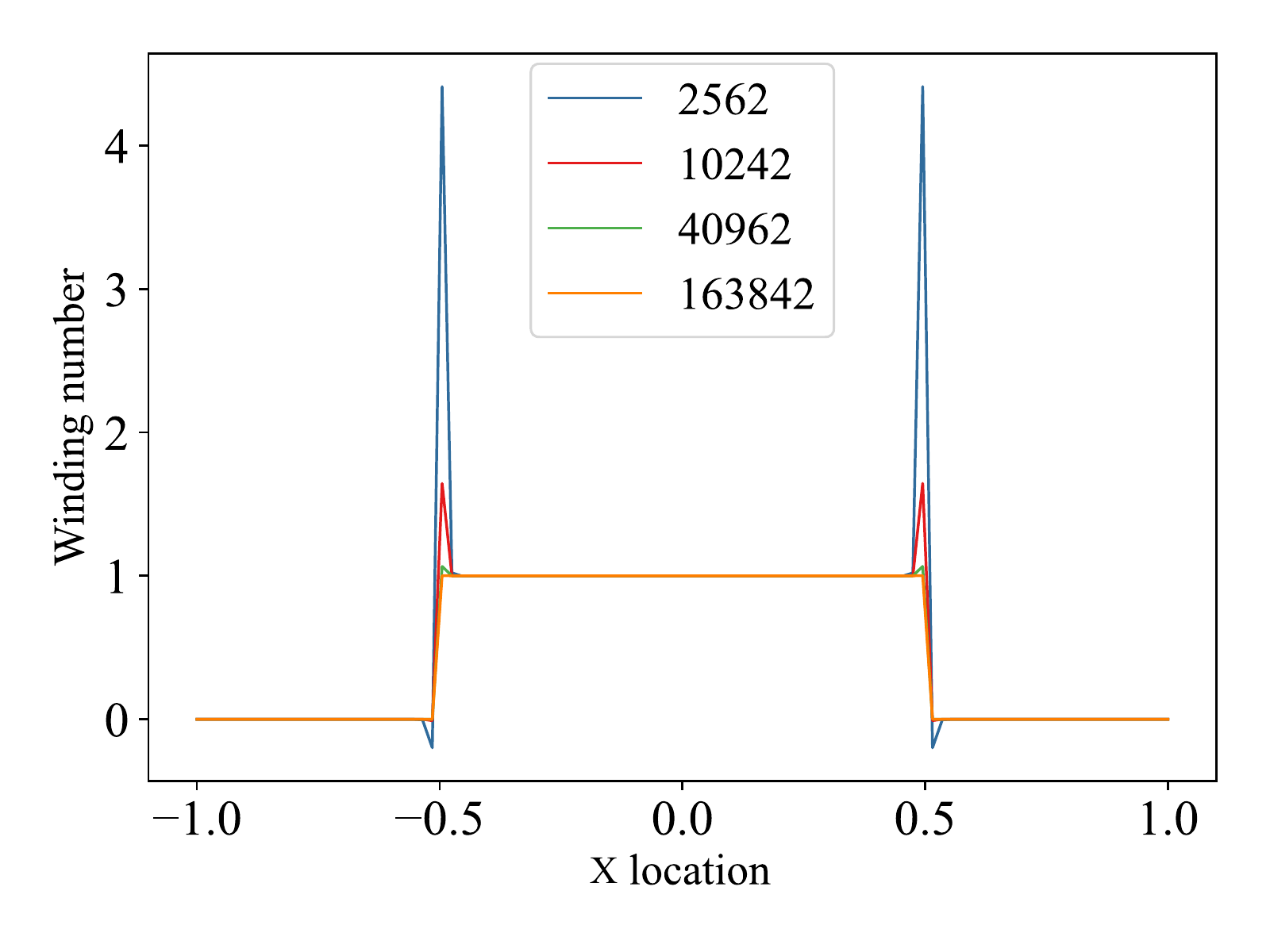}
        \caption{}
        \label{fig:winding-points}
    \end{subfigure}
    \begin{subfigure}[b]{0.48\linewidth}
        \includegraphics[width=\linewidth, trim={0pt, 0pt, 0pt, 0pt}, clip]{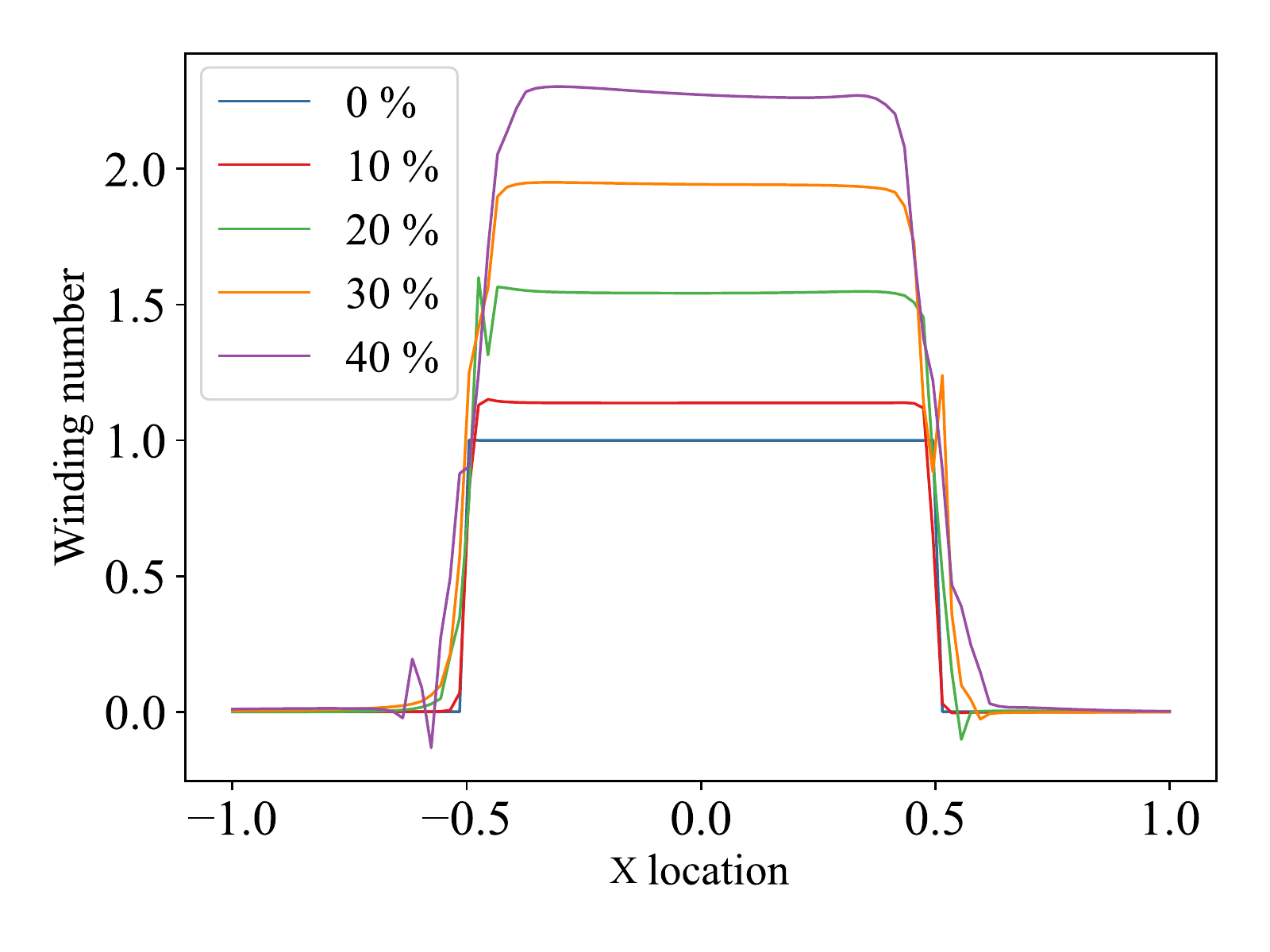}
        \caption{}
        \label{fig:winding-noise}
    \end{subfigure}
    \caption{Line-cuts of winding number along the $x$ axis for a sphere (a) with a different number of points in the point cloud, and (b) with an increase in the noise for the sampled point cloud.}
    \label{fig:winding-points-noise}
\end{figure}

Since the winding number is a scalar field, we analyze the behavior by plotting the winding number at one point along the $x$ axis. The ideal winding number for a sphere looks similar to the line-cut representing $163,842$ points as shown in \figref{fig:winding-points}. The winding number at $x<-0.5$ and $x>0.5$ for a sphere centered around origin with a diameter $1.0$ is $0$, and $1$ for $-0.5\leq x \leq 0.5$. When fewer points are sampled, the dipole moments (see \citet{barill2018fast} for a detailed discussion on dipole moments) do not cancel in the local region around the boundary, which leads to local self-intersections. However, we can resolve a watertight geometry without local self-intersections by using $0.5$ as a threshold to classify a given point to be within the geometry.

\subsubsection{Robustness to noise}

\begin{figure}[!t]
    \centering
    \begin{subfigure}[b]{0.48\textwidth}
        \includegraphics[width=\linewidth, trim={0pt, 0pt, 0pt, 0pt}, clip]{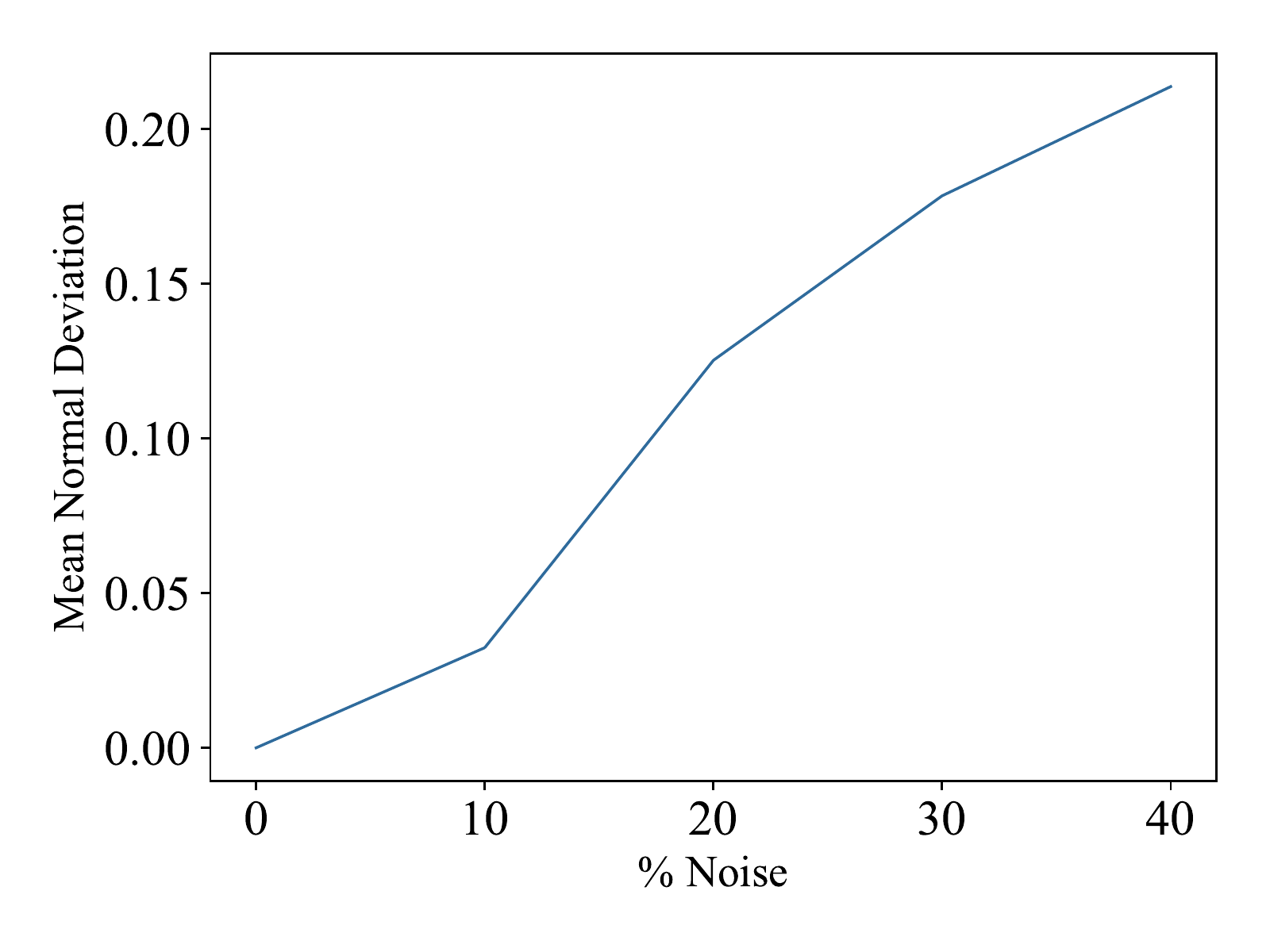}
        \caption{}
        \label{fig:points_noise-normals}
    \end{subfigure}
        \begin{subfigure}[b]{0.48\textwidth}
        \includegraphics[width=\linewidth, trim={0pt, 0pt, 0pt, 0pt}, clip]{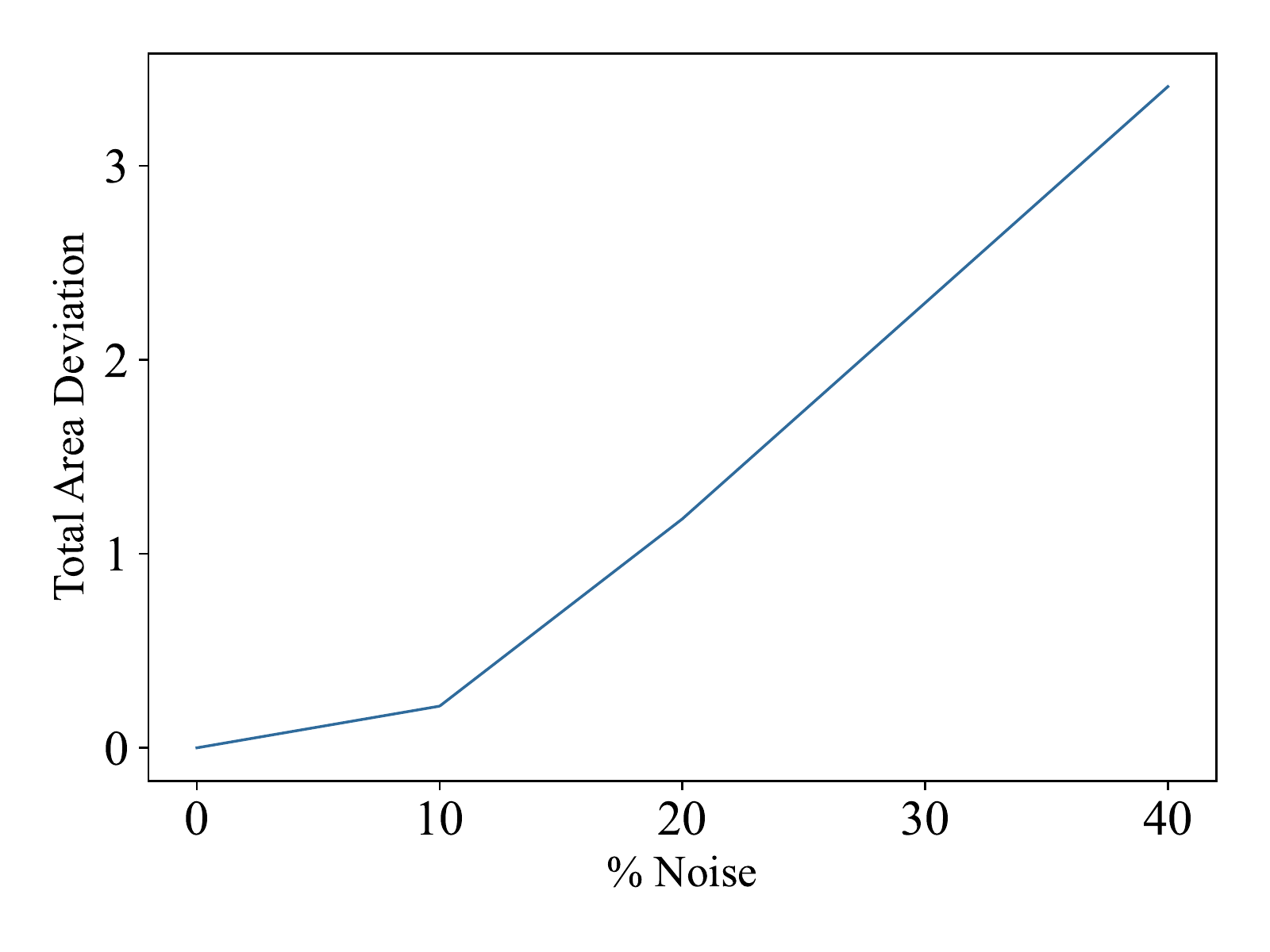}
        \caption{}
        \label{fig:points_noise-area}
    \end{subfigure}
    \caption{The degradation in the normal estimation and Voronoi area estimation with increasing variance of noise.}
    \label{fig:points_noise}
\end{figure}

\begin{figure}[!t]
     \centering
     \begin{subfigure}[b]{0.15\textwidth}
         \centering
        \includegraphics[width=\linewidth, trim={4.5in, 1.6in, 5.1in, 1.9in}, clip]{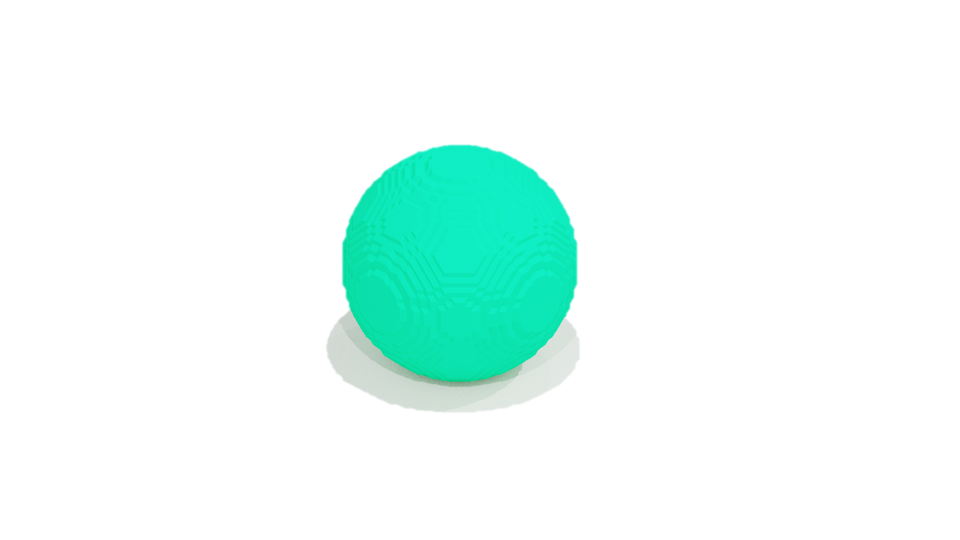}
         \caption{}
     \end{subfigure}
     \hfill
     \begin{subfigure}[b]{0.15\textwidth}
         \centering
         \includegraphics[width=\linewidth, trim={4.5in, 1.6in, 5.1in, 1.9in}, clip]{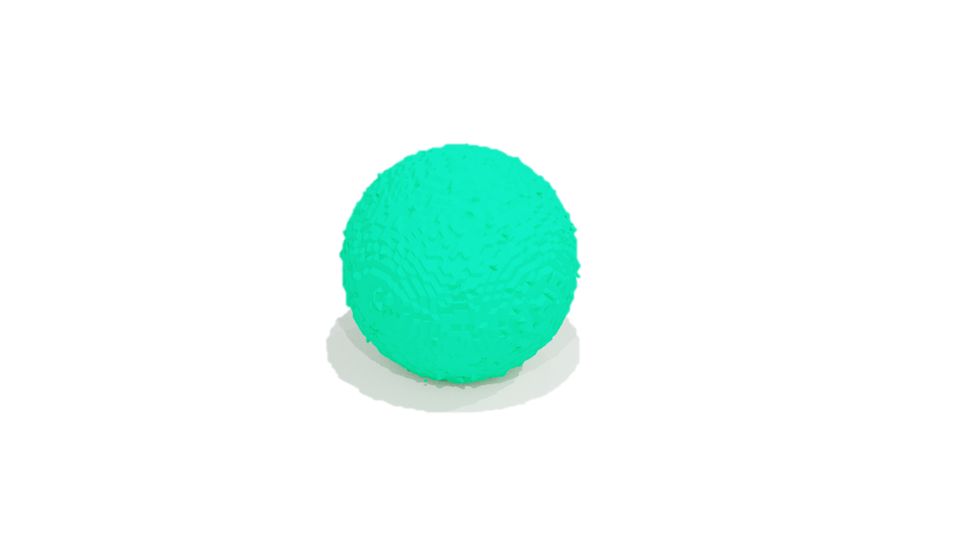}
         \caption{}
     \end{subfigure}
     \hfill
     \begin{subfigure}[b]{0.15\textwidth}
         \centering
         \includegraphics[width=\linewidth, trim={4.5in, 1.6in, 5.1in, 1.9in}, clip]{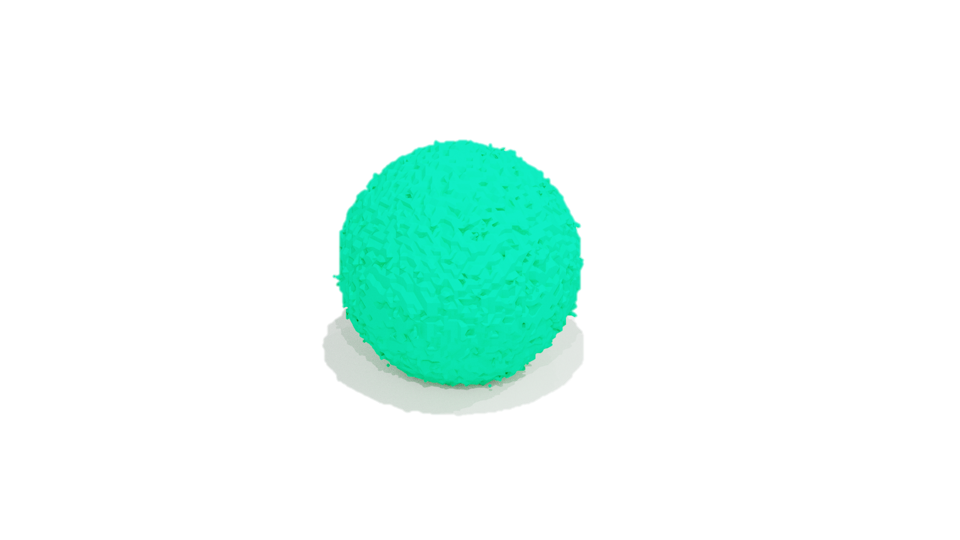}
         \caption{}
     \end{subfigure}
     \hfill
     \begin{subfigure}[b]{0.15\textwidth}
         \centering
         \includegraphics[width=\linewidth, trim={4.5in, 1.6in, 5.1in, 1.9in}, clip]{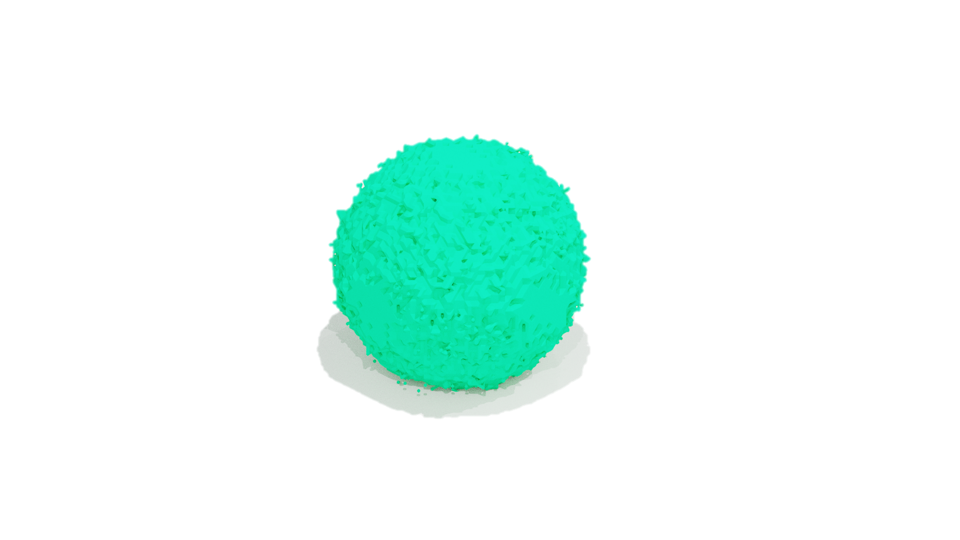}
         \caption{}
     \end{subfigure}
     \hfill
     \begin{subfigure}[b]{0.15\textwidth}
         \centering
         \includegraphics[width=\linewidth, trim={4.5in, 1.7in, 5.1in, 1.9in}, clip]{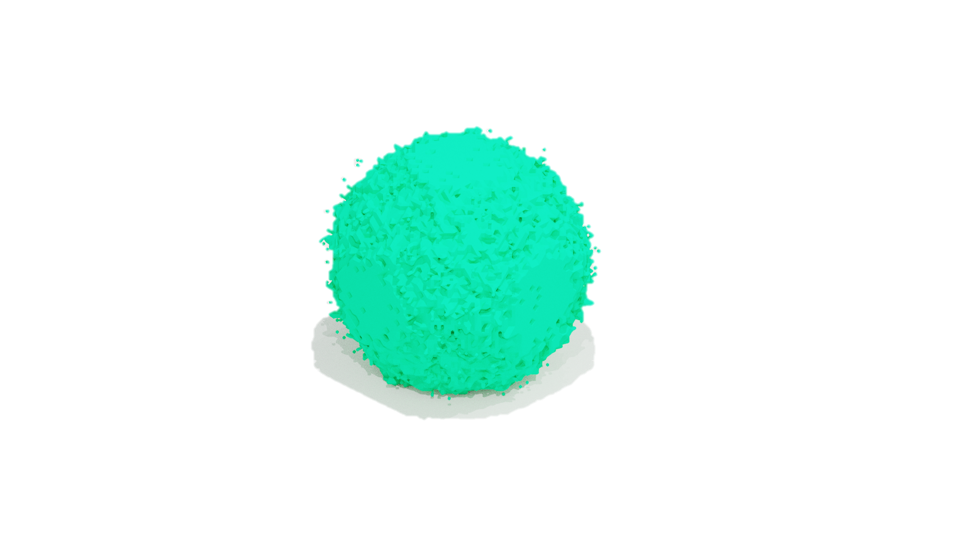}
         \caption{}
     \end{subfigure}
        \caption{Rendering of spheres obtained from reconstructed using winding number from (a) $0\%$, (b) $10\%$, (c) $20\%$, (d) $30\%$, and (e) $40\%$ Gaussian noise in the point cloud.}
    \label{fig:model_reconstruction}
\end{figure}

Often point clouds acquired from actual sensors have significant noise associated with them. While understanding the efficacy of noise removal approaches is not in the scope of this paper, we still would like to ensure that the proposed method is robust to noise. To understand the effect of noise, we add a specified Gaussian noise at each point. \figref{fig:points_noise} shows the increase in the deviation with an increase in the \% noise added to the original point cloud. To see how this affects the geometric reconstruction, we obtain line cuts of the winding number as earlier in \figref{fig:winding-noise}. As would be expected, adding more noise makes the reconstruction non-manifold. \figref{fig:model_reconstruction} shows the renderings of the reconstructed surface mesh of the sphere. With the increase in noise, many dimples are created on the surface that could change the aerodynamics of the shape; nevertheless, the shapes generated are all 2-manifold and watertight.

\subsection{Incompressible flow around a point cloud sphere}
\label{SubSec:SphereFlow}
We perform mesh convergence and validation studies using icosphere point clouds representing spheres. We create an icosphere by subdividing the faces of an icosahedron and projecting the resulting nodes onto the enclosing sphere. Though the icosphere does not achieve an ideal pattern of equilateral triangles, the generated geometry is equivalent to a well-defined 2-manifold surface mesh. From this tessellated icosphere, points are spawned at the center of each triangle element and collected into a point cloud. We begin with a description of the problem setup and background meshes used for the convergence studies and validations.

\begin{figure}[!t] 
    \centering
    \includegraphics[width=.68\linewidth]{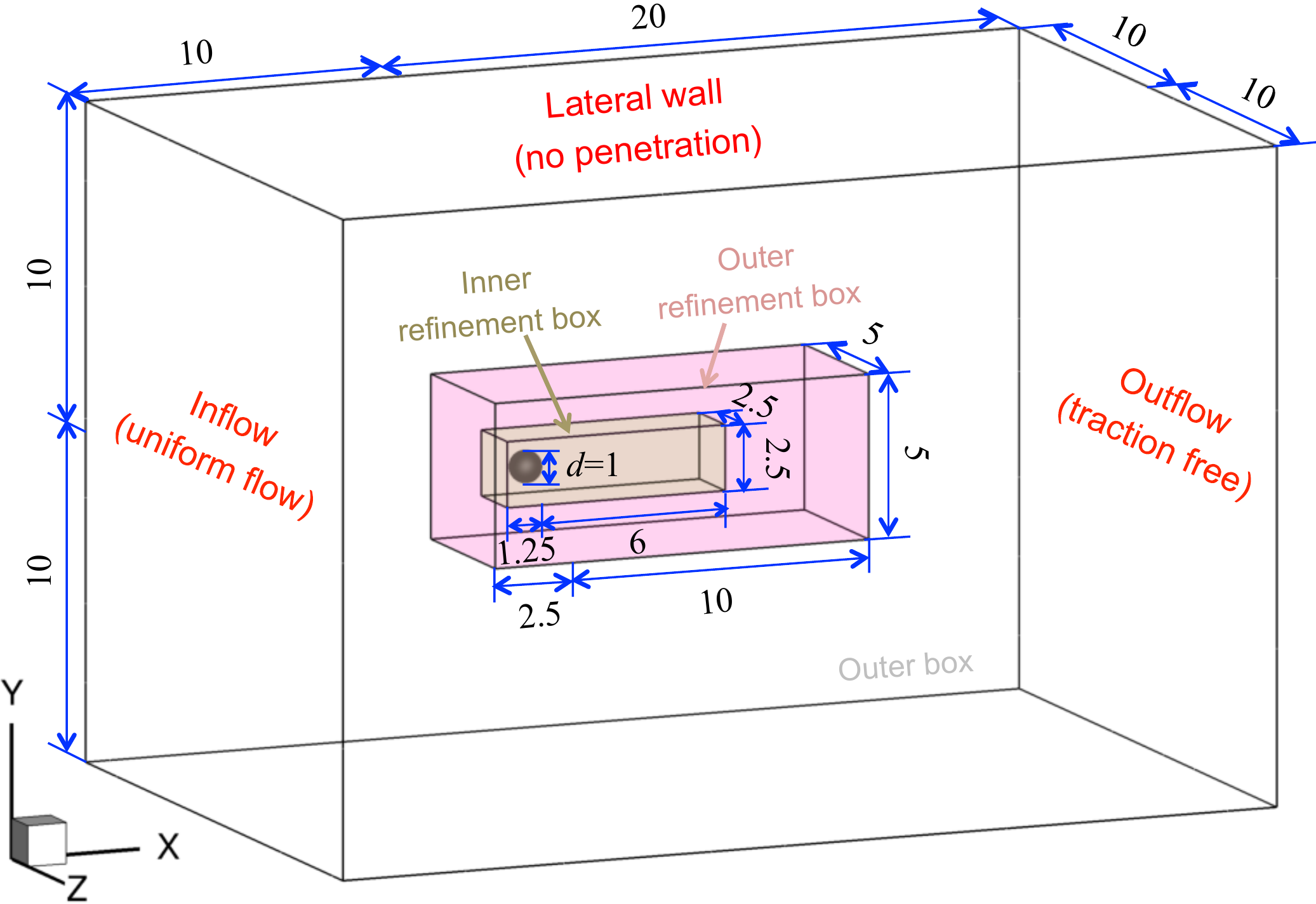}
    \caption{Computational domain for the icosphere flow validations. The outer frame bounds the fluid domain, enclosing an outer refinement region, which itself encloses an inner refinement region. The icosphere point cloud resides at the origin.}
    \label{fig:icosphere_domain}
\end{figure}

\subsubsection{Problem setup}

We follow the flow around a sphere study undertaken by \citet{Xu15ig} with the same domain dimensions and very similar background meshes.  Note that the problem is non-dimensional, and the icosphere has a diameter of one. The computational domain, boundary conditions, and immersed icosphere are shown in Figure~\ref{fig:icosphere_domain}. A uniform inflow velocity of one in the $x$-direction is set at the inlet, and the no-penetration condition is set at the lateral walls, both enforced strongly. The outflow boundary is traction free. The no-slip condition is enforced weakly on the sphere point cloud using the methods described in the previous sections. The density of the fluid is set to one, and the Reynolds number ($Re = \mu^{-1}$) is defined as the inverse of the viscosity. Non-dimensional mesh settings used for this problem are summarized in Table~\ref{tab:icosphere_meshes}. IM0, IM1, and IM2 are immersogeometric tetrahedral meshes of increasing mesh density. Note that IM2 has a comparable mesh resolution to the boundary-fitted mesh BM2 used in \citet{Xu15ig}, which we use as a reference. We consider $Re =$ 100, 300 and 3700 in this study, and the time step sizes in the simulations are set to $1.0\times10^{-2}$, $1.0\times10^{-3}$ and $1.0\times10^{-3}$, respectively. 

As point clouds lack continuous surfaces to recursively refine surface quadrature points, a certain spatial density of point cloud relative to the background mesh's cut element size is required to prevent flow leakage. \citet{Hsu16fqa} recommends a maximum ratio of 2 between the sizes of a tessellated surface element and the volume element cut by it. Before conducting validation studies, we derive an analogous ratio for point clouds with an icosphere geometry. This is found by converging flow quantities with increasingly dense point clouds on a fixed background mesh, specifically IM0.

\begin{table}[!t]
\centering
\caption{Element sizes in tetrahedral meshes for simulating flow around a sphere.}
\label{tab:icosphere_meshes}
\setlength\extrarowheight{3pt}
\begin{tabular}{l r r r r r}
    \hline
    Mesh & No. of elements & Cut element & Inner~region & Outer~region & Base~domain \\
    \hline
    \text{IM0} & $\text{304,330}$ & $0.020$ & $0.20$ & $0.8/\!\sqrt{2}$ & $1.2$ \\
    \text{IM1} & $\text{1,833,434}$ & $0.010$ & $0.10$ & $0.4/\!\sqrt{2}$ & $1.0$ \\
    \text{IM2} & $\text{9,041,302}$ & $0.005$ & $0.05$ & $0.2/\!\sqrt{2} $ & $0.8$ \\\hline
    \text{BM2} \cite{Xu15ig}& $\text{8,519,435}$ & $0.005$ & $0.05$ & $0.2/\!\sqrt{2} $ & $0.8$ \\
    \hline
\end{tabular}
\vspace{6pt}
\end{table}

\subsubsection{Point cloud convergence with fixed mesh}
\label{Sec:PointcloudConverge}

\begin{table}[!t]
\centering
\caption{Point cloud convergence for flow around a sphere with adaptive quadrature level 2 (AQ2) on background mesh IM0. Blank values indicate divergent solutions as a result of insufficient point cloud density.}
\label{tab:pcConverge}
\setlength\extrarowheight{3pt}
\begin{tabular}{l ll ll ll}
    \hline
    \multirow{2}{*}{Points} & \multicolumn{2}{c}{$Re = 100$} & \multicolumn{2}{c}{$Re = 300$} & \multicolumn{2}{c}{$Re = 3700$} \\
    \cline{2-3}   \cline{4-5}  \cline{6-7}
     & $C_D$         & $L/d$      & $\overline{C}_D$ & $St$        & $\overline{C}_D$ & $St$          \\ \hline
    642    & 3.531         & 1.627         &               &               &                &               \\
    2562   & 1.404         & 1.092         & 1.104         & 0.105         &                &               \\
    10242  & 1.087         & 0.982         & 0.690         & 0.125         & 0.643          & 0.062         \\
    40962  & 1.094         & 0.973         & 0.676         & 0.144         & 0.419          & 0.093         \\
    163842 & 1.094         & 0.973         & 0.676         & 0.144         & 0.419          & 0.093         \\
    655362 & 1.094         & 0.973         & 0.676         & 0.144         & 0.419          & 0.093         \\
    \hline
\end{tabular}
\end{table}

We simulate flow over six different densities of point clouds on IM0 for $Re =$ 100, 300, and 3700. The coarsest geometry has an average point spacing equal to about five times the cut element size, while the finest geometry has a point spacing of approximately $1/6$ of IM0's cut element size.  This tabulation aims to find the point cloud density at which our measured flow heuristics gain independence from point spacing. In \tabref{tab:pcConverge}, drag coefficient ($C_D$) and recirculation length ($L/d$) are listed for the steady-state $Re=100$ solutions, whereas we report time-averaged drag coefficient ($\overline{C}_D$) and Strouhal number ($St$) for vortex shedding cases of the two other Reynolds numbers. The drag coefficient is computed as $C_D=2F_D/(\rho U^2 A)$, where $F_D$ is the drag force, $\rho$ is the fluid density, $U$ is the inflow speed, and $A$ is the frontal area of the sphere. The drag force is evaluated using the variationally consistent conservative definition of traction~\cite{Bazilevs10d, Xu15ig}. The recirculation bubble length is computed as $L/d$, where $d$ is the diameter of the sphere, and $L$ is the length from the rear end of the sphere to the point where the velocity in $x$-direction changes sign. The Strouhal number is computed as $St=fd/U$, where $f$ is the frequency of vortex shedding.

As $Re$ increases, we observe an increase in the minimum point cloud density required to achieve stable flow solutions. As indicated by the blank cells in \tabref{tab:pcConverge}, the coarsest point clouds exhibit divergent solutions at higher Reynolds numbers, even with small time step sizes. As $Re$ increases, it becomes more necessary that each cut element encloses at least one icosphere point so that each cut element adequately ``feels'' the forcing contribution from the immersed point cloud to prevent flow leakage.

The results in \tabref{tab:pcConverge} show that convergence is consistently achieved when using a point cloud of 40,962 points. This density translates to an average point spacing valued at about $2/3$ of IM0's cut element size. This agrees with the spatial density recommendations of \citet{Xu15ig}, as this point cloud is the coarsest permutation that maintains a point cloud spacing smaller than IM0's cut element size. In subsequent simulations, we prescribe a point cloud density that ensures at least one point per cut element.

\subsubsection{Mesh convergence and flow validation}
Here, we perform convergence studies on mesh density and adaptive quadrature level and validate the flow quantities of interest at $Re =$ 100, 300, and 3700 against those reported in \citet{Xu15ig}. The adaptive quadrature refinement cases are denoted as AQ followed by the level number. One icosphere cloud, with its average point spacing less than the cut element size of the finest mesh IM2, is used in all simulations in this section. The same flow qualities as in \secref{Sec:PointcloudConverge} are evaluated and the results are shown in Tables~\ref{tab:pcConvergeRe100}--\ref{tab:pcConvergeRe3700}. For all cases, the convergence under adaptive quadrature refinement is clearly shown.

\begin{table}[!t]
\centering
\small
\caption{Convergence study for flow around a sphere at $Re$ = 100 with different mesh densities and adaptive quadrature levels.}
\label{tab:pcConvergeRe100}
\setlength\extrarowheight{3pt}
\newcommand{\tabincell}[2]{\begin{tabular}{@{}#1@{}}#2\end{tabular}}
\begin{tabular}{cllllll}
    \hline
                   \multirow{2}{*}{{Mesh}} & \multicolumn{3}{c}{${C}_D$} & \multicolumn{3}{c}{$L/d$} \\ \cline{2-7}
       & {IM0}  & {IM1}  & {IM2}  & {IM0}  & {IM1}  & {IM2} \\
    \hline
    AQ0         & 0.927         & 0.960         & 0.988         & 1.010         & 0.848         & 0.857        \\
    AQ1         & 1.078         & 1.080         & 1.092         & 0.997         & 0.846         & 0.859        \\
    AQ2         & 1.094         & 1.091         & 1.092         & 0.982         & 0.853         & 0.858        \\
    AQ3         & 1.094         & 1.091         & 1.092         & 0.973         & 0.854         & 0.858        \\
    \hline
    BM2~\cite{Xu15ig}    &     &  &1.093                    &  &  & 0.857                   \\
    \hline
\end{tabular}
\end{table}

\begin{table}[!t]
\centering
\small
\caption{Convergence study for flow around a sphere at $Re$ = 300 with different mesh densities and adaptive quadrature levels.}
\label{tab:pcConvergeRe300}
\setlength\extrarowheight{3pt}
\newcommand{\tabincell}[2]{\begin{tabular}{@{}#1@{}}#2\end{tabular}}
\begin{tabular}{cllllll}
    \hline
                    \multirow{2}{*}{{Mesh}} & \multicolumn{3}{c}{$\overline{C}_D$} & \multicolumn{3}{c}{$St$} \\ \cline{2-7}
    & {IM0}  & {IM1}  & {IM2}  & {IM0}  & {IM1}  & {IM2} \\
    \hline
    AQ0         & 0.602         & 0.620         & 0.649         & 0.139         & 0.144         & 0.139        \\
    AQ1         & 0.675         & 0.643         & 0.657         & 0.144         & 0.139         & 0.136        \\
    AQ2         & 0.677         & 0.658         & 0.662         & 0.142         & 0.136         & 0.135        \\
    AQ3         & 0.676         & 0.659         & 0.662         & 0.144         & 0.135         & 0.135        \\
    \hline
    BM2~\cite{Xu15ig} &  & & 0.661                   & & &0.135                 \\
    \hline
\end{tabular}
\end{table}

\begin{table}[!t]
\centering
\small
\caption{Convergence study for flow around a sphere at $Re$ = 3700 with different mesh densities and adaptive quadrature levels. BF denotes the boundary-fitted solution from \citet{Xu15ig}.}
\label{tab:pcConvergeRe3700}
\setlength\extrarowheight{3pt}
\newcommand{\tabincell}[2]{\begin{tabular}{@{}#1@{}}#2\end{tabular}}
\begin{tabular}{cllllll}
    \hline
                    \multirow{2}{*}{{Mesh}}& \multicolumn{3}{c}{$\overline{C}_D$} & \multicolumn{3}{c}{$St$} \\ \cline{2-7}
       & {IM0}  & {IM1}  & {IM2}  & {IM0}  & {IM1}  & {IM2} \\
    \hline
    AQ0         & 0.562         & 0.468         & 0.399         & 0.083         & 0.164         & 0.219        \\
    AQ1         & 0.516         & 0.407         & 0.395         & 0.089         & 0.163         & 0.218        \\
    AQ2         & 0.419         & 0.402         & 0.394         & 0.093         & 0.160         & 0.218        \\
    \hline
    BF~\cite{Xu15ig} & & &0.393              &   &   & 0.217 \\
    \hline
\end{tabular}
\end{table}

\begin{figure}[!t]
    \centering
    \includegraphics[width=1.0\linewidth]{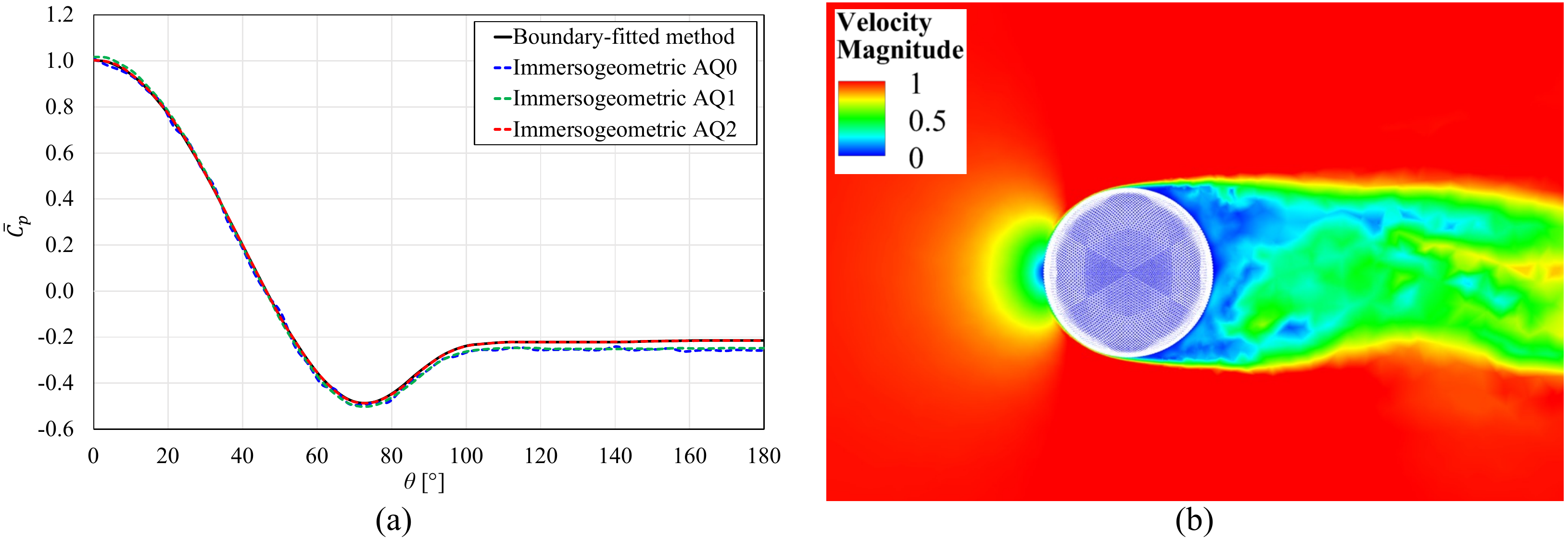}
    \caption{(a) Time-averaged pressure coefficient along the upper crown line of the sphere obtained using IM2 with different AQ levels. Immersogeometric results are compared against the reference boundary-fitted result \cite{Xu15ig}. (b) Velocity magnitude contour on a planar cut around the sphere.}
    \label{fig:cp_re3700}
\end{figure}

For $Re =$ 100 and 300, data obtained above and including IM1 and AQ2 refinement levels are in excellent agreement with the reference values (BM2) reported in \citet{Xu15ig}. The results clearly show mesh convergence and demonstrate the importance of capturing the point cloud geometry in cut cells when integrating the background elements. For $Re=3700$, the results show that IM2 is essential to obtaining accurate solutions. For this configuration and flow condition, there occurs a laminar flow separation near the equator of the sphere and a transition to turbulence in the wake of the object~\cite{rodriguez11direct}. Compared to the drag coefficient, the Strouhal number appears particularly sensitive to mesh density, unable to reach a closer value until IM2. In \figref{fig:cp_re3700}, the time-averaged pressure coefficient ($\overline{C}_p$) along the sphere's upper crown line is plotted for the IM2 case. $\theta = 0^\circ$ corresponds to the stagnation point and $\theta = 180^\circ$ is the trailing point of the sphere. AQ0 shows oscillatory behavior in $\overline{C}_p$, signifying that further refined quadrature points are required in cut elements. AQ1 remedies the oscillation but largely follows the same pressure coefficient curve as AQ0. Finally, AQ2 produces a result that is in excellent agreement with the boundary-fitted reference \cite{Xu15ig}. It should be noted that the refinement levels, solution accuracy, and convergence behavior presented in this section are in full agreement with those in \citet{Xu15ig}.

\subsection{Buoyancy-driven flow}

\begin{figure}[!t]
    \centering
    \includegraphics[width=0.8\linewidth]{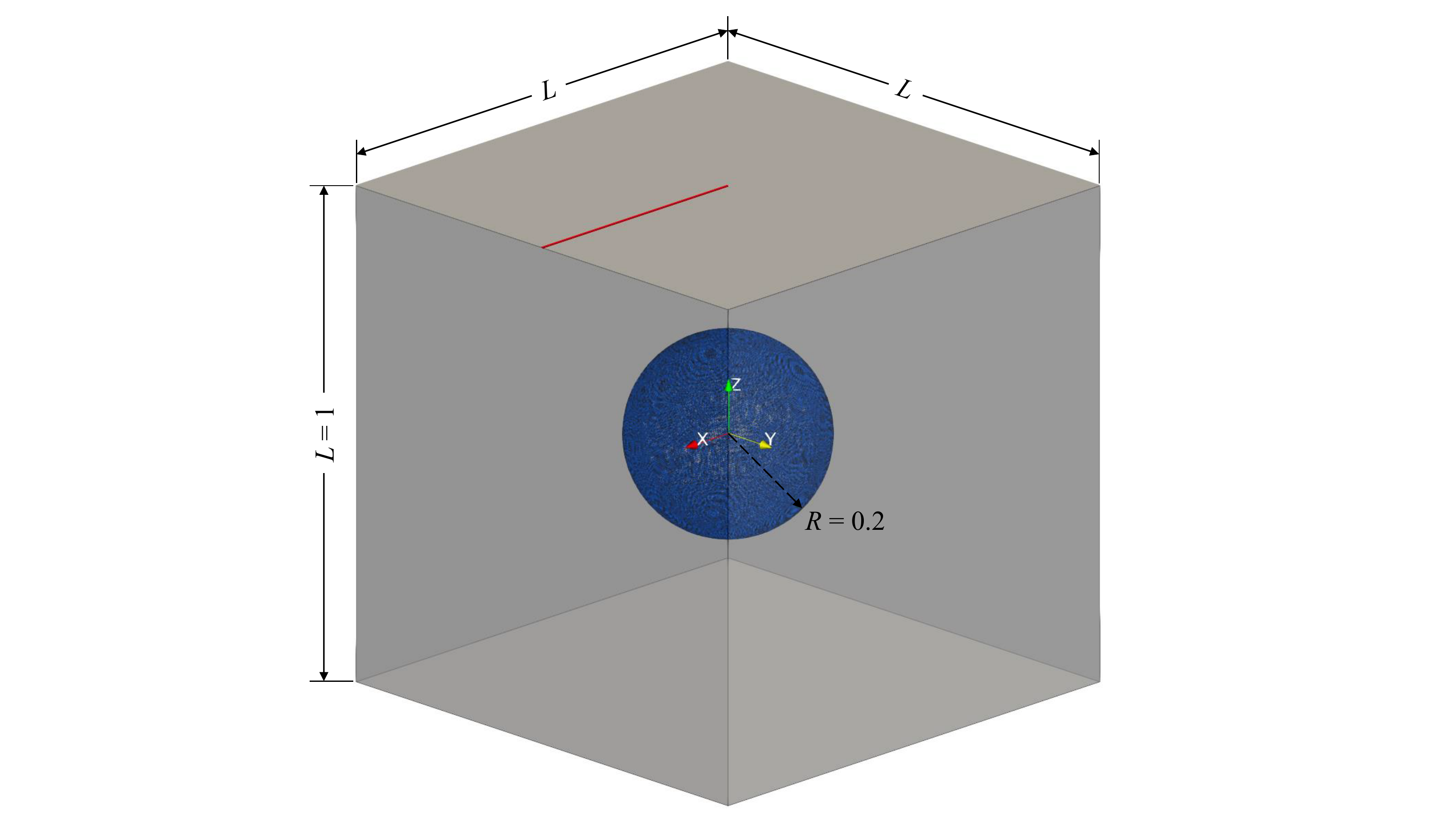}
    \caption{Computational domain for the problem of natural convection in a sealed cube enclosure with a heated sphere. The red line indicates where Nusselt number is plotted to compare against reference data.}
    \label{fig:therm_domain}
\end{figure}

After the validation of isothermal fluid flow, we validate the thermal IMGA by simulating buoyancy-driven flows. Following the problem of natural convection in an enclosure with a heated sphere investigated by \citet{Yoon10}, a static spherical boundary is situated within a sealed cubical enclosure filled with air, as shown in~\figref{fig:therm_domain}. Buoyant flow is driven by a temperature differential between the sphere and the enclosure walls. This case is chosen as it requires simulation of heat transfer involving immersed Dirichlet boundaries and accurate resolution and application of buoyant forces on fluids. The sphere's boundary, with a non-dimensional radius of $R=0.2$, is immersed within the cubical simulation mesh. This cube has non-dimensional edge lengths of 1, with the coordinate system at the domain's center and coordinate axes parallel to the domain's edges. The gravity acts in the $-z$ direction. All boundaries are treated as no-slip walls. A high temperature $T_\text{h} =1$ is applied on the sphere, while a low temperature $T_\text{c}=0$ is applied on the cube domain surfaces. The normalized temperature can be defined as $\Theta=(T-T_\text{c})/(T_\text{h}-T_\text{c})$.

We simulate two cases at Rayleigh number $Ra=1000$ with $\delta=0$ and $\delta=0.25$. Rayleigh number is defined as $Ra = g \beta L^3 (T_\text{h}-T_\text{c})/(\nu \alpha)$, where $g=1$ is the gravitational acceleration, $\beta=1$ is the thermal expansion coefficient, $L=1$ is the length of the enclosure, $\nu$ is the kinematic viscosity, and $\alpha$ is the thermal diffusivity. $\delta$ refers to the $z$ coordinate of the sphere where $x=0$ and $y=0$. For both cases, the Prandtl number, $Pr = \nu/\alpha$, of air is used ($Pr = 0.7$). With the aforementioned problem setup, one can adjust the value of $\nu$ or $\alpha$ to achieve the desired $Ra$ number. Three meshes are tested with each of the two $\delta$ cases, with non-dimensional element sizes listed in \tabref{tab:buoyancymesh}. The mesh sizes on the cube domain surfaces and near the immersed sphere are the ``base elements'' and ``cut elements'' sizes, respectively, and a smooth transition is achieved in the space between the boundaries. A time step size of $5.0\times10^{-3}$ is used in all simulations.

\begin{table}[!t]
\centering
\caption{Non-dimensional element size settings of each simulation mesh for buoyancy-driven flow validation.}
\label{tab:buoyancymesh}
\setlength\extrarowheight{3pt}
\begin{tabular}{lr r}
    \hline
    {Mesh} & {Cut elements} & {Base elements} \\
    \hline
    \text{Coarse} & $0.020$ & $0.040$ \\ 
    \text{Medium}& $0.010$ & $0.020$ \\ 
    \text{Fine} & $0.005$ & $0.010$ \\ 
    \hline
\end{tabular}
\end{table}

\begin{figure}[!t] \centering
    \includegraphics[width=0.49\linewidth,trim={0.8in 0.9in 0.8in 0.8in},clip]{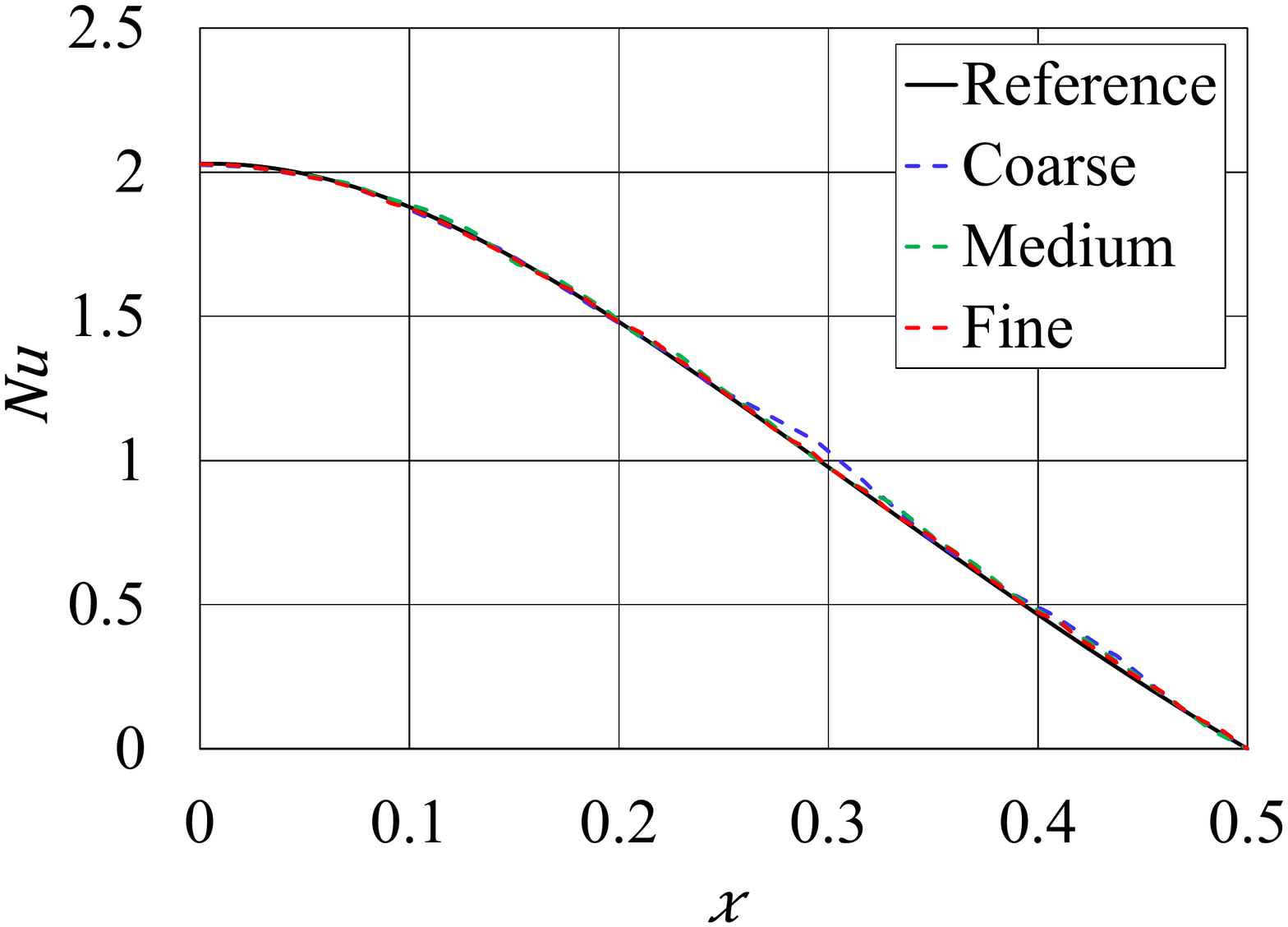}
    \includegraphics[width=0.49\linewidth,trim={0.8in 0.9in 0.8in 0.8in},clip]{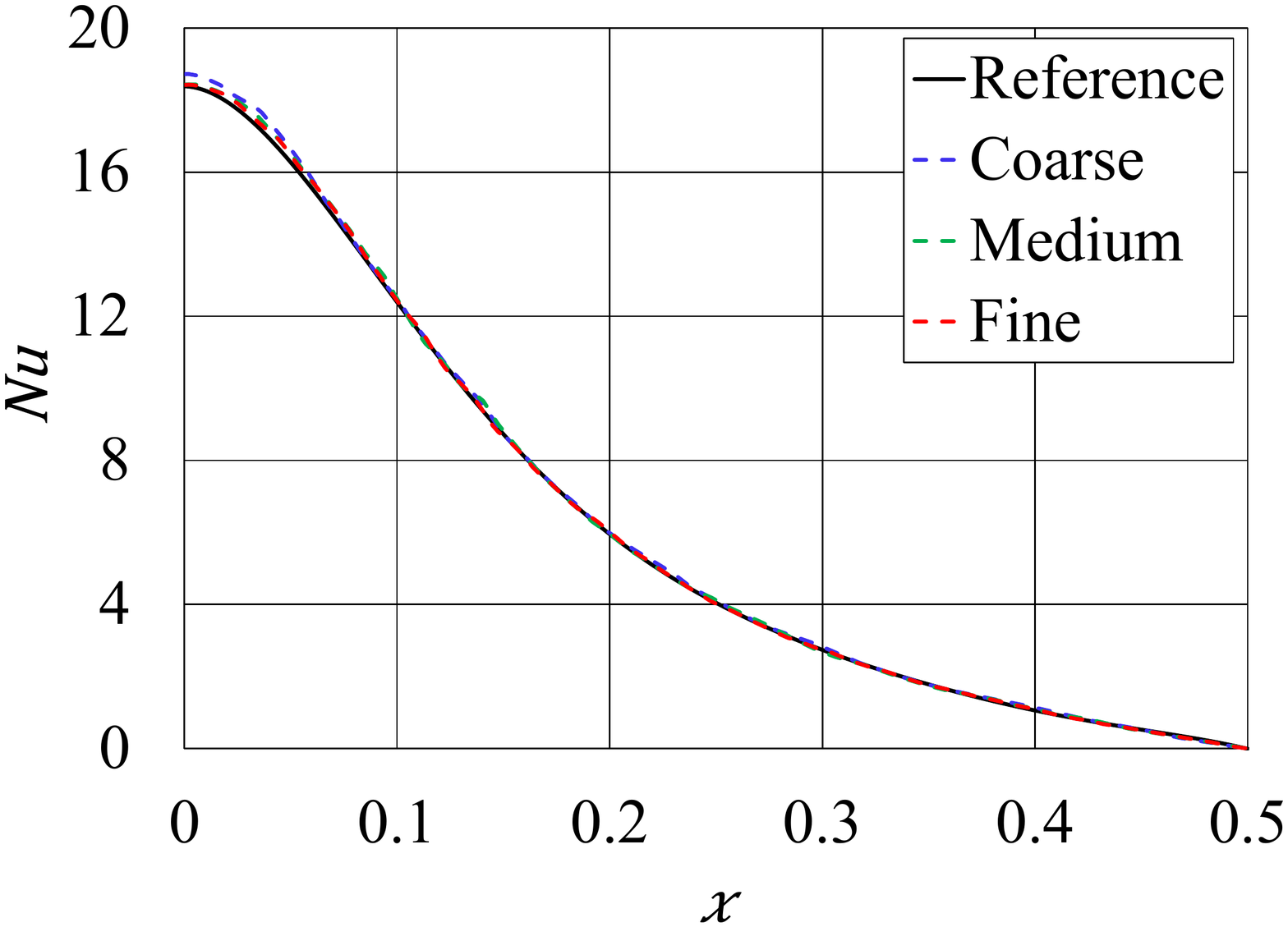}
    \caption{Nusselt number along the centerline of the top surface of the sealed enclosure for $\delta=0$ (left) and for $\delta=0.25$ (right).}
    \label{fig:therm}
\end{figure}

For each of the six simulation case permutations, the Nusselt number ($Nu$) is calculated at the top surface of the enclosure, referring to the wall perpendicular to the $z$-axis located at $z=0.5$. The Nusselt number can be calculated by $Nu =\nabla\Theta\cdot\mathbf{n}$,
where $\mathbf{n}$ is the unit normal vector to the wall. In \figref{fig:therm}, we plot this value along a line between $\left(0.0, 0.0, 0.5\right)$ and $\left(0.5, 0.0, 0.5\right)$ and compare our results with the reference values reported by \citet{Yoon10}.

In the case of $\delta=0$, for the most part, even the coarse mesh is sufficient in reproducing the temperature gradient curve. All meshes accurately simulate boundary values relative to the reference solution. Unsurprisingly, the coarse mesh produces some notable solution oscillations. Grid independence is apparent after the medium mesh, where the medium and fine meshes improve the solution accuracy and track the reference data to a strong degree. Similarly, all three meshes with $\delta=0.25$ produce comparable solutions near the edge of the top surface. The coarse mesh solution noticeably deviates from the reference solution at the center of the top surface. Again, the coarse mesh exhibits some oscillations throughout the solution, whereas the medium and fine meshes are successful in reproducing the reference curve.

\subsection{Compressible flow over a torpedo-shaped body}

\begin{figure}[!t] \centering
    \includegraphics[width=0.9\linewidth]{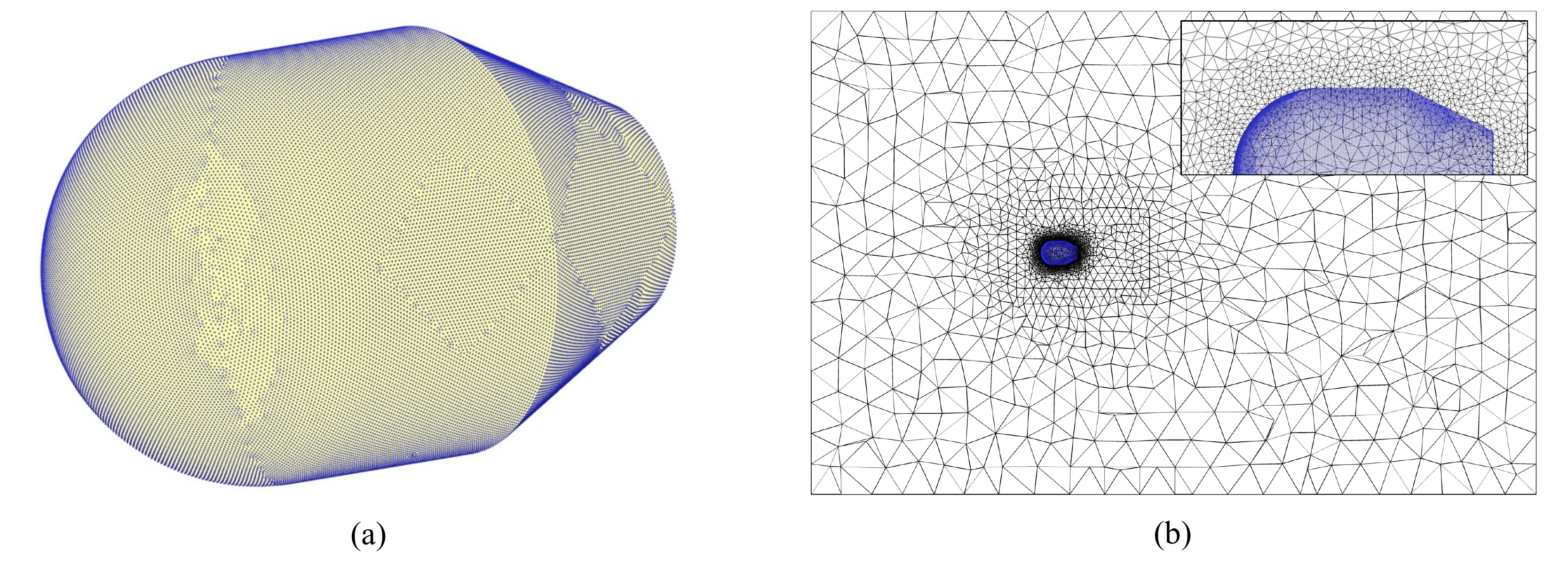}
    \caption{(a) Point cloud representation of the torpedo-shaped body. (b) IM0 mesh with a zoom on the region near the point cloud representation of a torpedo-shaped body.}
    \label{fig:bullet_mesh}
\end{figure}

\begin{figure}[!t] \centering
    \includegraphics[width=.95\linewidth]{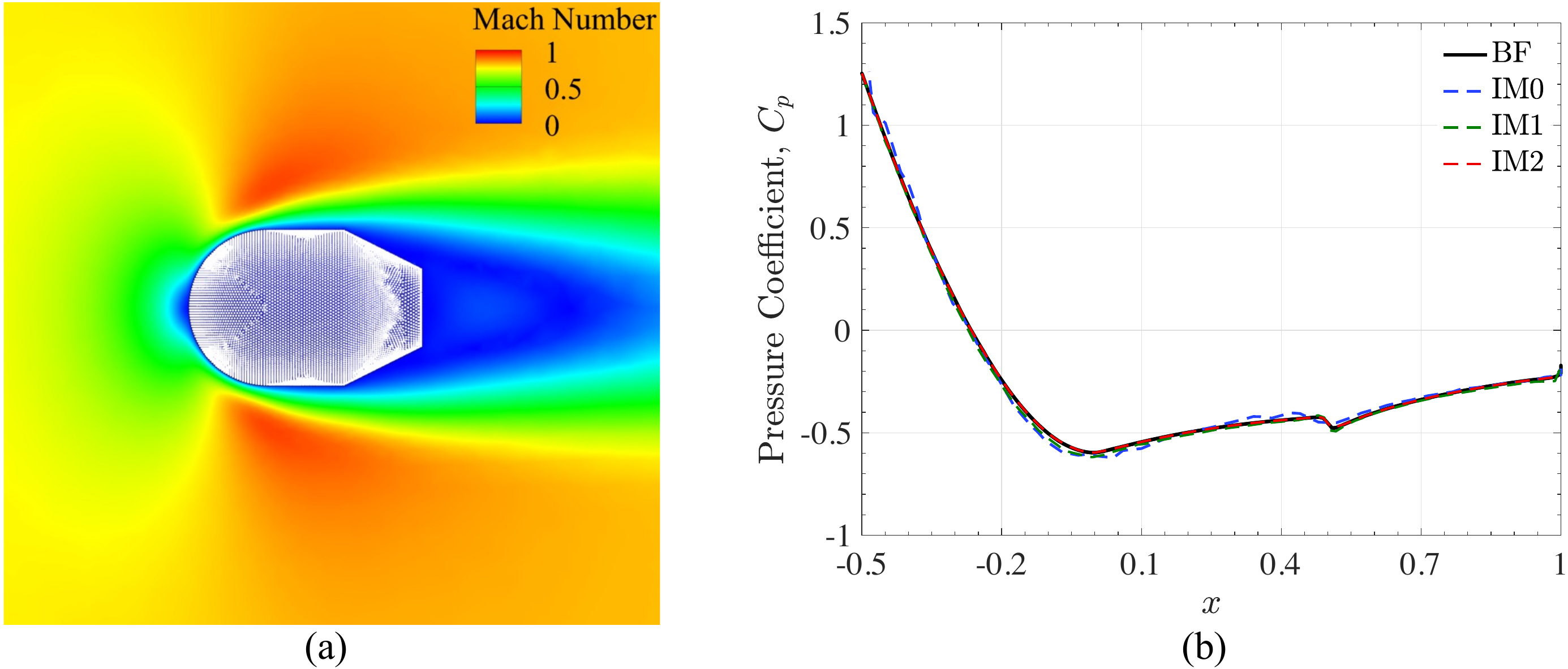}
    \caption{(a) Mach number contours for the subsonic flow ($M=0.8$) around a torpedo-shaped body. (b) Pressure coefficient along the upper crown line of the torpedo-shaped body as a function of the streamwise coordinate.}
    \label{fig:bullet_pressure}
\end{figure}

In this section, we simulate laminar flow around a torpedo-shaped body represented by the point cloud data in both the subsonic and supersonic regimes. The dimension of the geometry and the computational domain can be found in \citet{Wang17gp}. The point cloud representation of the torpedo-shaped body is shown in \figref{fig:bullet_mesh}{a}. To perform the simulation at subsonic speed of $M = 0.8$, the inflow quantities are set to $p=1.1161$, $\|\mathbf{u}\|=1.0$, and $T=3.8713\times 10 ^{-3}$. The dynamic viscosity $\mu$ is set to a constant value of 0.01. For the supersonic case of $M = 2.0$, the inflow quantities are set to $p=0.1786$, $\|\mathbf{u}\|=1.0$, and $T=6.1941\times 10 ^{-4}$. The dynamic viscosity is determined from Sutherland's law: $\mu = ({C_1 T^{\frac{3}{2}}})/({T+S})$, where $S = 1.406 \times 10^{-4}$ and $C_1=0.906$. For both the subsonic and supersonic cases, no-penetration and zero-heat flux boundary conditions are enforced on all the lateral boundaries of the domain. The outlet boundary is set to have the same total traction as the inlet. On the point cloud object, the velocity is set to zero, and the temperature is set as the stagnation temperature determined by $T_\text{D} = (1+0.5(\gamma-1)M^2)T$; both conditions are enforced weakly. The heat capacity ratio $\gamma$ is 1.4, the ideal gas constant $R$ = 288.293, and the Prandtl number $Pr$ is 0.72. Note that all quantities are dimensionless.

\begin{figure}[!t] \centering
    \includegraphics[width=.95\linewidth]{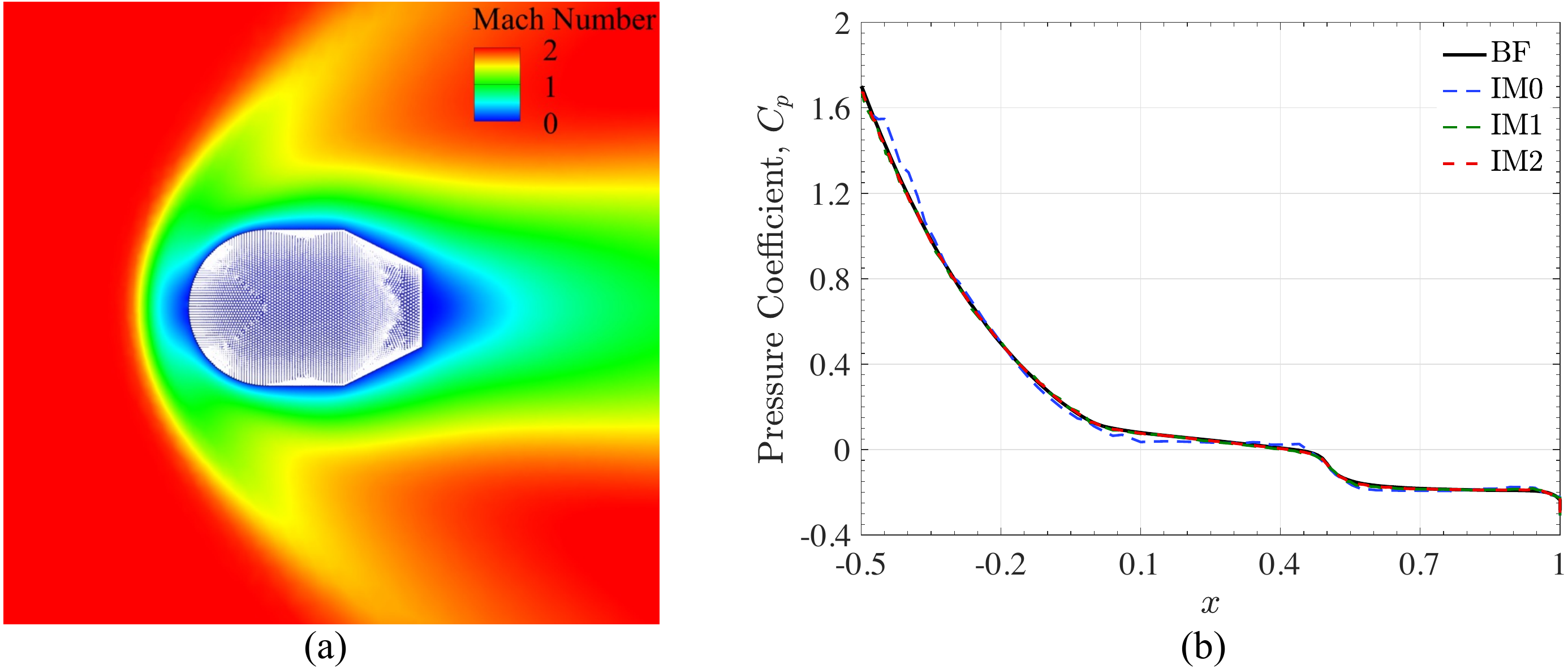}
    \caption{(a) Mach number contours for the supersonic flow ($M=2.0$) around a torpedo-shaped body. (b) Pressure coefficient along the upper crown line of the torpedo-shaped body as a function of the streamwise coordinate.}
    \label{fig:bullet_pressure_sup}
\end{figure}

We perform a mesh refinement study to assess the performance of the compressible-flow point-cloud IMGA formulation. Simulations are carried out on three meshes named IM0, IM1, and IM2, from coarser mesh to finer mesh, respectively (see \citet{Wang17gp} for the notation and statistics of these meshes). To illustrate the mesh design, we plot a planar cut through the center of the coarsest mesh IM0 in \subfigref{fig:bullet_mesh}{b}. The simulations are performed using a time step size of 0.005 until a steady state is reached. \subfigref{fig:bullet_pressure}{a} and \subfigref{fig:bullet_pressure_sup}{a} show the Mach number contour plots computed on IM2 for the subsonic and supersonic cases, respectively. The pressure coefficient distributions along the upper crown line of the torpedo-shaped body as a function of the streamwise coordinate for different meshes for the subsonic and supersonic cases are plotted in \subfigref{fig:bullet_pressure}{b} and \subfigref{fig:bullet_pressure_sup}{b}, respectively. The pressure coefficient results are also compared with the boundary-fitted computations using a comparable mesh resolution for both cases. The results demonstrate that IM1 and IM2 meshes produce converged solutions and show excellent agreement with the boundary-fitted computations. They are also in excellent agreement with those reported in \citet{Xu19ct}. Note that a two-level recursive adaptive quadrature rule is employed to faithfully capture the immersed geometry and produce an essentially converged solution.

\section{Flow over an industrial vehicle}
\label{Sec:Loader}

Extending upon the practical application of IMGA to industrial geometry in B-reps and tessellated formats~\cite{Xu15ig, Hsu16fqa, Wang17gp, Xu19ct}, we demonstrate the compatibility of this method with remarkably detailed geometry in a point cloud format. The workflow for computational analysis of an industrial product design typically begins with a B-rep CAD model representing an industrial object to be manufactured. This B-rep model primarily exists for purposes other than CFD, meaning that the geometry is often unsuitable for boundary-fitted analysis. Significant effort is invested to either ``clean up'' the existing B-rep or create a surrogate version with limited complexity and airtight topology. Only then may boundary-fitted mesh generation begin, accompanied by its own set of difficulties. \citet{Hsu16fqa} and \citet{Wang17gp} subverted these computational analysis roadblocks using IMGA on an agricultural tractor and a tractor-trailer truck in trimmed NURBS and analytic surface formats. Compared to the tractor and truck, the geometry of this demonstration is unique in that it is overwhelmingly composed of finite-thickness shells, one type of topology targeted by CAD cleanup operations.

The geometry in this demonstration represents a medium-sized construction vehicle: a John Deere 544K Wheel Loader. Given that this is a low-speed utility vehicle, there is little concern for aerodynamic forces experienced by the vehicle, which is contrary to common vehicular CFD analysis, like that of a tractor-trailer truck. For this type of vehicle in an actual product analysis application, fluid simulation seeks to forecast how components, especially the engine and electronics, remain within thermal constraints throughout their continuous operation. Physics-coupled simulations are particularly useful to predict temperatures that the key components of the vehicle experience. Regarding the 544K demonstration that we subsequently exhibit, fluid flow similarly transpires within this low-speed incompressible flow regime, though we purposely avoid boundary conditions expressing resemblance to conditions within the operating range of the real-life vehicle. In this section, we demonstrate the utility of point-cloud IMGA applied to an industrial analysis workflow.

\begin{figure}[!t]
    \centering
    \includegraphics[width=0.99\linewidth]{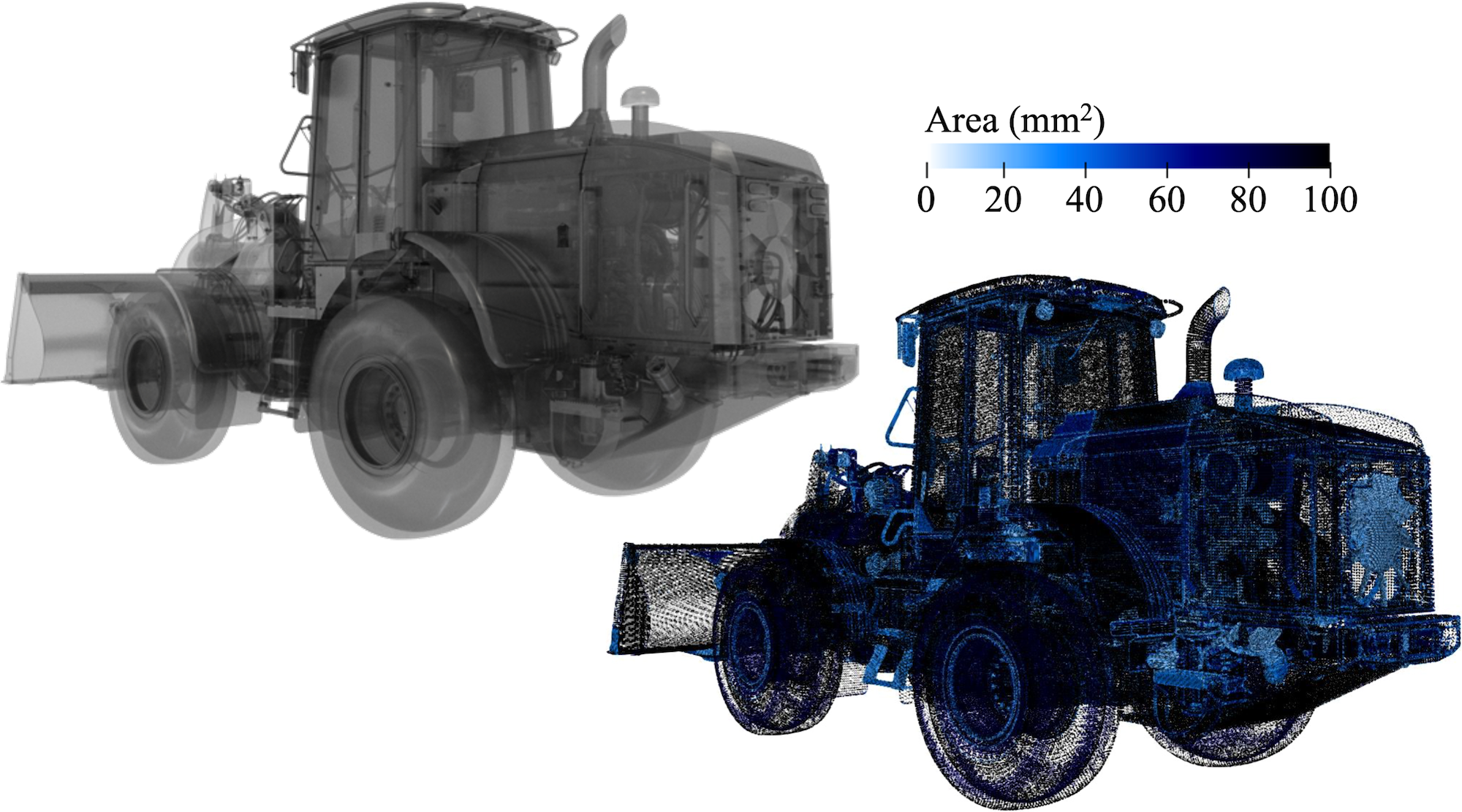}
    \caption{Translation of John Deere 544K construction vehicle's surface representation to point cloud format. (Left) Translucent surfaces exposing complex details within the vehicle. (Right) Sampling of surface representation into points, each point colored by its scalar area.}
    \label{fig:544K_surfacetopoints}
\end{figure}

The generality of the point cloud format affords flexibility regarding the types of CAD formats that can be analyzed. The \emph{as-manufactured} analysis advantage of point-cloud IMGA is fully realized with the virtualization of physical objects, but true \emph{as-designed} analysis still presents worthwhile advantages over conventional analysis of simplified geometries. Lacking a practical 3D scan, we begin with the comprehensive CAD assembly used by Deere~\&~Company for design and manufacturing purposes. This B-rep is ill-suited for boundary-fitted analysis as it is non-manifold with gaps, intersections, and collocations. Instead of converting it to watertight NURBS, analytic, or tessellated surfaces, the assembly is sampled into points with normal vectors and area scalars as in \figref{fig:544K_surfacetopoints} in a matter of minutes. A significant amount of manual labor is saved by circumventing the geometry simplification process typically required for boundary-fitted analysis of such complex geometry. All impermeable features within the vehicle are retained; in the absence of support for sub-scale porosity, this preliminary demonstration omits porous screens and radiators. It should be noted that all parameters utilized to simulate the 544K are unrepresentative of reality, and presented simulation results make no claims regarding the performance of the actual product as it exists physically. This exercise demonstrates point-cloud IMGA's compatibility with multiphysics simulation using complex point clouds.

\begin{figure}[!t]
    \centering
    \includegraphics[width=0.6\linewidth,trim={0in 0.3in 0in 0.3in},clip]{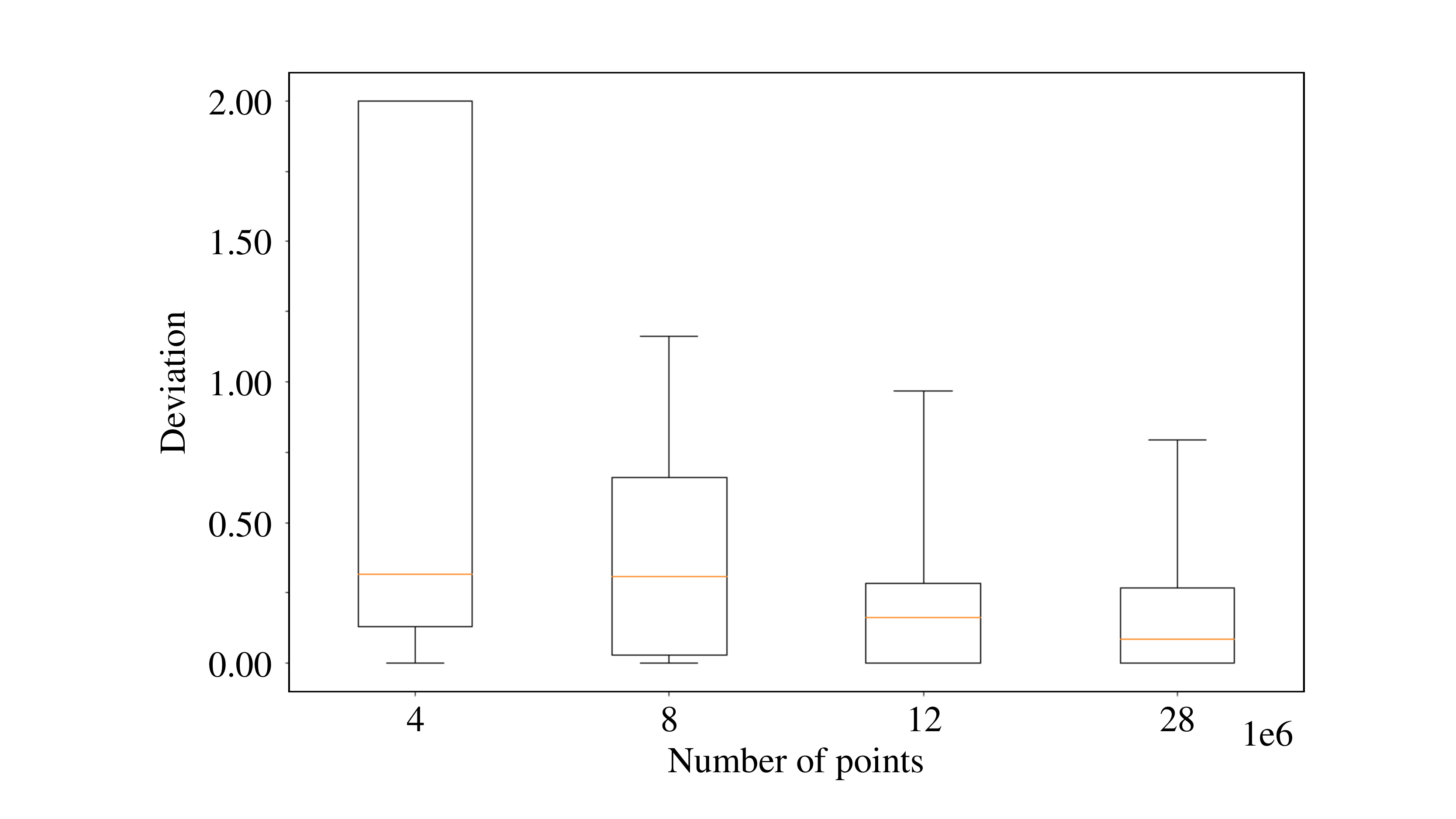}
    \caption{Box plot of the deviation of calculated normals against analytical normal vectors of the John Deere 544K. The same procedure is performed on four point clouds of the 544K with varying point density.}
    \label{fig:pointcloud_deviation}
\end{figure}

\subsection{Geometric pre-processing accuracy}
The vehicle's point cloud is discretized by sampling its B-rep surface model. We begin with a 0.025\,m spacing between points, which returns a total of 4,041,875 point cloud elements. This operation filters small geometric features, automating an otherwise intensive manual task. Similar to the icosphere, sampling of the surface geometry provides normal vectors and area scalars as analytical values to compare the quality of pre-processing approximations. Point spacing is decreased to produce three more point clouds with greater element populations of 8,135,975 points, 11,855,474 points, and 27,937,410 points. Normals estimation is performed using 4 neighboring points, and a box plot of each point cloud's collection of normal vector deviations is displayed in \figref{fig:pointcloud_deviation}.

Here, the performance of jet-fitting resembles the normal estimation on the Fandisk. A deviation of two indicates flipped normals, suggesting that the point cloud with four million elements contains an excess of incorrectly aligned normal vectors. \citet{cazals2005estimating} warn that the ambiguity of normal vectors on flat surfaces is a weakness of normal vector calculation in its current form. Using a larger number of neighboring points for the normal estimation increases the error, which we theorize is due to the ubiquity of thin plates in this geometry. As the number of neighbors increases for a point on a thin plate, points on one side of the plate are grouped with points on the other side, malevolently affecting calculations for points on both sides. As the density of the point cloud increases, a point on one side of a plate is more likely to be in a neighborhood exclusively with points also on the same side of the plate. As will be discussed later, we find that the two coarsest point clouds are too sparse to produce reliable normals for stable flow simulations of John Deere 544K using point-cloud IMGA. The vehicle point cloud with 12 million elements, which has an average distance of 0.00625\,m between points, is therefore used for the subsequent studies.

\begin{figure}[!b]
    \centering
    \includegraphics[width=0.99\linewidth]{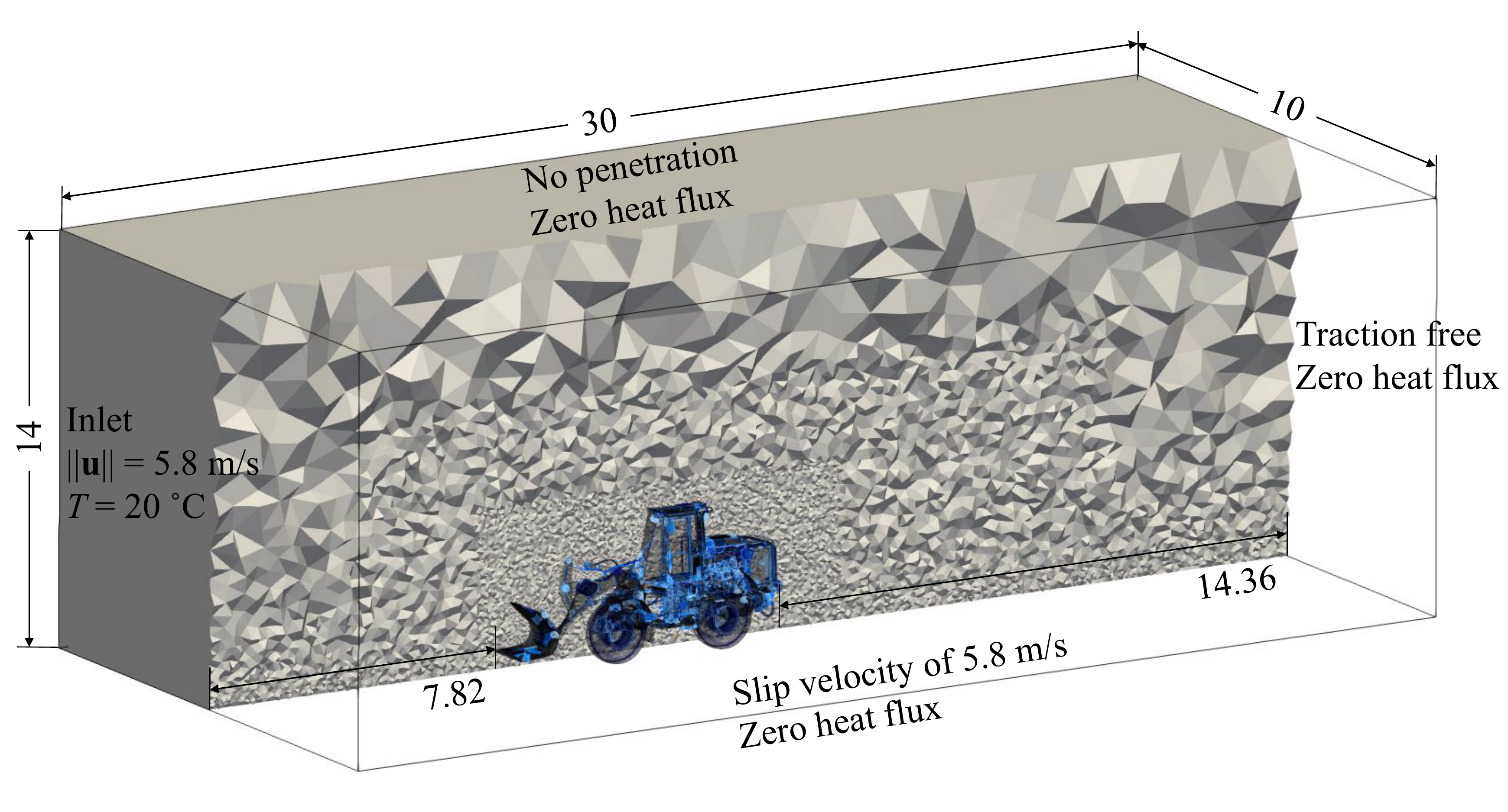}
    \caption{Mesh refinement, domain dimensions (in meters), and boundary conditions for the flow analysis of John Deere 544K driving forwards.}
    \label{fig:544K_mesh}
\end{figure}

\subsection{Problem setup}

\figref{fig:544K_mesh} visualizes a mid-plane clip of the domain mesh along with the boundary condition setup. A uniform flow velocity of 5.8\,m/s is applied at the inlet, and the outlet is traction free. A slip velocity of 5.8\,m/s is applied on the stationary ground boundary to produce the ground effect of a forward-moving vehicle in a stationary fluid domain. No-penetration conditions are applied on the top and lateral walls. The wheels of the construction vehicle are modeled as rotating no-slip walls, with the angular velocity matching the tire surface velocity with the floor's slip velocity at their common interface. The fan located rearward of the vehicle is modeled similarly and rotates at 600\,RPM to draw air out of the engine compartment. All other elements of the point cloud are treated as stationary no-slip walls. These no-slip velocity boundary conditions are weakly enforced on the vehicle's point cloud.

In this incompressible airflow simulation, we use standard properties of air at atmospheric pressure and temperature $T = 20\,^{\circ}$C. The density is $1.204\,\text{kg}/\text{m}^{3}$, the dynamic viscosity is $ 1.825~\text{kg}/(\text{m\,s})$, the specific heat is $1007\,\text{J}/(\text{kg\,K})$, and the thermal conductivity is $0.02514\,\text{W}/(\text{m\,K})$. Points belonging to the engine, transmission, and axles are weakly enforced to a temperature of $T = 100\,^{\circ}$C. For all other elements of the vehicle point clouds, zero-heat flux thermal conditions are assumed. The domain is initialized with an ambient temperature of 20\,$^\circ$C, which is also applied strongly at the inlet as a reference. Zero-heat flux conditions are prescribed on all other surfaces of the fluid domain. The time step size for the simulation is $\Delta t = 1.0 \times 10^5$.

\subsection{Immersed mesh generation}

Only in rare cases is an open-source meshing tool such as Gmsh \cite{geuzaine2009gmsh} suitable for boundary-fitted volume meshing of production-ready product assemblies. However, owing to the topological simplicity of generating a parallelepiped volume mesh, notable examples of open-source mesh generation software typically provide more than enough features to set up an IMGA simulation case such as this example. Here, the functionalities accessed through Gmsh allow us to create the simulation domain in a streamlined manner. 2D and 3D element size parameters help control mesh density in the flow regions of interest. For targeted refinement of tetrahedra near the John Deere 544K point cloud, Gmsh's  ``Attractor'' field trivializes the import of point coordinates to define smooth refinement regions surrounding every point cloud element. Continuing the theme of open-source workflow, sampling discrete points from CAD surfaces is similarly straightforward with Open\,CASCADE~\cite{opencascade} and freely available Python packages libigl~\cite{koch2019geometric} and trimesh~\cite{Trimesh}.

Visible in \figref{fig:544K_mesh}, the meshing strategy includes two rectangular prism refinement regions with the ``outer'' region completely enclosing the ``inner'' region, and elements adjacent to the floor have element size equal to that of the ``inner'' region's elements. A spherical refinement zone originates from each point cloud element of the coarsest case ($\sim$4 million points), prescribing a tetrahedral element size of 0.025\,m to comfortably retain an average of at least one point cloud element per volume boundary element. This heuristic ratio is chosen to negate leakage of physical domain flow into the quiescent fictitious domain, based on the study conducted in Section \ref{SubSec:SphereFlow}. The size of the elements that contain (cut cells) or are near the point cloud is 0.025\,m. The element size in the inner refinement zone and close to the floor is 0.1\,m. The element sizes in the outer refinement zone and in the far field are 0.4\,m and 1.0\,m, respectively. Including both the physical and fictitious domains, the immersed mesh consists of 19,803,968 tetrahedral elements. A large majority of elements within the interior compartment must be refined, which increases the overall element count. 

As mentioned earlier, the average distance between the points sampled on the B-rep model for the coarsest case is 0.025\,m. Point spacing is decreased to produce three more point clouds with denser point distributions. For IMGA flow simulations, we observe that the estimated normals on the two coarsest point clouds are not sufficiently accurate to produce stable flow solutions, even though they satisfy the one-point-per-element criterion defined earlier. In the evaluation of weak-boundary-condition formulation, oscillatory normals can lead to significant instability. We find that the point cloud with 12 million points, which has an average spacing between points to be around 0.00625\,m, has enough accuracy in the normal estimation and can produce a stable flow solution for our demonstration purposes. It should be noted again that requiring higher point cloud density here is not due to the point-cloud IMGA formulation but because of the need to obtain better normals estimation.

The parallelization strategy proposed by~\citet{HsuAkk11a} is employed for the IMGA simulations presented in this paper. In this strategy, the problem mesh is partitioned into subdomains by balancing the number of elements in each partition; each subdomain is then assigned to a processing core. However, this approach can sometimes create a highly unbalanced distribution of quadrature points in each partition since quadrature points aggregate in the cut elements due to the use of adaptive quadrature. In this work, we use the strategy proposed by \citet{Xu19ct} and weigh each element by the number of quadrature points it contains. This is then used as the metric in the graph/mesh partitioning package METIS~\cite{Karypis98} for determining mesh partitions by balancing the summation of user-defined weights. This approach produces a more balanced computational load on each processing core.

\subsection{Simulation results}

\begin{figure}[!t]\centering
    \includegraphics[width=\textwidth]{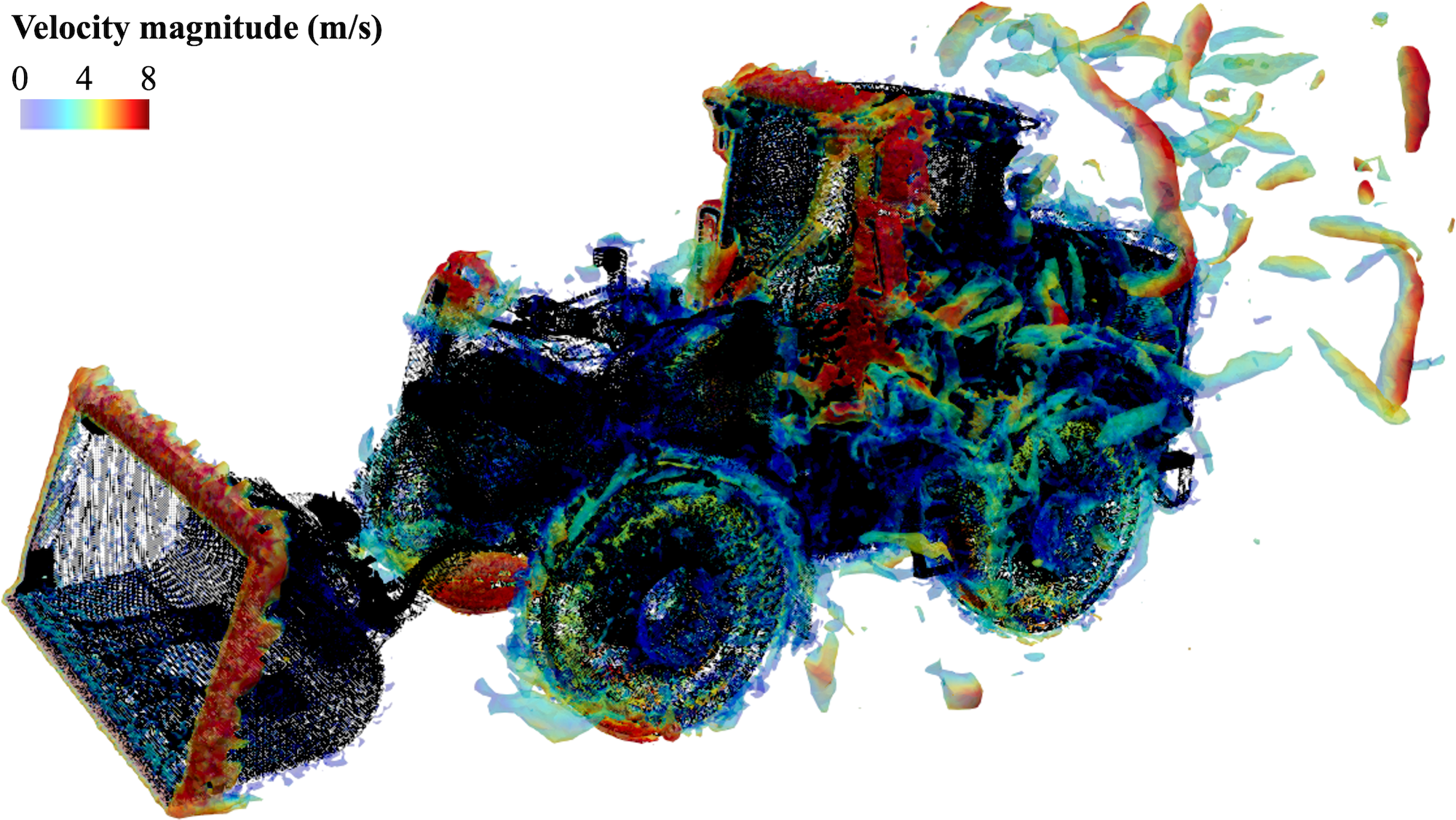}
    \caption{Visualization of the instantaneous vortical structures of turbulent flow around the John Deere 544K colored by the velocity magnitude. The point cloud representation of the vehicle is used directly to perform IMGA.}  
    \label{fig:544K_vortex}
    \end{figure}  

\begin{figure}[!t]
    \centering
    \includegraphics[width=0.9\linewidth]{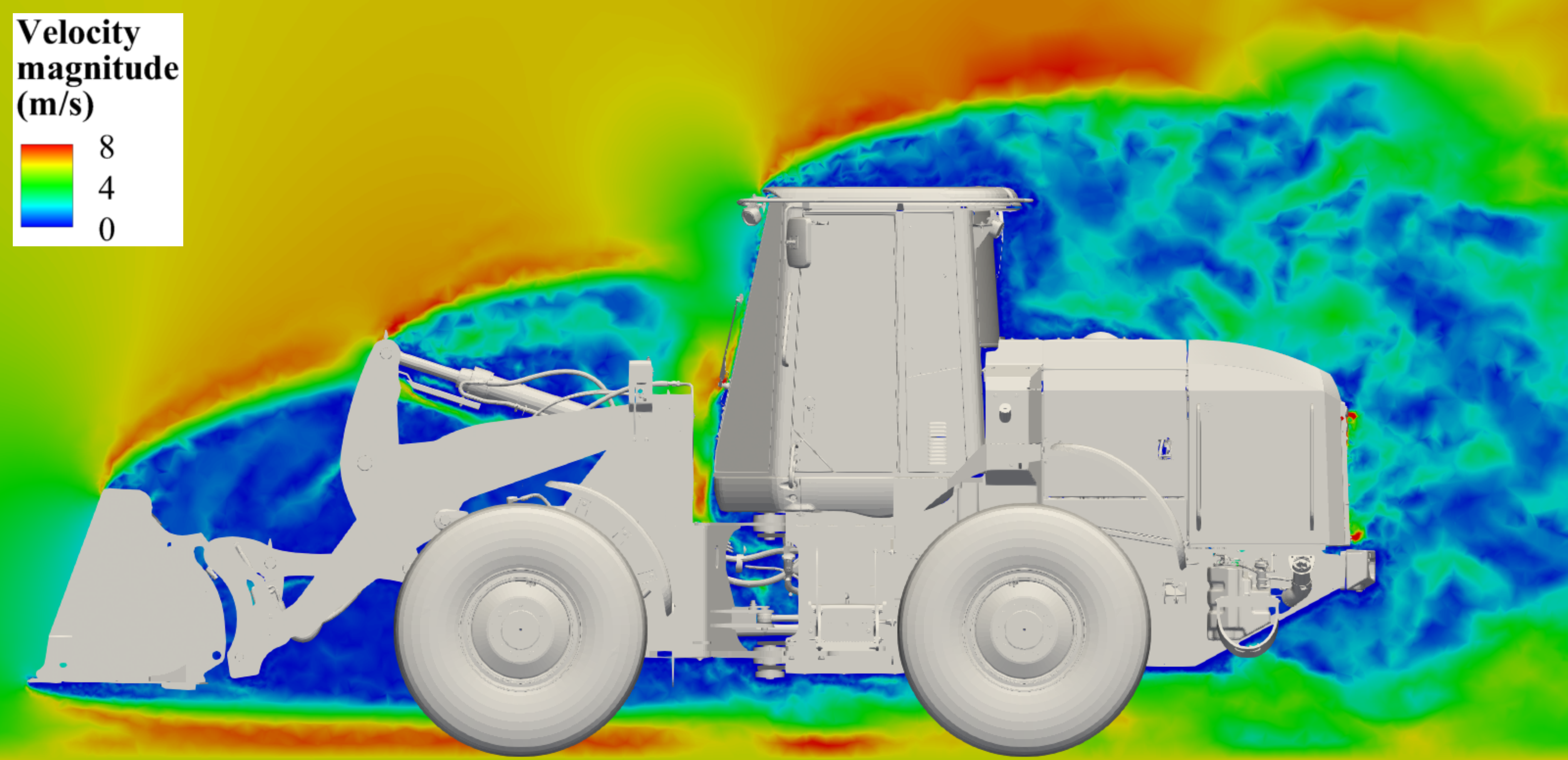}
    \caption{Planar slice of the velocity field down the vehicle's center line.}
    \label{fig:544K_side}
\end{figure}

\begin{figure}[!t]
    \centering
    \includegraphics[width=0.9\linewidth]{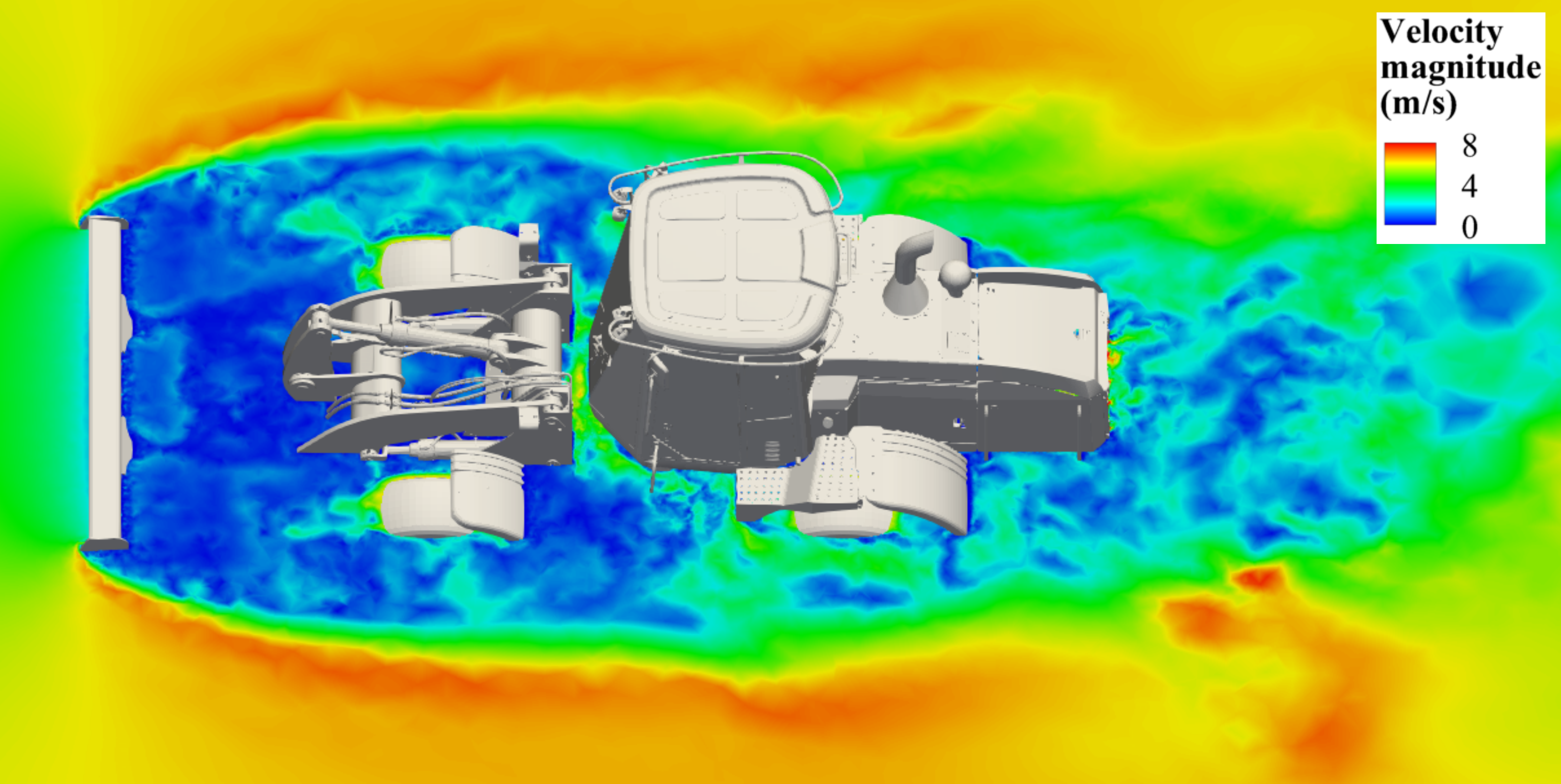}
    \caption{Slice of the velocity field on a plane parallel to the ground.}
    \label{fig:544K_iso}
\end{figure}

We solve for the airflow around the construction vehicle with the aforementioned boundary conditions and mesh. In this section, we observe an instantaneous snapshot of the flow solution. We focus on the velocity field, confirming that external airflow conforms to the immersed boundaries constructing the outer shell of the vehicle. We also look further inside the vehicle, especially in the engine compartment, to reveal the treatment of complex thin shell boundaries by inside-outside testing based on winding number. The latter portion of this section examines the vehicle interior, focusing on heat conduction and convection in the flow temperature field.

As is expected at $Re=3\times10^{6}$, external flow is fully turbulent across the vehicle. \figref{fig:544K_vortex} shows the visualization of the instantaneous vortical structures colored by the velocity magnitude of turbulent flow around the John Deere 544K. The construction vehicle is visualized using a point cloud that was used to directly perform immersogeometric fluid flow and heat transfer analysis. To distinguish internal airflow from external airflow in subsequent visualizations, we substitute the surface CAD model in place of the point cloud utilized for simulation. \figref{fig:544K_side} shows that the front bucket attachment forces the brunt of flow movement, simulating a recirculation bubble above and downstream of the bucket. The driver cabin appears to direct some flow into the lower engine compartment, whereas flow over the top of the cabin immediately separates and heightens the large wake region behind the 544K. Thicker boundary layers are apparent at the rear of the vehicle due to the adverse pressure gradients present there. Finally, the fan extracts air from the engine compartment and adds axial flow to the turbulent wake.

\begin{figure}[!t]
    \centering
    \includegraphics[width=0.9\linewidth]{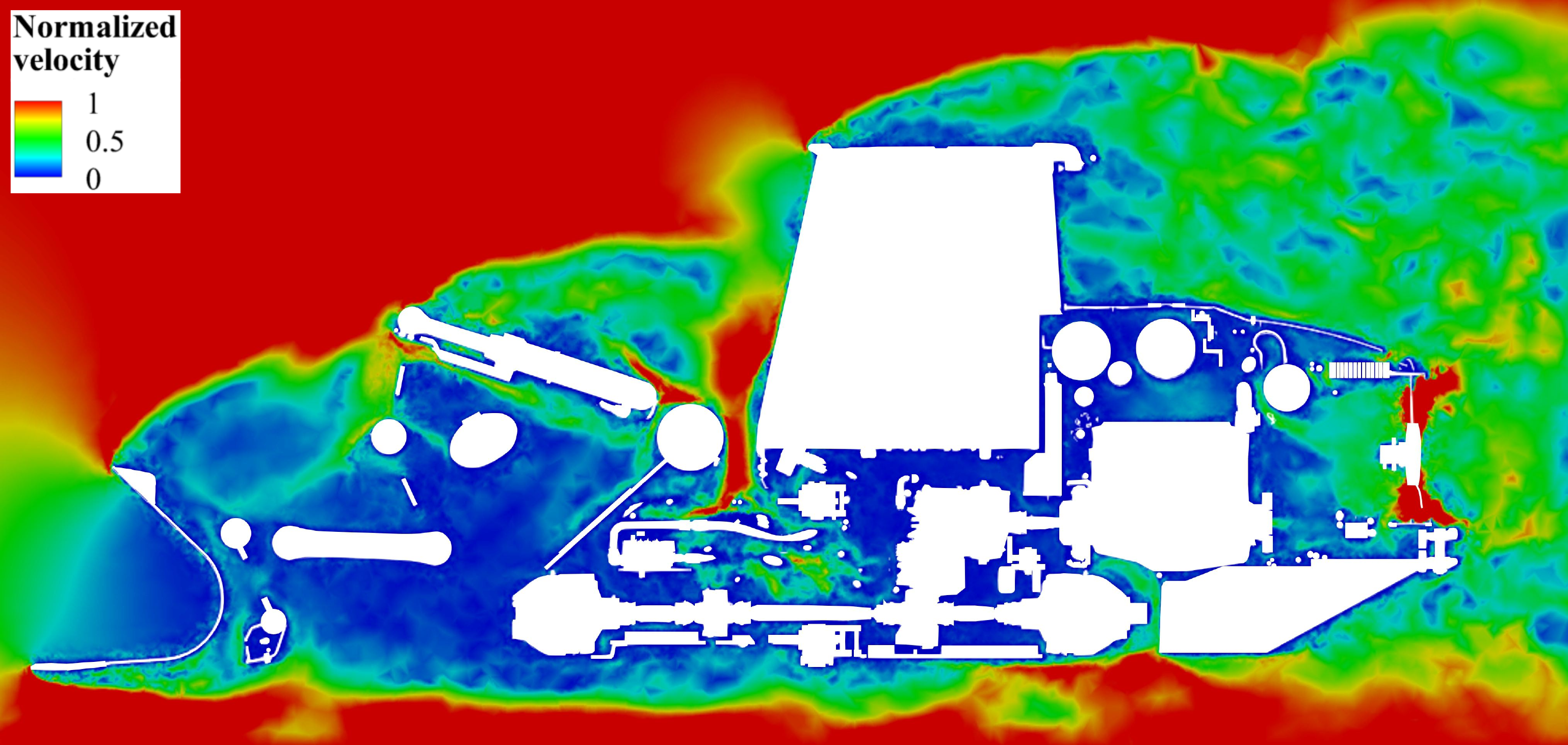}
    \caption{Planar slice of both the normalized velocity field and 544K along the vehicle's center line. The velocity magnitude is normalized by the freestream velocity.}
    \label{fig:544K_U_inside}
\end{figure}

\begin{figure}[!t]
    \centering
    \includegraphics[width=0.9\linewidth]{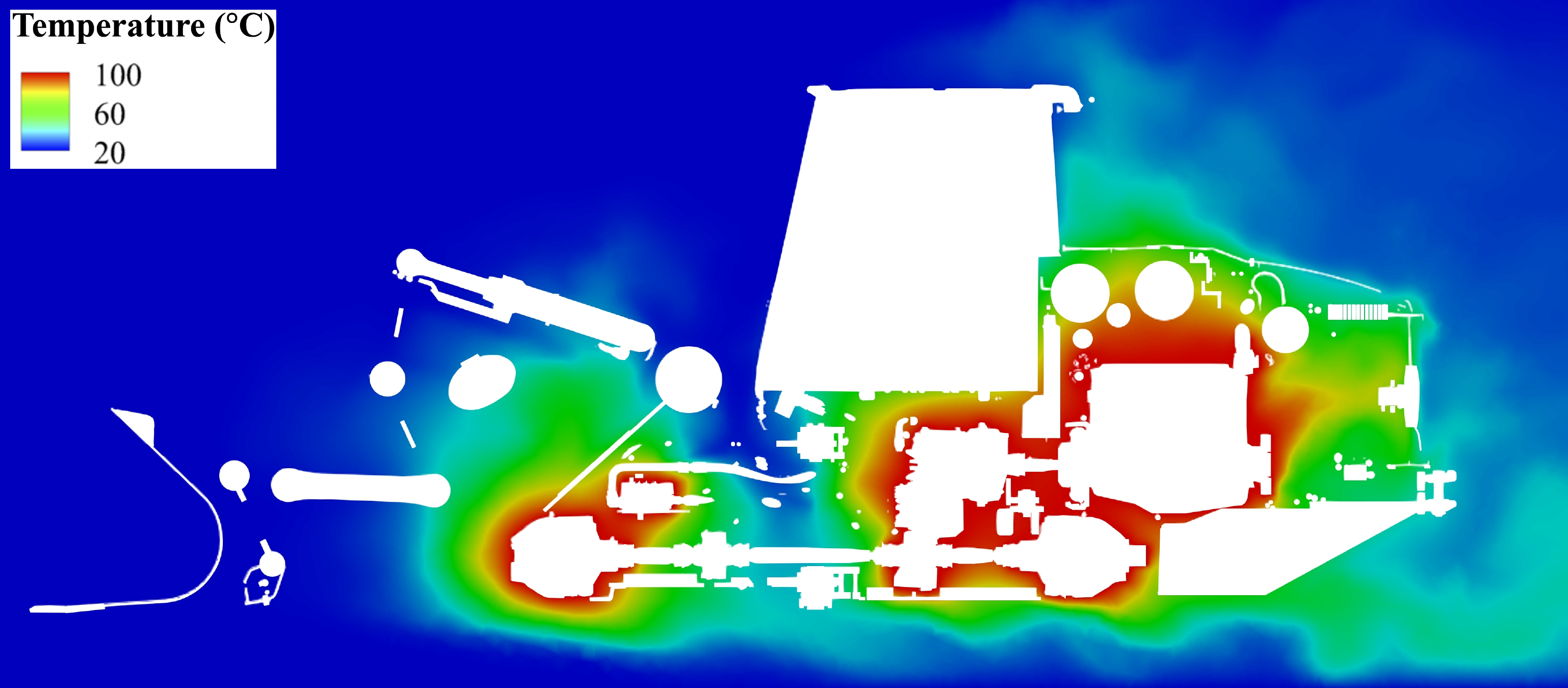}
    \caption{Planar slice of both the temperature field ($T$, in $^\circ$C) and 544K along the vehicle's center line.}
    \label{fig:544K_T_inside}
\end{figure}

Looking at another sample of the same velocity field snapshot in \figref{fig:544K_iso}, the rotating wheel boundary conditions are evident. The surface velocity of each wheel's contact patch is equal to the freestream velocity, applied as a weak boundary condition to elements containing pertinent elements of the point cloud. The tires create minor recirculation bubbles downstream, but vortex shedding appears to be deterred by the presence of mudguards. We see another dimension of flow movement by the front bucket attachment to the degree that the vehicle's entire width is engulfed by the bucket's wake region. From this angle of the velocity field slice, the fan shows a greater level of flow extraction and interacts with the weak vortex street trailing the vehicle. Moving onto the interior compartment in \figref{fig:544K_U_inside}, we feature the same velocity field sample as in \figref{fig:544K_side} but with 2D slicing of the 544K and visualization of individual mesh elements. The bulk of airflow enters the interior section through the gap between the cabin and the hydraulic assembly. We observe moving air in the inner region, desirable for convecting heat away from critical components into the colder ambient air.

We conclude this solution analysis by looking at the temperature field in \figref{fig:544K_T_inside}, which is purely demonstrative of point-cloud IMGA's heat transfer capabilities and unrepresentative of the real-life vehicle's actual performance. A temperature boundary condition of value $T = 100\,^{\circ}$C is weakly imposed on points belonging to the front axle hub, electronics housing, transmission case, rear axle hub, and engine. The long diagonal plate above the front axle hub impedes heat convection in that direction. As expected, the group of components in the primary engine compartment emits a greater heat load into the vehicle's wake region. The fan extracts high-temperature air that mixes with cooler air around the extremities of the compartment, pushing this air into the vehicle's wake. Again, the plates above the engine demonstrate heat conduction without air penetration across the boundaries.

\section{Conclusions}
\label{Sec:Conclusions}

We have presented a new method for immersogeometric fluid flow and heat transfer analysis that directly uses point cloud representation of objects. Analogous to previous work using analytic surface equations to generate surface Gaussian quadrature points, our method here repurposes point cloud elements into quadrature points. We employed computationally cheaper methods of normal vector and area generation using spatially local neighbors of points, proving that the results are appreciably accurate compared to full surface reconstruction. Prominent difficulties associated with geometry cleanup are avoided entirely, and the point cloud format trivializes immersed mesh generation using even basic meshing codes. Simulation results obtained using immersogeometric point clouds are in excellent agreement with reference solutions. Our method is painlessly applied to perform flow analysis of an incredibly complex industrial geometry, a large construction vehicle, incorporating moving point clouds and thermal fluid boundary conditions to demonstrate the utility of point-cloud IMGA in commercial and industrial applications.

\vspace{0.5in}

{
\small
\section*{Acknowledgments}
J. Khristy was partly supported by Deere~\&~Company as a part-time employee. M.-C. Hsu was partially supported by the National Heart, Lung, and Blood Institute of the National Institutes of Health under award number R01HL129077. A. Krishnamurthy was partially supported by the NSF Grant No.~LEAP-HI-2053760 and OAC-1750865. This support is gratefully acknowledged. We also thank the Texas Advanced Computing Center (TACC) at the University of Texas at Austin for providing high-performance computing resources that contributed to the results presented in this paper.
}

\vspace{0.5in}

\section*{Appendix}

\appendix 
\section{Definitions of Euler Jacobian and diffusivity matrices}\label{App:AppendixA}

In this appendix, we use the strong form of governing equations of compressible flows in 3D space (i.e. $d=3$) to illustrate the definitions of Euler Jacobian and diffusivity matrices. The solution variables of compressible flows can be written using conservation variables $\mathbf{U}$ or pressure-primitive variables $\mathbf{Y}$, defined as
\begin{align}
\label{conservation_var}
\mathbf{U} = 
\begin{bmatrix}
 \rho \\
 \rho u_1 \\
 \rho u_2 \\
 \rho u_3 \\
 \rho e \\
\end{bmatrix}\text{, and }
\mathbf{Y} = 
\begin{bmatrix}
 p \\
 u_1 \\
 u_2 \\
 u_3 \\
 T \\
\end{bmatrix}\text{,}
\end{align}
where $\rho$ is the density, $u_i$ is the $i^{\rm th}$ velocity component, $i=1,...,d$ with $d = 2$ or $3$ being the space dimension, $e$ is the specific internal energy, $p$ is the pressure, and $T$ is the temperature. The pressure, density, and temperature are related through the ideal gas equation of state, $p = \rho R T$, where $R$ is the ideal gas constant.
Furthermore, we assume a calorically perfect gas in this work and define the specific internal energy as $e = c_\text{v} T$, where $c_\text{v} = R/(\gamma-1)$ is the specific heat at constant volume, and $\gamma$ is the heat capacity ratio.
The governing equations of compressible flows can then be written as
\begin{align}
\label{ALE-conserve}
\mathbf{U}_{,t} +\mathbf{F}^{\text{adv}}_{i,i} +\mathbf{F}^{\text{sp}} - \mathbf{F}^{\text{diff}}_{i,i} - \mathbf{S} = \mathbf{0}  \text{,}
\end{align}where $\mathbf{F}^{\text{adv}}_{i}$ and $\mathbf{F}^{\text{diff}}_i$ are the vectors of convective and diffusive fluxes, respectively, defined as 
\begin{align}
\label{eq:convective_flux}
\mathbf{F}^{\text{adv}}_i =
\begin{bmatrix}
\rho u_i\\
\rho u_i u_1\\
\rho u_i u_2\\
\rho u_i u_3 \\
\rho  u_i e  \\
\end{bmatrix}+
\begin{bmatrix}
0\\
p\delta_{1i}\\
p\delta_{2i}\\
p\delta_{3i}\\
0\\
\end{bmatrix}\text{, and }
\mathbf{F}^\text{diff}_i= 
\begin{bmatrix}
0\\
\tau_{1i}\\
\tau_{2i}\\
\tau_{3i}\\
-q_i\\
\end{bmatrix}\text{.}
\end{align}
$\mathbf{F}^{\text{sp}}$ is the contribution of stress--power in the energy equation, defined as
\begin{align}
\label{eq:stress_power}
\mathbf{F}^\text{sp} = 
\begin{bmatrix}
0 \\
0 \\
0 \\
0\\
p u_{i,i} - \tau_{ij}u_{j,i}\\
\end{bmatrix}\text{,}
\end{align}
and $\mathbf{S}$ is the source term. In Eqs.~\eqref{eq:convective_flux}--\eqref{eq:stress_power}, $\delta_{ij}$ is the Kronecker delta, and $\tau_{ij}$ and $q_i$ are the viscous stress and heat flux, respectively, given by
\begin{align}
\tau_{ij} &= \lambda u_{k,k}\delta_{ij}+\mu\left(u_{i,j}+u_{j,i}\right)\text{,} \\
 q_i &= - \kappa T_{,i}\text{,}
\end{align}
where $\mu$ is the dynamic viscosity, $\lambda$ is the second coefficient of viscosity ($\lambda = - 2 \mu /3$ based on Stokes' hypothesis), and $\kappa$ is the thermal conductivity. We further split the convective flux into $\mathbf{F}^{\text{adv}}_{i} = \mathbf{F}^{\text{adv} \backslash p}_{i} +\mathbf{F}^p_{i}$, where $\mathbf{F}^{\text{adv} \backslash p}_{i}$ and $\mathbf{F}^p_{i}$ are the first and second terms, respectively, on the right-hand side of $\mathbf{F}^{\text{adv}}_{i}$'s definition in Eq.~\eqref{eq:convective_flux}. 

With these preliminaries being defined, the Euler Jacobian and diffusivity matrices can be defined as follows: $\hat{\mathbf{A}}_i = \dfrac{\partial \mathbf{F}^\text{adv}_i}{\partial \mathbf{U}}$, $\hat{\mathbf{A}}^\text{sp}_i$ is such that $\hat{\mathbf{A}}^\text{sp}_i \mathbf{U}_{,i} = \mathbf{F}^\text{sp}$, $\hat{\mathbf{K}}_{ij}$ is such that $\hat{\mathbf{K}}_{ij}\mathbf{U}_{,j} = \mathbf{F}^\text{diff}_i$, $\mathbf{A}_0 = \dfrac{\partial \mathbf{U}}{\partial \mathbf{Y}}$, $\mathbf{A}_i = \dfrac{\partial \mathbf{F}^\text{adv}_i}{\partial \mathbf{Y}} = \dfrac{\partial \mathbf{F}^\text{adv}_i}{\partial \mathbf{U}}\dfrac{\partial \mathbf{U}}{\partial \mathbf{Y}}=\hat{\mathbf{A}}_i\mathbf{A}_0$, $\mathbf{A}^\text{sp}_i$ is such that $\mathbf{A}^\text{sp}_i \mathbf{Y}_{,i} = \mathbf{F}^\text{sp}$, and $\mathbf{K}_{ij}$ is such that  $\mathbf{K}_{ij}\mathbf{Y}_{,j} = \mathbf{F}^\text{diff}_i$. Based on the splitting of $\mathbf{F}^{\text{adv}}_{i}$ into $\mathbf{F}^{\text{adv} \backslash p}_{i}$ and $\mathbf{F}^p_{i}$, we can further split $\mathbf{A}_i$ as  $\mathbf{A}_i = \mathbf{A}^{\text{adv} \backslash p}_i+\mathbf{A}^p_i$ to separate the pressure term from the convective flux. Detailed expressions for the matrices appearing in the quasi-linear forms can be found in Appendix A of~\citet{Xu2017c}. 

\clearpage
\bibliographystyle{unsrtnat}
\bibliography{refs-imga}

\end{document}